# The Theory of Heating of the Solar Corona and Launching of the Solar Wind by Alfvén Waves

## Albert Maksimov Varonov

*Department of Theoretical Physics, "St. Clement of Ohrid" University of Sofia*

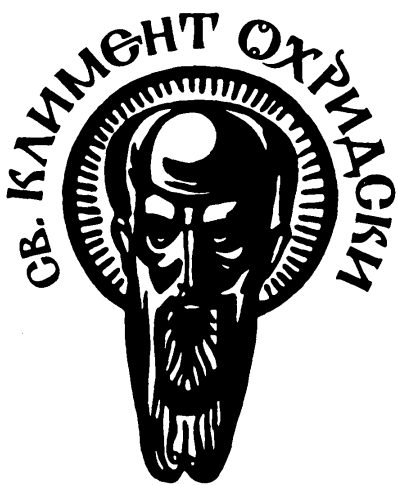

Thesis submitted for the degree of Doctor of Philosophy
of the "St. Clement of Ohrid" University of Sofia

## Jury

Assoc. Prof. Stanimir Kolev, PhD — Faculty of Physics, Sofia University
Prof. Vassil M. Vasilev, PhD — Institute of Mechanics, Bulgarian Academy of Sciences
Assoc. Prof. Bojidar Srebrov, PhD — National Institute of Geophysics, Geodesy and Geography, Bulgarian Academy of Sciences
Assoc. Prof. Dantchi Koulova, PhD — Institute of Mechanics, Bulgarian Academy of Sciences

### Thesis Advisor

Prof. Todor Mishonov, PhD, DSc — Faculty of Physics, Sofia University

Version: 19/07/2019

# CONTENTS

# **1**

# INTRODUCTION

The Sun, our nearest star, is a vital energy source for all living organisms on our planet Earth. Visible during clear sky day, our nearest star outshines every other celestial object. But very rarely it happens the Sun not to be visible during the day and then a day turns into a night for several minutes. The cause for this is a total solar eclipse – an event occurring when the Earth's only natural satellite the Moon passes between our planet and the Sun. The distance from the Earth to the Moon is approximately 400 Moon radii and the distance from the Earth to the Sun is approximately 400 solar radii too, meaning that the visible from the Earth angular diameters of both the Moon and the Sun are almost equal. This allows the Moon to cover the entire Sun for an Earth-based observer during a total solar eclipse (when the Moon is positioned between the Earth and the Sun). And during a total solar eclipse the solar atmosphere is revealed in the "night" sky around the eclipsed Sun. This is the only natural way to observe the solar atmosphere since its brightness is much lower than the brightness of both the solar photosphere and the daylight sky. As normal sunlight from the photosphere is blocked by the Moon and therefore no sunlight reaches the Earth to be scattered by its atmosphere – it's nighttime for a little while.

Now, let us briefly follow the long history of the discovery of our contemporary problem, which starts more than 200 years ago.

## 1.1 History of Solar Atmosphere Observations

Fraunhofer invented the spectroscope and was the first to observe the solar spectrum in total 574 lines [Fraunhofer (1814)] and a set of spectral lines are named Fraunhofer lines. Several decades later Kirchoff [Kirchoff (1859)] discovered that the Fraunhofer lines belong to chemical elements and together with Bunsen [Kirchoff & Bunsen (1860)] began studying the spectra of chemical elements in their laboratory.

Astronomers also quickly made use of the newly invented spectroscope, fitted it to their telescopes and began spectroscopic research in their vast laboratory, too. The most interesting and studied object in this laboratory was our nearest star, as it is nowadays too.

### 1.1.1 The bright green spectral line in the solar corona

During the total solar eclipse of 7 August 1869 W. Harkness and Charles Young discovered a bright green line designated 1474K on the used Kirchoff scale at that time in the solar corona, the outer layer of the solar atmosphere [Young (1869), Claridge (1937)]. Young identified the 1474 line to be an iron line put down by both Kirchoff and Angström but he did not believe in his discovery: *Should it turn out that this line in the aurora does actually coincide with 1474, it will be of interest to inquire whether we are to admit the existence of iron vapor in and above our atmosphere, or whether in the spectrum of iron this line owes its presence to some foreign substance, probably some occluded gas as yet unknown, and perhaps standing in relation to the magnetic powers of that metal.* [Young (1869), p. 378]. Later Young found out that this green line was actually a doublet and recognised that the less refrangible line coincides with a line from the iron spectrum that is visible only from a Leiden jar spark: *The more refrangible line is undoubtedly the real corona line, and the other belongs to the spectrum of iron, the close coincidence being merely accidental.* [Young (1876)]. Lieveing and Dewar compared their laboratory iron spectrum with Young's solar spectrum and also discovered that the iron line at 5316.07 Å corresponds with *with the less refrangible of the two solar lines at this place* [Lieveing & Dewar (1881)], however their work was not recognised [Claridge (1937)].

Subsequent measurements with constantly improving equipment showed that the more refrangible green line is at 5303 Å and in addition more coronal lines were discovered. A hypothetical element called "coronium" was introduced to explain the origin of these lines but the emphasis was put on the green line at 5303 Å [Claridge (1937)], which later Walter Grotrian found it is stronger than all other coronal lines combined [Grotrian (1933), Claridge (1937)]. Alongside the eclipse observations, Bernard Lyot constructed the first coronagraph and set it up at the Observatory of the Pic du Midi to study the solar corona, as shown in Fig. 1.1, without the need for total solar eclipses and the expeditions to observe them [Lyot (MNRAS 1939), Eddy (1979)]. The coronagraph allowed Lyot precisely and much more intensively to measure the green coronal line at 5302.86 Å [Lyot (1932)], many other coronal lines [Eddy (1979), Chap. 2] and even take films of solar prominences and filaments in 1935 [Lyot (1939), Lyot (MNRAS 1939)]. Grotrian found that Fe X ($Fe^{9+}$)[1] and Fe XI ($Fe^{10+}$) term separations determined by Bengt Edlén [Edlén (1937), Swings (1943)] coincided with two coronal lines: red at 6374 Å and 7892 Å [Grotrian (1939), Swings (1943), Edlén (1945)]. After this discovery, Edlén studied the forbidden lines of Fe XIII ($Fe^{12+}$), Fe XIV ($Fe^{13+}$), Ni XII ($Ni^{11+}$) and in 1941 announced his results – the bright green line in the solar corona spectrum is due to a (magneto-dipole) transition of Fe XIV[2] in the solar corona [Edlén (1942), Swings (1943), Edlén (1945)].

---

[1] Fe X, $^2P_{1/2} \rightarrow {}^2P_{3/2}$, $3s^23p^5$, ground state isoelectronic of Cl I or [Ar]$3p^{-1}$, red line.

[2] Fe XIV, $^2P_{3/2} \rightarrow {}^2P_{1/2}$, $3s^23p^1$, ground state isoelectronic of Al I, green line.

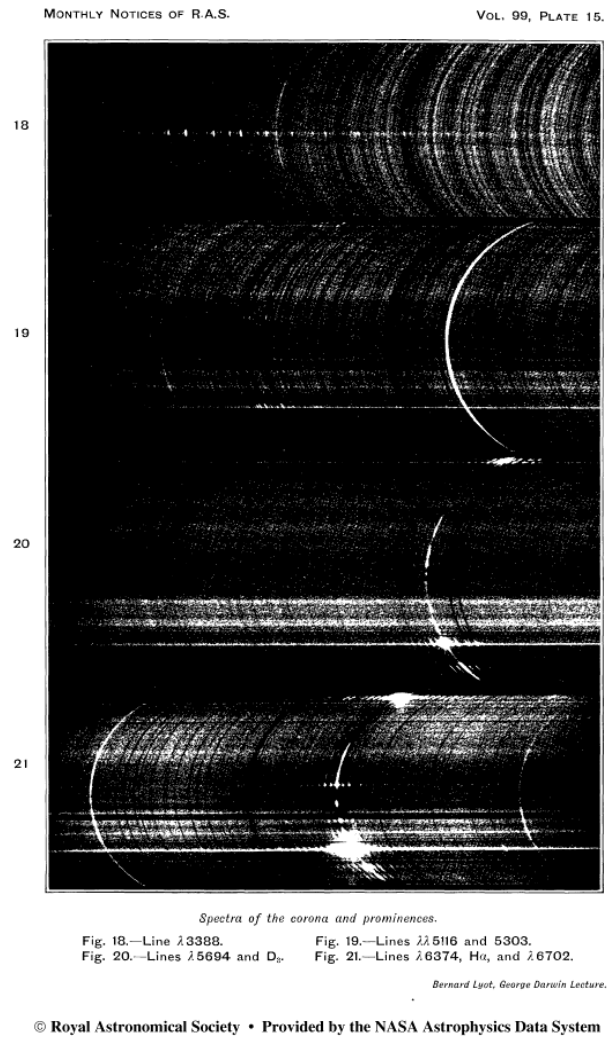

Figure 1.1: B. Lyot's plate from [Lyot (MNRAS 1939), Plate 15] showing spectra of the solar corona and prominences. All lines have the form of the slit, which is an arc. On the second row from top to bottom Ni XIII ($Ni^{12+}$) line at 5116 Å is barely visible on the left and the bright green coronal line of Fe XIV ($Fe^{13+}$) at 5303 Å is on the right. On the bottom row is the red part of the spectrum, from left to right the lines are 6374 Å of Fe X ($Fe^{9+}$), H$\alpha$ overexposed and 6702 Å.

The 72 years old problem had finally been resolved. The mysterious coronal green line, the strongest of all, is a result from a $Fe^{13+}$ magneto-dipole transition. There were so many hints for the solution, we may now even say that it was obvious but nevertheless, it took these wise and very skillful scientists 72 years to find the source of the bright green coronal line at 5303 Å.

But the solution of this 72 years old problem posed another even more difficult 77 years old problem. A problem from 1941 that until nowadays has no solution [Sakurai (2017)]. Stripping 13 electrons from an iron nucleus requires a temperature of million degrees. How come a several thousand degree (kK) solar photosphere produce a million-degree (MK) outer solar atmosphere? The importance and influence of this new problem to the future research at that time was beyond doubt: *Edlén's identification of the coronal lines has opened an immense new field in solar and stellar physics. In normal times Edlén's discovery would have already inspired many other theoretical investigations. There is not the slightest doubt that it will affect the whole orientation of solar research for years to come.* [Swings (1943)]. Let us remember that these cited words were written in 1943, when the World War II was raging across the globe and the Manhattan project, where many of the world's leading scientists worked, had already been started.

### 1.1.2   The new problem and its first proposed theoretical solution

Shortly after the discovery of Edlén, Hannes Alfvén suggested the existence of magnetohydrodynamic (MHD) waves and their importance in solar physics [Alfvén (1942)]. He was the first one to offer a theoretical explanation of the MK temperature of the solar corona: *A first theoretical attempt at explaining the origin of the high-energy coronal particles has recently been published by H. Alfvén* [Swings (1943)]. The first idea by Alfvén is that MHD waves, now called Alfvén waves (AW), created by the turbulence in the photosphere are transmitted upwards to the chromosphere and their damping in the inner corona produces the MK temperature of the solar corona [Alfvén (1947)]. The absorption proportional to $\omega^2$ of AW causes the heating and therefore the high frequency AW are absorbed and not observed in the corona.

Alfvén's idea for the viscous heating of plasma by absorption of AW was analyzed in the theoretical work by Heyvaerts [Heyvaerts & Priest (1983)]. In support of this idea is the work by Chitta [Chitta et al. (2012), Figs. 8 and 9]. The authors came to the conclusion that the spectral density of AW satisfies a power law with an index of 1.59. This gives a strong hint that this scaling can be extrapolated in the nearest spectral range for times less than 1 s and frequencies in the Hz range.

But despite these studies, again history repeated itself and Alfvén's theoretical explanation remained mostly unrecognised, as well as most of his work cf. [Dessler (1970), Alfvén (1988), Peratt (1988), Fälthammar & Dessler (1995)], until the launch of the Hinode spacecraft Fig. 1.2. The Hinode spacecraft[3] provided the necessary data Alfvènic type MHD modes to be observed [Tomczyk et al. (2007), De Pontieu et al. (2007), Okamoto et al. (2007, Katsukawa et al. (2007), Day (2009), Jess et al. (2009), Erdélyi & Fedun (2007)] and the well-forgotten spatially and temporally ubiquitous waves in the solar corona [Tomczyk et al. (2007)] came again into the limelight and gave strong support for the idea of Alfvén and the frequency of the notion AW as a keyword significantly increased. A clear presence of outward and inward propagating waves in the corona was noted and $k - \omega$ diagnostics revealed coronal wave power spectrum with an exponent of $\approx -3/2$ (cf. Fig. 2 of [Tomczyk & McIntosh (2009)]). The low frequency AW, on the other hand, reach the Earth orbit and thanks to the magnetometers on the various satellites we "hear" the basses of the great symphony of solar turbulence.

---

[3] https://hinode.msfc.nasa.gov/

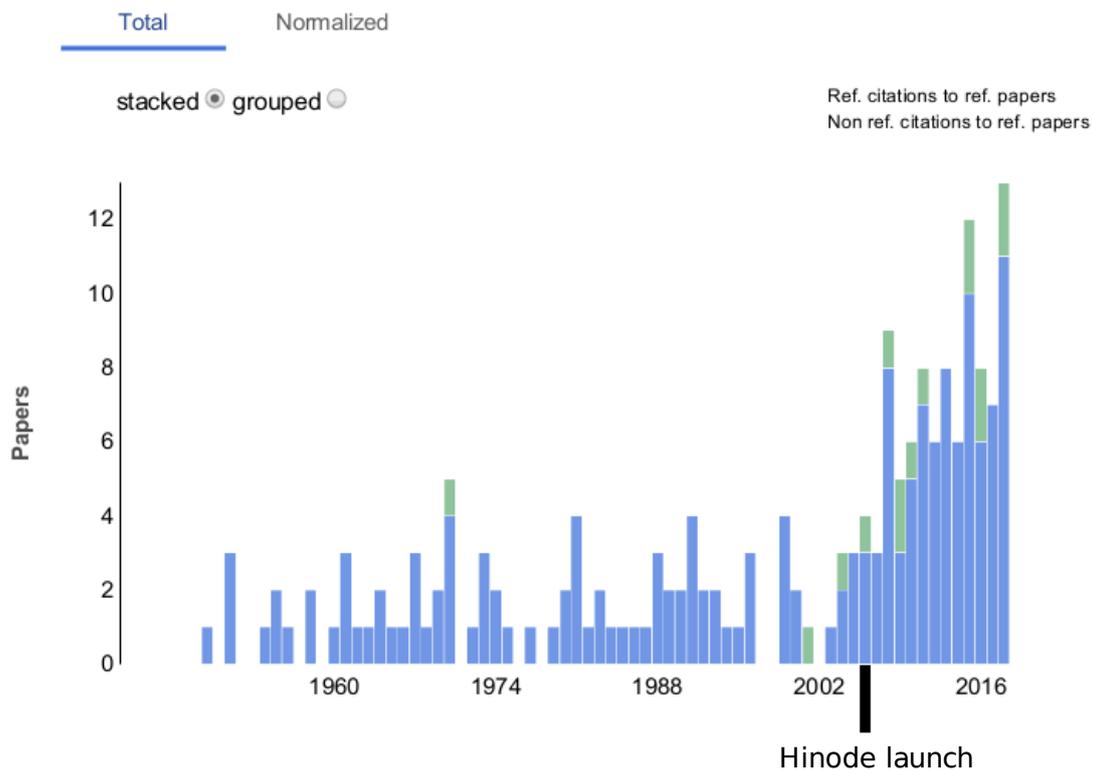

Figure 1.2: Papers citing of H. Alfvén's work [Alfvén (1947)] per year taken from the NASA Astrophysics Data System (ADS) here on 18 June 2018 with the year of the launch of Hinode added. Note the almost exponential increase of papers citing [Alfvén (1947)] after the launch of the Hinode spacecraft in 2006 and the first published results in 2007, cited in the text.

Later Hinode observations confirmed the damping of AW at low heights in the solar atmosphere [Hahn (2013), Hahn & Sawin (2014), Gupta (2017)], which is the last necessary ingredient to prove that Hannes Alfvén was right with his first theoretical idea of the heating of the solar corona, the same way Charles Young was right with his first idea of the origin of the bright coronal green line. AW are present in the solar atmosphere, have enough power to heat the solar corona [De Pontieu et al. (2007), Srivastava et al. (2017)] and are being absorbed meaning that Alfvén's predictions have already been confirmed, except for the only one – AW absorption heats up the solar corona to MK.

The latest space mission for studying the Sun the Parker Solar Probe [Fox et al. (2016)] launched in August 2018 is planned to approach the Sun within 10 solar radii. Its main scientific goals are to *understand the heating of the solar corona and to explore what accelerates the solar wind*,[4] meaning that the mechanism of the solar corona heating is still unknown.

Here it is demonstrated that Alfvén's last prediction is also confirmed, AW absorption do indeed heats up the solar corona. But before moving to the final solution to this problem, let us take a closer look at the composition and characteristics of the solar atmosphere.

### 1.1.3   The solar atmosphere

It is widely known that observations made from the Earth are limited by its atmosphere. In addition to weather, the atmospheric gases absorb the solar radiation with shorter wavelengths (most ultraviolet and X-ray) [Eddy (1979)], otherwise no life on our planet could arise. First attempts for solar studies in these "hidden" from ground observers wavelengths were made with balloons, planes and rockets. But detailed data were obtained with the first full scale manned astronomical observatory in space – the Apollo Telescope Mount (ATM) on Skylab [Eddy (1979), Doschek (1997)]. ATM contained eight solar instruments, which recorded data from X-ray to visible light on photographic film. The instrument that mainly revealed the region just below the MK solar corona is the Naval Research Laboratory ultraviolet spectrograph (ATM experiment S-082B) covering wavelengths between 970 Å and 3940 Å. A simplified schematic representation of the original optical system of this ultraviolet spectrograph [Bartoe et al. (1977)] is given in Fig. 1.3 with explanation of the instrument operation in the figure caption. At least 65 new emission lines were identified from ultraviolet spectra taken from Skylab, which was more than twice the number identified by that time from rocket exploration [Eddy (1979), Chap. 6]. These lines in details revealed the distinct parts of the solar atmosphere based on physical quantities, shown in Fig. 1.4 [Eddy (1979), Fig on page 2]. The ordinate axis is height, measured from the solar photosphere in red, where sunspots are seen. Above the photosphere is the chromosphere in orange, the transition region (TR) in dark orange and the corona in gray. The yellow and orange peaks are chromospheric spicules that permeate into the corona. There are two abscissa axes – a temperature one at the bottom and a density one at the top of the graphic. The continuous line shows the temperature and the dashed line shows the mass density. At the photosphere the temperature is about 6 kK and within the low chromosphere it begins to drop. After this initial drop, a sharp minimum occurs, followed by a rise and a roughly constant temperature in the upper chromosphere. Entering the thin transition region, the temperature increases more than 10 times in step-like manner. The temperature increase continues in the solar corona, however more gradually reaching MK temperatures. Evidently from the structure of the solar atmosphere, the key to the long standing solar corona heating problem is "hidden" in the tiny solar TR.

---

[4] http://parkersolarprobe.jhuapl.edu/index.php

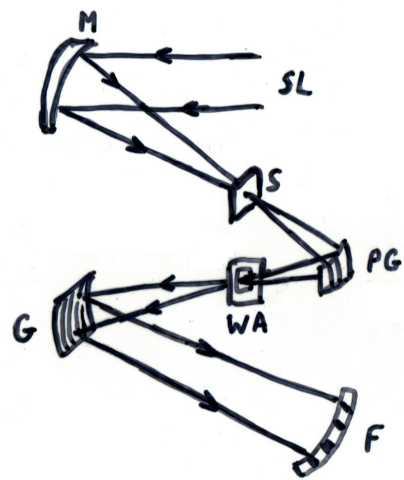

Figure 1.3: A simplified schematic diagram of the ultraviolet spectrograph ATM experiment S-082B on Skylab adopted from the original paper [Bartoe et al. (1977)]. Sunlight SL is reflected from the telescope off-axis paraboloid mirror M to a slit plate S (some of the light is reflected to a pointing reference system, which is not drawn here). The passing sunlight through the slit is directed to one of two predisperser grating PG. The produced spectrum by the PG passes through waveband aperture WA to block parts of the spectrum outside the desired range. Finally the spectrum that has passed through the WA is dispersed by a main grating G and is focused on a photographic film F.

## 1.1.4 The solar transition region

Referring several times to a tiny in width transition region up to now, it is time to reveal what this means. In Fig. 1.5 [Eddy (1979), Fig on page 36] our planet Earth is placed across the solar disk to scale the solar atmosphere. The width of the tiny TR is of the order of the 1970's linear size of metropolitan Los Angeles, which nowadays should be roughly equal to the linear size of Sofia.

Since the time of Skylab, there have been many more experimental data processing and new unmanned solar space telescopes observing the Sun, for instance NASA's SDO[5] (Solar Dynamics Observatory) and IRIS mission[6] (Interface Region Imaging Spectrograph). The calculated height dependent temperature profiles of the solar atmosphere from observations after Skylab do not differ too much between each other [Gabriel (1976), Withbroe & Noyes(1977), Vernazza et al.(1981), Nicholas et al. (1981), Mariska (1992), Golub & Pasachoff (1997), Dermendjiev (1997),Golub & Pasachoff (2002),Peter (2004),Aschwanden (2005),Avrett & Loeser (2008),Tian et al. (2010)], the width of the TR is still of the order of few kilometres. A single day walking distance compared to the Earth's radius is negligible and therefore compared to the Sun's radius is vanishingly diminutive. Yet, there is the key to the 77-years old problem of the heating of the solar corona.

---

[5]https://sdo.gsfc.nasa.gov/

[6]https://www.nasa.gov/mission_pages/iris/index.html

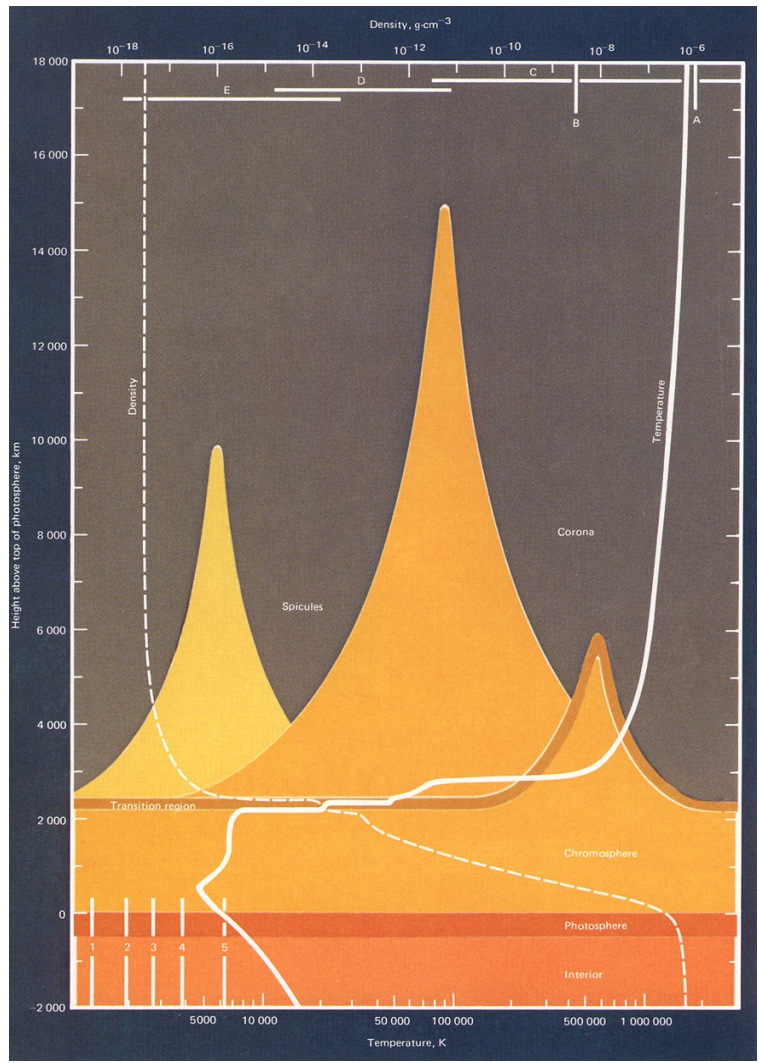

Figure 1.4: Height dependent detailed structure of the solar atmosphere with temperature and density [Eddy (1979), Fig on page 2]. Height in kilometres increases upward (ordinate scale), measured from the top of the photosphere. Continuous line shows temperature with corresponding bottom abscissa scale and dashed line shows mass density with corresponding upper abscissa scale. The distinct parts of the solar atmosphere are shown in different colors: the chromosphere is in yellow, the transition region is in dark yellow and the corona is in gray. Note the abrupt temperature rise more than 10 times in the very thin transition region.

### 1.1.5 Other theoretical proposals

The calamity of the ideas for the solar corona heating still on the arena are a lot, a search in the NASA astrophysics data system[7] for the expression "coronal heating" between 1947–2019 (21.06 to be exact) results in 9924 documents, while the expression "solar corona heating" matches a little bit less – 7081 documents. The same searches performed in ordinary Internet (Google search engine) yield correspondingly approximately 145 000 and 6120. But all these documents have one thing in common – in none of them the temperature height profile has been calculated based on physics (or first) principles and therefore no estimate of the TR width has been made, which is the main purpose of the current research.

---

[7] https://ui.adsabs.harvard.edu/

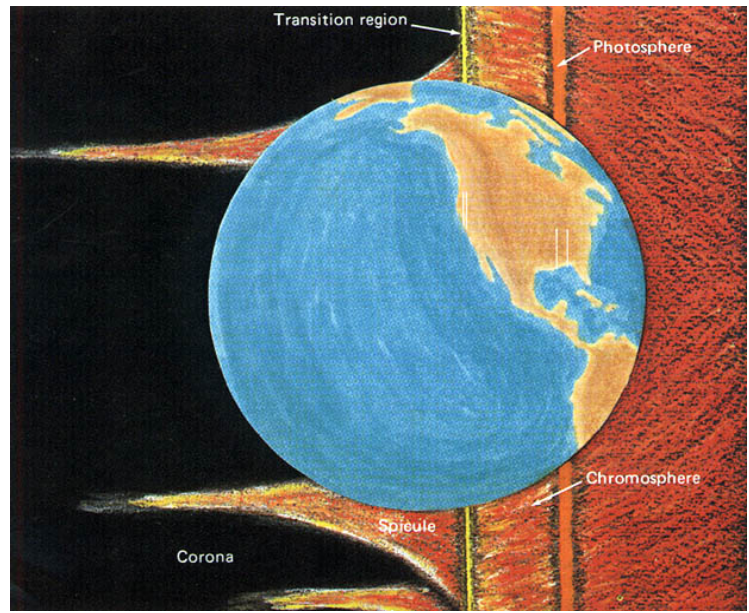

Figure 1.5: Our tiny American-oriented Earth as a yardstick to scale the solar atmosphere [Eddy (1979), Fig on page 36]. The photosphere (orange layer) is about 400 km thick, corresponding to the Alabama state width, the chromosphere (red-orange layer) spans several thousand kilometres, corresponding to the Alabama – Los Angeles distance on the Earth scale and the tiny transition region corresponds in width to metropolitan Los Angeles. Spicules are roughly with the size of the Earth's diameter, while the corona is too large for this terrestrial scale used; it spans several solar radii outwards.

A thorough review of the proposed solar corona heating mechanisms can be found in [Aschwanden (2005), chap. 9], while a brief one in [Erdélyi & Ballai (2007)]. In general, the coronal heating models are divided into DC (Direct Current) and AC (Alternating Current) types, *which characterize the electromechanic coronal response to the photospheric driver that provides the ultimate energy source for heating* [Aschwanden (2005), sec. 9.2]. The proposed physical processes that both heating model types are based on are more than 10 [Aschwanden (2005), chap. 9] and here we will mention only the most popular ones.

Magnetic reconnection was one of the most (if not the most) popular heating mechanisms of the solar corona [Aschwanden (2005), chap. 10]. This is a process occurring in solar flares – large explosions in the solar atmosphere caused by rearrangement of magnetic field lines that release lots of energy in the form of accelerated ionised particles and radiation in almost all wavelengths.

A solar flare indeed heats its surroundings, however, it is an occasional event with varying magnitude, while the TR exists on a permanent basis around the Sun, meaning that it cannot sustain a continuous supply for the heating of the whole TR. A solar flare is often followed by a coronal mass ejection launching solar plasma through the interplanetary space [Srebrov (2003)]. The absence of regular solar flares led further to the hypothesis that smaller flares, classified as microflares and nanoflares [Parker (1988)] to be discussed as a possible heating mechanism but today it seems they cannot provide enough energy [Sakurai (2017)]. These and other eruptive solar processes are described in [Filippov (2007)], where an equilibrium loss is explained with mathematical catastrophe theory (Subsec. 4.3.3). One of the latest proposed heating scenarios is a strong preferential ion heating from the principal investigators of the Parker Solar Probe but unfortunately its heating mechanism is still unspecified according to its authors [Kasper & Klein (2019)].

There are theoretical calculations that come up with a height dependent temperature profile, which is simulated to match the observational ones. For instance, in [Hansteen & Leer(1995), Eq. (32)] a coronal heating function, which is a coronal energy input consistent with a mechanical flux is used. The authors also discuss that because the lack of knowledge of the coronal heating process, they adopt a certain distribution of the heating functions to protons and electrons as a reference. Similar approach is used in [Pinto & Rouillard (2017), Eq. (6)], where it is discussed that the actual heating processes are still under debate. A kinetic treatment also makes use of model heating functions [Lie-Svendsen & Esser (2005), Eq. (22)], which are no consequence from fundamental physical laws and trying to avoid numerical problems, the heating is ramped up gradually. Authors discuss that the ion heating mechanism is unknown and there is no explanation for the casting of the heating function. In another study authors use a prescribed external energy flux to form a TR [Grappin et al. (2010)] and in order to prevent possibly insufficient damping in the shock region below the TR, an additional kinematic viscosity is used. A detailed temperature height profile of the TR is present in [Cranmer et al. (2007)] but the authors have adopted a phenomenological form for the MHD turbulence damping rate and finally conclude that *future work must involve more physical realism for the models, expanded comparisons with existing observations, and predictions of as-yet unobserved quantities that may be key discriminators between competing theoretical models*. In order to fit the calculated temperature profile to observational data in full neglect of the viscous friction taking only Spitzer Ohmic resistivity as a dissipation mechanism, it is necessary to add to the *ab initio approach* artificial velocity drivers, which is one of the main goals by [Gudiken (2004), Gudiksen & Nordlund (2005)]. There are numerous works like the aforementioned ones and we can continue this discussion almost forever but now again we return to the first theoretical idea of Hannes Alfvén.

### 1.1.6 The theoretical solution continued

Any idea attempting to explain the heating of the solar corona must include a calculation (or at least an estimate calculation) of the width of the solar TR $\lambda$. In order to qualitatively explain this tiny width of the TR, the idea of self-induced opacity of the plasma for AW was introduced [Mishonov et al. (2007)]. Within an MHD calculation the width of the solar TR $\lambda$ is given by [Mishonov et al. (2011)]

$$\frac{1}{\lambda} = \max_x \frac{1}{T}\frac{\mathrm{d}T}{\mathrm{d}x} \equiv \max_x \frac{\mathrm{d}}{\mathrm{d}x}\langle \ln T(x) \rangle, \tag{1.1}$$

where $T$ is the temperature in energy units and $x$ is height of the solar atmosphere. Using contemporary computing systems, the maximum reached heating of the solar corona was "only" 30 times with respect to the temperature of the photosphere [Zahariev and Mishonov (2010), Mishonov et al. (2011), Topchiyska et al. (2013), Mishonov et al. (2015)], which is just 2-3 times short of the measured one. The current thesis is a continuation of this more than a decade lasting research with some new effects taken into account and most importantly: a final solution to the problem of self-induced opacity of AW heating the solar corona, which is precisely the first theoretical idea of Hannes Alfvén .

A similar idea was also analysed by [Suzuki & Inutsuka (2005), Suzuki (2008), Shoda et al. (2018)] but both calculations go up to 100 solar radii away from the Sun therefore no details of the TR are available. There are lots of other similar papers [Davila (1987), Ofman & Davila (1995), Ofman & Davila (1998), Nakariakov et al. (2000), Ofman (2010), Farahani et al. (2012)] for instance, however, none of these is helpful for comparison with our work because there is no height dependent temperature profile in any of them.

The remainder of this chapter is devoted to a thorough theoretical introduction in MHD, starting from hydrodynamics and a brief dimensional introduction in Kinetics, all of them needed for the solution of the problem of the heating of the solar corona.

## 1.2 Hydrodynamics

This section is devoted to the basic theory of hydrodynamics or fluid mechanics – the theory of motion of liquids and gases.

### 1.2.1 Equation of motion in hydrodynamics

We start our theoretical treatment with the equation of continuity in hydrodynamics [Landau & Lifshitz (1987), Eq. (1.2)]

$$\partial_t \rho + \mathrm{div}(\rho \mathbf{v}) = 0, \quad \partial_t \equiv \frac{\partial}{\partial t}, \tag{1.2}$$

where $\rho$ is the mass density and $\mathbf{v}$ is the velocity of the fluid. Next we include the equation of motion of the fluid

$$\partial_t(\rho v_i) = -\partial_k \Pi_{ik}, \quad \partial_k \equiv \frac{\partial}{\partial x_k}, \tag{1.3}$$

where $\Pi_{ik}$ is the momentum flux density tensor of the fluid [Landau & Lifshitz (1987), Sec. 15], which consists of an ideal part $\Pi_{ik}^{(\mathrm{id})}$ and viscous part $\Pi_{ik}^{(\mathrm{visc})}$, which is equal to the viscous stress tensor with opposite sign [Landau & Lifshitz (1987), Eq. (15.3)]

$$\Pi_{ik} = \Pi_{ik}^{(\text{id})} + \Pi_{ik}^{(\text{visc})}, \tag{1.4}$$

$$\Pi_{ik}^{(\text{id})} = \rho v_i v_k + P \delta_{ik}, \tag{1.5}$$

$$\Pi_{ik}^{(\text{visc})} = -\eta \left( \partial_i v_k + \partial_k v_i - \frac{2}{3} \delta_{ik} \partial_l v_l \right) - \zeta \delta_{ik} \partial_l v_l, \tag{1.6}$$

where $P$ is the fluid pressure, $\eta$ is dynamic or shear viscosity, $\zeta$ is second or bulk viscosity and $\delta_{ik}$ is the Kronecker delta. The divergence of the ideal momentum flux density tensor is

$$\partial_k \Pi_{ik}^{(\text{id})} = v_i \partial_k (\rho v_k) + \rho v_k \partial_k (v_i) + \partial_i P, \tag{1.7}$$

and substituting it in Eq. (1.3), we obtain the Euler equation in ideal hydrodynamics [Landau & Lifshitz (1987), Eq. (2.3)]

$$(\partial_t + \mathbf{v} \cdot \nabla) \mathbf{v} = -\frac{\nabla P}{\rho}. \tag{1.8}$$

The divergence of the viscous momentum flux tensor is

$$\partial_k \Pi_{ik}^{(\text{visc})} = -\partial_k \left\{ \eta \left( \partial_i v_k + \partial_k v_i - \frac{2}{3} \delta_{ik} \partial_l v_l \right) \right\} - \partial_i \left( \zeta \partial_l v_l \right), \tag{1.9}$$

after differentiating the left hand side, the equation of motion Eq. (1.3) can be written in the form

$$v_i \partial_t \rho + \rho \partial_t v_i = -\partial_k \Pi_{ik}^{(\text{id})} - \partial_k \Pi_{ik}^{(\text{visc})}. \tag{1.10}$$

Substitution of the divergences of the ideal Eq. (1.7) and non-ideal momentum flux density tensor Eq. (1.9) gives

$$v_i [\partial_t \rho + \partial_k (\rho v_k)] + \rho (\partial_t + v_k \partial_k) v_i = -\partial_i P + \partial_k \left\{ \eta \left( \partial_i v_k + \partial_k v_i - \frac{2}{3} \delta_{ik} \partial_l v_l \right) \right\} + \partial_i \left( \zeta \partial_l v_l \right), \tag{1.11}$$

after some rearrangement. The first two terms on left hand side cancel each other by means of the equation of continuity Eq. (1.2), where $\partial_t \rho = -\partial_k (\rho v_k)$ and in this way we obtain the most general form of the equation of motion of a viscous fluid [Landau & Lifshitz (1987), Eq. (15.5)]

$$\rho \mathrm{d}_t v_i = -\partial_i P + \partial_k \left\{ \eta \left( \partial_i v_k + \partial_k v_i - \frac{2}{3} \delta_{ik} \partial_l v_l \right) \right\} + \partial_i \left( \zeta \partial_l v_l \right), \tag{1.12}$$

where

$$\mathrm{d}_t \equiv \frac{\mathrm{d}}{\mathrm{d}t} \equiv \partial_t + \mathbf{v} \cdot \nabla \equiv \partial_t + v_k \partial_k \tag{1.13}$$

is the substantial time derivative [Landau & Lifshitz (1987), Eq. (2.2)].

### 1.2.2 Energy flux in hydrodynamics

First we will consider the energy flux of an ideal fluid, that is a fluid whose thermal conductivity and viscosity are negligible.

The absence of viscous friction and heat exchange means that the motion of an ideal fluid is adiabatic or the entropy of any particle of the fluid remains constant. Therefore for the entropy per unit mass $s$ we have the condition

$$\mathrm{d}_t s \equiv \partial_t s + \mathbf{v} \cdot \nabla s = 0. \tag{1.14}$$

Multiplying this equation with $\rho$, adding and subtracting the term $s\partial_t \rho$, we get

$$\rho \partial_t s + s \partial_t \rho - s \partial_t \rho + \rho \mathbf{v} \cdot \nabla s = 0. \tag{1.15}$$

The first two terms are total derivative, the third one can be substituted from the equation of continuity Eq. (1.2), which alongside the last term on the left hand side form a total derivative again and in this way, the "equation of continuity" for entropy is obtained [Landau & Lifshitz (1987), Eq. (2.7)]

$$\partial_t(\rho s) + \mathrm{div}(\rho s \mathbf{v}) = 0. \tag{1.16}$$

Let us now recall some necessary and important thermodynamic relations. The differential of the energy $E$ of a body is [Landau & Lifshitz (1980), Eq. (12.2)]

$$\mathrm{d}E = T\mathrm{d}S - P\mathrm{d}V, \tag{1.17}$$

where $T$ is the temperature, $S$ is the entropy and $V$ is the volume of the body. Let us add the last term on the right hand side to a total differential and move it to the left hand side to get

$$\mathrm{d}(E + PV) \equiv \mathrm{d}H = T\mathrm{d}S + V\mathrm{d}P, \tag{1.18}$$

where the quantity $H \equiv E + PV$ is enthalpy or heat function of the body [Landau & Lifshitz (1980), Eqs.(14.2), (14.3)]. Now we divide both equations Eq. (1.17) and Eq. (1.18) by a unit mass to obtain equations for the differentials of the energy $\varepsilon$ and enthalpy per unit mass $h$ [Landau & Lifshitz (1987), Sec. 6]

$$\mathrm{d}\varepsilon = T\mathrm{d}s - P\mathrm{d}\left(\frac{1}{\rho}\right) = T\mathrm{d}s + \frac{P}{\rho^2}\mathrm{d}\rho. \tag{1.19}$$

$$\mathrm{d}h \equiv \mathrm{d}\left(\varepsilon + \frac{P}{\rho}\right) = T\mathrm{d}s + \frac{\mathrm{d}P}{\rho}. \tag{1.20}$$

The energy of a unit volume of a fluid consists of its kinetic energy of the motion $\rho v^2/2$ ($v \equiv |\mathbf{v}|$) and its internal energy per unit volume $\rho\varepsilon$. The change in time of this energy is

$$\partial_t \left(\frac{1}{2}\rho v^2 + \rho\varepsilon\right). \tag{1.21}$$

The time differentiation of the fluid kinetic energy is

$$\partial_t \left( \frac{1}{2} \rho v^2 \right) = \frac{1}{2} v^2 \partial_t \rho + \rho \mathbf{v} \cdot \partial_t \mathbf{v}. \tag{1.22}$$

The time derivative of the mass density is substituted from the continuity equation Eq. (1.2) $\partial_t \rho = -\nabla \cdot (\rho \mathbf{v})$ and the time derivative of the velocity is substituted from equation of motion Eq. (1.12), omitting the viscous terms $\rho \partial_t \mathbf{v} = -\rho \mathbf{v} \cdot \nabla \mathbf{v} - \nabla P$. In this way we obtain

$$\partial_t \left( \frac{1}{2} \rho v^2 \right) = -\frac{1}{2} v^2 \nabla \cdot (\rho \mathbf{v}) - \mathbf{v} \cdot (\rho \mathbf{v} \cdot \nabla \mathbf{v} - \nabla P) = -\frac{v^2}{2} \nabla \cdot (\rho \mathbf{v}) - \rho \mathbf{v} \cdot \nabla \left( \frac{v^2}{2} \right) - \mathbf{v} \cdot \nabla P. \tag{1.23}$$

The first two terms form a total derivative and the pressure gradient term is expressed from the equation for the differential of the enthalpy per unit mass Eq. (1.20) $\nabla P = \rho \nabla h - \rho T \nabla s$ and substituted in the time derivative of the kinetic energy

$$\partial_t \left( \frac{1}{2} \rho v^2 \right) = -\nabla \cdot \left( \rho \mathbf{v} \frac{v^2}{2} \right) - \rho \mathbf{v} \cdot \nabla h + \rho T \mathbf{v} \cdot \nabla s. \tag{1.24}$$

The time differentiation of the fluid internal energy is

$$\partial_t (\rho \varepsilon) = \varepsilon \partial_t \rho + \rho \partial_t \varepsilon. \tag{1.25}$$

Using $\partial_t \rho = -\nabla \cdot (\rho \mathbf{v})$ from the equation of continuity Eq. (1.2), $\varepsilon = h - p/\rho$ from Eq. (1.20) and

$$\partial_t \varepsilon = T \partial_t s + \frac{P}{\rho^2} \partial_t \rho \tag{1.26}$$

from the equation for the differential energy Eq. (1.19) and substituting them in the time derivative of the internal energy, we obtain

$$\partial_t (\rho \varepsilon) = -h \nabla \cdot (\rho \mathbf{v}) - \frac{P}{\rho} \partial_t \rho + \rho T \partial_t s + \frac{P}{\rho} \partial_t \rho = -h \nabla \cdot (\rho \mathbf{v}) + \rho T \partial_t s. \tag{1.27}$$

Combining the time derivatives of the kinetic energy Eq. (1.24) and internal energy Eq. (1.27), the total change in time of the energy of an ideal fluid is

$$\partial_t \left( \frac{1}{2} \rho v^2 + \rho \varepsilon \right) = -\nabla \cdot \left( \rho \mathbf{v} \frac{v^2}{2} \right) - \rho \mathbf{v} \cdot \nabla h + \rho T \mathbf{v} \cdot \nabla s - h \nabla \cdot (\rho \mathbf{v}) + \rho T \partial_t s. \tag{1.28}$$

The second and fourth terms on the right hand side form the total derivative $-\nabla \cdot (\rho \mathbf{v} h)$, while the third and fifth terms could be rearranged in $\rho T (\partial_t s + \mathbf{v} \cdot \nabla s)$, where it is evident that they cancel each other because of the constant entropy $s$ condition Eq. (1.14). The energy conservation law for an ideal fluid is

$$\partial_t \left( \frac{1}{2} \rho v^2 + \rho \varepsilon \right) = -\nabla \cdot \left[ \rho \mathbf{v} \left( \frac{v^2}{2} + h \right) \right] = -\nabla \cdot \mathbf{q}, \quad \mathbf{q} = \rho \mathbf{v} \left( \frac{v^2}{2} + h \right), \tag{1.29}$$

where the $\mathbf{q}$ is the energy flux density vector [Landau & Lifshitz (1987), Eqs.(6.1), (6.3)].

Let us now consider the time derivative of the energy of a viscous fluid. The entropy $s$ naturally increases due to the friction processes taking place in the fluid and the condition Eq. (1.14) does not hold any more. The time derivative of the internal energy for an ideal fluid Eq. (1.27) is the same but the time derivative of the kinetic energy Eq. (1.24) is not. For clarity let us rewrite Eq. (1.22) again

$$\partial_t \left(\frac{1}{2}\rho v^2\right) = \frac{1}{2}v^2\partial_t\rho + \rho\mathbf{v}\cdot\partial_t\mathbf{v}.$$

Again, we substitute $\partial_t\rho = -\nabla(\rho\mathbf{v})$ from the continuity equation Eq. (1.2) and $\rho\partial_t\mathbf{v}$ from the equation of motion Eq. (1.12) with the viscous terms. We can write the equation of motion Eq. (1.12) in the form

$$\rho(\partial_t + \mathbf{v}\cdot\nabla)\mathbf{v} = -\nabla P - \nabla\cdot\mathbf{\Pi}^{(\text{visc})}, \tag{1.30}$$

and express the time derivative of the velocity

$$\rho\partial_t\mathbf{v} = -\rho\mathbf{v}\cdot\nabla\mathbf{v} - \nabla P - \nabla\cdot\mathbf{\Pi}^{(\text{visc})}. \tag{1.31}$$

In this way for the time derivative of the kinetic energy of a viscous fluid we obtain

$$\partial_t\left(\frac{1}{2}\rho v^2\right) = -\frac{1}{2}v^2\nabla\cdot(\rho\mathbf{v}) - \mathbf{v}\cdot(\rho\mathbf{v}\cdot\nabla\mathbf{v} - \nabla P - \nabla\cdot\mathbf{\Pi}^{(\text{visc})}). \tag{1.32}$$

The first three terms are rearranged in exactly the same way as in the case of an ideal fluid

$$\partial_t\left(\frac{1}{2}\rho v^2\right) = -\nabla\cdot\left(\rho\mathbf{v}\frac{v^2}{2}\right) - \rho\mathbf{v}\cdot\nabla h + \rho T\mathbf{v}\cdot\nabla s - \mathbf{v}\cdot(\nabla\cdot\mathbf{\Pi}^{(\text{visc})}), \tag{1.33}$$

while the last term with the dissipative momentum flux density tensor is expressed as a total derivative

$$\partial_t\left(\frac{1}{2}\rho v^2\right) = -\nabla\cdot\left(\rho\mathbf{v}\frac{v^2}{2}\right) - \rho\mathbf{v}\cdot\nabla h + \rho T\mathbf{v}\cdot\nabla s - \nabla\cdot(\mathbf{v}\cdot\mathbf{\Pi}^{(\text{visc})}) + \mathbf{\Pi}^{(\text{visc})}\cdot\nabla\mathbf{v}. \tag{1.34}$$

Adding the expressions for the time derivatives of the kinetic energy per unit volume of a viscous fluid Eq. (1.34) and the internal energy per unit volume Eq. (1.27) and rearranging the divergence terms together, we obtain

$$\partial_t\left(\frac{1}{2}\rho v^2 + \rho\varepsilon\right) = -\nabla\cdot\left[\rho\mathbf{v}\left(\frac{v^2}{2} + h\right) + \mathbf{v}\cdot\mathbf{\Pi}^{(\text{visc})}\right] + \rho T(\partial_t + \mathbf{v}\cdot\nabla)s + \mathbf{\Pi}^{(\text{visc})}\cdot\nabla\mathbf{v}. \tag{1.35}$$

A comparison with the energy conservation law of an ideal fluid Eq. (1.29) shows that the energy conservation law of a viscous fluid is

$$\partial_t\left(\frac{1}{2}\rho v^2 + \rho\varepsilon\right) = -\nabla\cdot\mathbf{q}, \quad \mathbf{q} = \rho\mathbf{v}\left(\frac{v^2}{2} + h\right) + \mathbf{v}\cdot\mathbf{\Pi}^{(\text{visc})}, \tag{1.36}$$

where the term $\mathbf{v} \cdot \mathbf{\Pi}^{(\mathrm{visc})}$ is the flux due to internal friction processes [Landau & Lifshitz (1987), Sec. 49] and the equation for evolution of entropy is

$$\rho T(\partial_t + \mathbf{v} \cdot \nabla)s = -\Pi_{ik}^{(\mathrm{visc})} \partial_k v_i. \tag{1.37}$$

Finally we must take into account the heat conduction within the fluid. The energy flux density due to thermal conduction within the fluid always flows from places with higher to places with lower temperatures. Therefore the thermal conduction energy flux given by $-\varkappa \nabla T$, where $\varkappa$ is the thermal conductivity of the fluid, is added to the energy conservation law Eq. (1.36)

$$\partial_t \left(\frac{1}{2}\rho v^2 + \rho\varepsilon\right) = -\mathrm{div}\,\mathbf{q}, \quad \mathbf{q} = \rho\mathbf{v}\left(\frac{v^2}{2} + h\right) + \mathbf{v} \cdot \mathbf{\Pi}^{(\mathrm{visc})} - \varkappa\nabla T, \tag{1.38}$$

and to the equation for evolution of entropy Eq. (1.37)

$$\rho T(\partial_t + \mathbf{v} \cdot \nabla)s = -\Pi_{ik}^{(\mathrm{visc})} \partial_k v_i + \nabla \cdot (\varkappa\nabla T). \tag{1.39}$$

Eq. (1.38) is the general law of conservation of energy [Landau & Lifshitz (1987), Eq. (49.2)] and Eq. (1.39) is the general equation of heat transfer [Landau & Lifshitz (1987), Eq. (49.4)], where the first term is the dissipated energy through viscosity, and the second term is the conducted heat within the volume in interest.

## 1.3    Magneto-hydrodynamics

If a fluid is conductive and moves in a magnetic field, this movement will induce electric currents in the fluid. The magnetic fields of these electric currents change the external magnetic field and in this way a complex interaction between hydrodynamic and magnetic phenomena arises.

The magnetic force per unit volume $\mathbf{f}$ acting on the conductive fluid can be calculated by taking the divergence of the Maxwell stress tensor $\mathbf{\Pi}^{(\mathrm{Maxw})}$ [Landau & Lifshitz (1957), Eq. (15.2)]

$$f_i = -\partial_k \Pi_{ik}^{(\mathrm{Maxw})}, \quad \Pi_{ik}^{(\mathrm{Maxw})} = -\frac{1}{\mu_0}\left(B_i B_k - \frac{1}{2}\delta_{ik}B^2\right), \tag{1.40}$$

where $\mu_0$ is the vacuum permeability and the $\Pi_{ik}^{(\mathrm{Maxw})}$ only for external magnetic field is written in SI units. A conversion to CGS units is easily achieved by replacing $\mu_0$ with $4\pi$, [Landau & Lifshitz (1971), Eq. (33.3)] in the case of zero electric fields. The divergence of Maxwell stress tensor is

$$\partial_k \Pi_{ik}^{(\mathrm{Maxw})} = -\frac{1}{\mu_0}\left(B_k\partial_k B_i - \frac{1}{2}\partial_i B^2\right) = -\frac{1}{\mu_0}\left((\mathbf{B} \cdot \nabla)\mathbf{B} - \frac{1}{2}\nabla B^2\right), \tag{1.41}$$

where we have used Gauss's law

$$\nabla \cdot \mathbf{B} = 0. \tag{1.42}$$

Using the well-known formula from vector analysis

$$\mathbf{B} \times (\nabla \times \mathbf{B}) = \frac{1}{2}\nabla B^2 - (\mathbf{B} \cdot \nabla)\mathbf{B}, \tag{1.43}$$

we have

$$\partial_k \Pi_{ik}^{(\text{Maxw})} \equiv \nabla \cdot \mathbf{\Pi}^{(\text{Maxw})} = \frac{1}{\mu_0}\mathbf{B} \times (\nabla \times \mathbf{B}), \tag{1.44}$$

and for the magnetic force per unit volume we obtain

$$\mathbf{f} = -\nabla \cdot \mathbf{\Pi}^{(\text{Maxw})} = -\frac{1}{\mu_0}\mathbf{B} \times (\nabla \times \mathbf{B}). \tag{1.45}$$

## 1.3.1 Equation of motion in magneto-hydrodynamics

In magneto-hydrodynamics (MHD) the volume density of the Lorentz force $\mathbf{f}$ has to be added to the equation of motion of the fluid in ordinary hydrodynamics Eq. (1.12)

$$\rho \mathrm{d}_t v_i = -\partial_i P + \partial_k \left\{ \eta \left( \partial_i v_k + \partial_k v_i - \frac{2}{3}\delta_{ik}\partial_l v_l \right) \right\} + \partial_i \left( \zeta \partial_l v_l \right) - \frac{1}{\mu_0}[\mathbf{B} \times (\nabla \times \mathbf{B})], \tag{1.46}$$

which can also be written in more compact form using Eq. (1.30), Eq. (1.40) and according to Eq. (1.3)

$$\rho(\partial_t + \mathbf{v} \cdot \nabla)\mathbf{v} = -\nabla P - \nabla \cdot \mathbf{\Pi}^{(\text{visc})} - \nabla \cdot \mathbf{\Pi}^{(\text{Maxw})}. \tag{1.47}$$

Here it is useful to write down the total momentum flux density [Landau & Lifshitz (1957), Eq. (51.8)]

$$\mathbf{\Pi} = \mathbf{\Pi}^{(\text{id})} + \mathbf{\Pi}^{(\text{visc})} + \mathbf{\Pi}^{(\text{Maxw})}, \tag{1.48}$$

$$\Pi_{ik}^{(\text{id})} = \rho v_i v_k + P\delta_{ik}, \tag{1.49}$$

$$\Pi_{ik}^{(\text{visc})} = -\eta \left( \partial_i v_k + \partial_k v_i - \frac{2}{3}\delta_{ik}\partial_l v_l \right) - \zeta\delta_{ik}\partial_l v_l, \tag{1.50}$$

$$\Pi_{ik}^{(\text{Maxw})} = -\frac{1}{\mu_0} \left( B_i B_k - \frac{1}{2}\delta_{ik}B^2 \right). \tag{1.51}$$

Alongside the equation of motion of the conductive fluid, we need an equation for the evolution of the magnetic field. We start from Ampere's law

$$\nabla \times \mathbf{B} = \mu_0 \mathbf{j}_e, \tag{1.52}$$

where we have neglected the small displacement current and $\mathbf{j}_e$ is the electrical current in the fluid. For a reference frame $K'$ moving with the fluid (the fluid is at rest in $K'$) Ohm's law is

$$\mathbf{j}_e = \sigma \mathbf{E}', \tag{1.53}$$

where $\sigma$ is the electrical conductivity of the fluid and $\mathbf{E}'$ is the electric field in $K'$, which is given in terms of the electric field $\mathbf{E}$ and the magnetic field $\mathbf{B}$ of the reference frame in rest $K'$ by the Lorentz transformations [Landau & Lifshitz (1957), Sec. 49]

$$\mathbf{E}' \equiv \mathbf{E} + \mathbf{v} \times \mathbf{B}, \tag{1.54}$$

with the assumption that $v \ll c$, where $c$ is the velocity of light. In this way for the electric current we have

$$\mathbf{j}_e = \sigma(\mathbf{E} + \mathbf{v} \times \mathbf{B}), \tag{1.55}$$

[Landau & Lifshitz (1957), Eq. (49.2)]. Substituting Eq. (1.55) for the electric current $\mathbf{j}_e$ into Ampere's law, after a little rearrangement we obtain

$$\frac{\nabla \times \mathbf{B}}{\mu_0\,\sigma} = (\mathbf{E} + \mathbf{v} \times \mathbf{B}). \tag{1.56}$$

Taking rotation of this expression and substituting the rotation of the electric field $\mathbf{E}$ from Faraday's law of induction

$$\nabla \times \mathbf{E} = -\partial_t \mathbf{B}, \tag{1.57}$$

we get

$$\nabla \times \left( \frac{\nabla \times \mathbf{B}}{\mu_0\,\sigma} \right) = -\partial_t \mathbf{B} + \nabla \times (\mathbf{v} \times \mathbf{B}), \tag{1.58}$$

[Landau & Lifshitz (1957), Eq. (49.5)]. Using the vector identity

$$\nabla \times (\sigma^{-1} \mathbf{A}) = \sigma^{-1} \nabla \times \mathbf{A} - \mathbf{A} \times \nabla \sigma^{-1} \tag{1.59}$$

for $\mathbf{A} = \nabla \times \mathbf{B}$ and after some rearrangement of the terms, we obtain an equation for evolution of the magnetic field

$$\partial_t \mathbf{B} = \nabla \times (\mathbf{v} \times \mathbf{B}) - (\mu_0\,\sigma)^{-1} \nabla \times (\nabla \times \mathbf{B}) + (\nabla \times \mathbf{B}) \times \nabla(\mu_0\sigma)^{-1}. \tag{1.60}$$

Using the relation $\varepsilon_0 \mu_0 = 1/c^2$ between the vacuum susceptibility $\varepsilon_0$ and $\mu_0$ we can express

$$\frac{1}{\mu_0\,\sigma} = c^2 \varepsilon_0 \varrho \equiv \nu_{\mathrm{m}}, \quad \varrho = \frac{1}{\sigma}, \tag{1.61}$$

where $\varrho$ is the Ohmic resistivity and $\nu_{\mathrm{m}}$ is the magnetic diffusivity. In this way, we can write the magnetic field evolution equation

$$\partial_t \mathbf{B} = \nabla \times (\mathbf{v} \times \mathbf{B}) + \nu_{\mathrm{m}} \Delta \mathbf{B} + (\nabla \times \mathbf{B}) \times \nabla \nu_{\mathrm{m}}, \tag{1.62}$$

where we have used

$$\nabla \times (\nabla \times \mathbf{B}) = -\Delta \mathbf{B} - \nabla(\nabla \cdot \mathbf{B}) \tag{1.63}$$

and Gauss's law Eq. (1.42). We can expand the first term on the right hand side

$$\nabla \times (\mathbf{v} \times \mathbf{B}) = (\mathbf{B} \cdot \nabla)\mathbf{v} - \mathbf{B}(\nabla \cdot \mathbf{v}) - (\mathbf{v} \cdot \nabla)\mathbf{B}, \tag{1.64}$$

where the third term together with the partial derivative of $\mathbf{B}$ form the substantial derivative of the magnetic field and equation for the evolution of the magnetic field can be written

$$\mathrm{d}_t \mathbf{B} = (\mathbf{B} \cdot \nabla)\mathbf{v} - \mathbf{B} \operatorname{div}\mathbf{v} + \nu_{\mathrm{m}}\Delta\mathbf{B} - \operatorname{grad}\nu_{\mathrm{m}} \times \operatorname{rot}\mathbf{B}. \tag{1.65}$$

As far as we know, this equation has not been published so far because the last term in it is negligible for most plasma physics treatments. Omitting this term, i.e. for constant magnetic magnetic diffusivity, we can write this equation as

$$\partial_t \mathbf{B} = \operatorname{rot}(\mathbf{v} \times \mathbf{B}) + \nu_{\mathrm{m}}\Delta\mathbf{B}, \tag{1.66}$$

[Landau & Lifshitz (1957), Eq. (51.2)], as $\nu_{\mathrm{m}}$ is easily convertible to the CGS system, where $\varepsilon_0 = 1/4\pi$ and therefore

$$\nu_{\mathrm{m}} \equiv c^2 \varepsilon_0 \varrho \equiv \frac{c^2}{4\pi\sigma}. \tag{1.67}$$

### 1.3.2 Energy flux in magneto-hydrodynamics

We have derived the energy conservation law for dissipative hydrodynamics Eq. (1.38), where the change in energy with time is

$$\partial_t \left(\frac{1}{2}\rho v^2 + \rho\varepsilon\right), \tag{1.68}$$

where the first term is the kinetic energy per unit volume and the second term is the internal energy per unit volume. In MHD the energy density of the magnetic field $B^2/2\mu_0$ ($H^2/8\pi$ in CGS units) [Landau & Lifshitz (1971), Eq. (31.5)] must also be added

$$\partial_t \left(\frac{1}{2}\rho v^2 + \rho\varepsilon + \frac{B^2}{2\mu_0}\right), \quad B \equiv |\mathbf{B}|. \tag{1.69}$$

We will derive the energy analogously to the ideal and dissipative case. There is no change in the time derivative of the second term $\rho\varepsilon$ and we already have its result in Eq. (1.27). The time derivative of the kinetic energy, however, must be derived again. Starting from Eq. (1.22), which we write down again

$$\partial_t \left(\frac{1}{2}\rho v^2\right) = \frac{1}{2}v^2\partial_t\rho + \rho\mathbf{v}\cdot\partial_t\mathbf{v}, \tag{1.70}$$

we substitute $\partial_t\rho = -\nabla\cdot(\rho\mathbf{v})$ from the continuity equation Eq. (1.2) and $\rho\partial_t\mathbf{v}$ from the MHD equation of motion Eq. (1.47)

$$\rho\partial_t\mathbf{v} = -\rho(\mathbf{v}\cdot\nabla)\mathbf{v} - \nabla P - \nabla\cdot\boldsymbol{\Pi}^{(\mathrm{visc})} - \nabla\cdot\boldsymbol{\Pi}^{(\mathrm{Maxw})}, \tag{1.71}$$

and in this way we obtain

$$\partial_t \left(\frac{1}{2}\rho v^2\right) = -\frac{v^2}{2}\nabla\cdot(\rho\mathbf{v})v^2 - \rho\mathbf{v}\cdot(\mathbf{v}\cdot\nabla)\mathbf{v} + \mathbf{v}\cdot\left(-\nabla P - \nabla\cdot\boldsymbol{\Pi}^{(\mathrm{visc})} - \nabla\cdot\boldsymbol{\Pi}^{(\mathrm{Maxw})}\right). \tag{1.72}$$

The subsequent substitutions and rearrangements are analogous to these already done with Eq. (1.32) to obtain Eq. (1.34) with the last term being the only difference here

$$\partial_t \left( \frac{1}{2} \rho v^2 \right) = - \nabla \cdot \left( \rho \mathbf{v} \frac{v^2}{2} \right) - \rho \mathbf{v} \cdot \nabla h + \rho T \mathbf{v} \cdot \nabla s - \nabla \cdot (\mathbf{v} \cdot \mathbf{\Pi}^{(\text{visc})}) + \mathbf{\Pi}^{(\text{visc})} \cdot \nabla \mathbf{v}$$
$$- \mathbf{v} \cdot (\nabla \cdot \mathbf{\Pi}^{(\text{Maxw})}). \tag{1.73}$$

We have already calculated the divergence of the Maxwell stress tensor to obtain an expression for the magnetic force per unit volume Eq. (1.45) therefore the last term can rewritten as

$$\mathbf{v} \cdot (\nabla \cdot \mathbf{\Pi}^{(\text{Maxw})}) = \mathbf{v} \cdot \frac{1}{\mu_0} [\mathbf{B} \times (\nabla \times \mathbf{B})] = \frac{1}{\mu_0} (\mathbf{v} \times \mathbf{B}) \cdot (\nabla \times \mathbf{B}), \tag{1.74}$$

using the property of the dot interchange of a triple product. Now the time derivative of the kinetic energy per unit volume has the form

$$\partial_t \left( \frac{1}{2} \rho v^2 \right) = - \nabla \cdot \left( \rho \mathbf{v} \frac{v^2}{2} \right) - \rho \mathbf{v} \cdot \nabla h + \rho T \mathbf{v} \cdot \nabla s - \nabla \cdot (\mathbf{v} \cdot \mathbf{\Pi}^{(\text{visc})}) + \mathbf{\Pi}^{(\text{visc})} \cdot \nabla \mathbf{v}$$
$$- \frac{1}{\mu_0} (\mathbf{v} \times \mathbf{B}) \cdot (\nabla \times \mathbf{B}). \tag{1.75}$$

The time derivative of the energy density of the magnetic field is

$$\partial_t \left( \frac{B^2}{2\mu_0} \right) = \frac{1}{\mu_0} \mathbf{B} \cdot \partial_t \mathbf{B}. \tag{1.76}$$

The time derivative of the magnetic field we substitute from the equation of evolution of the magnetic field Eq. (1.62) and Eq. (1.63) for the second derivative of the magnetic field

$$\partial_t \left( \frac{B^2}{2\mu_0} \right) = \frac{1}{\mu_0} \mathbf{B} \cdot [\nabla \times (\mathbf{v} \times \mathbf{B}) - \nu_{\text{m}} \nabla \times (\nabla \times \mathbf{B}) + (\nabla \times \mathbf{B}) \times \nabla \nu_{\text{m}}]. \tag{1.77}$$

Next we expand the parentheses and express terms the first two terms as total derivatives of divergences

$$\partial_t \left( \frac{B^2}{2\mu_0} \right) = - \frac{1}{\mu_0} \{ \nabla \cdot [\mathbf{B} \times (\mathbf{v} \times \mathbf{B})] - (\mathbf{v} \times \mathbf{B}) \cdot (\nabla \times \mathbf{B}) \} \tag{1.78}$$
$$+ \frac{\nu_{\text{m}}}{\mu_0} \{ \nabla \cdot [\mathbf{B} \times (\nabla \times \mathbf{B})] - (\nabla \times \mathbf{B}) \cdot (\nabla \times \mathbf{B}) \} + \frac{1}{\mu_0} \mathbf{B} \cdot (\nabla \times \mathbf{B}) \times \nabla \nu_{\text{m}},$$

according to the divergence product rule

$$\nabla \cdot [\mathbf{B} \times (\mathbf{v} \times \mathbf{B})] = (\mathbf{v} \times \mathbf{B}) \cdot (\nabla \times \mathbf{B}) - \mathbf{B} \cdot [\nabla \times (\mathbf{v} \times \mathbf{B})], \tag{1.79}$$

where for the second we just replace $\mathbf{v}$ with the operator $\nabla$. Replacing the dot and cross operators of the triple product of the last term, we obtain

$$\frac{1}{\mu_0} \mathbf{B} \cdot (\nabla \times \mathbf{B}) \times \nabla \nu_{\text{m}} = \frac{1}{\mu_0} [\mathbf{B} \times (\nabla \times \mathbf{B})] \cdot \nabla \nu_{\text{m}} = \nabla \cdot \mathbf{\Pi}^{(\text{Maxw})} \cdot \nabla \nu_{\text{m}}, \tag{1.80}$$

according to Eq. (1.44). Now adding the time derivatives internal energy per unit volume Eq. (1.27), the MHD kinetic energy per unit volume Eq. (1.75) and the energy density of the magnetic field Eq. (1.78) we obtain the cumbersome expression

$$\partial_t \left( \frac{1}{2}\rho v^2 + \rho\varepsilon + \frac{B^2}{2\mu_0} \right) = -h\nabla \cdot (\rho\mathbf{v}) + \rho T\partial_t s - \nabla \cdot \left( \rho\mathbf{v}\frac{v^2}{2} \right) - \rho\mathbf{v} \cdot \nabla h + \rho T\mathbf{v} \cdot \nabla s$$
$$- \nabla \cdot (\mathbf{v} \cdot \boldsymbol{\Pi}^{(\mathrm{visc})}) + \boldsymbol{\Pi}^{(\mathrm{visc})} \cdot \nabla\mathbf{v} - \frac{1}{\mu_0}(\mathbf{v} \times \mathbf{B}) \cdot (\nabla \times \mathbf{B})$$
$$- \frac{1}{\mu_0}\nabla \cdot [\mathbf{B} \times (\mathbf{v} \times \mathbf{B})] + \frac{1}{\mu_0}(\mathbf{v} \times \mathbf{B}) \cdot (\nabla \times \mathbf{B})$$
$$+ \frac{\nu_\mathrm{m}}{\mu_0}\nabla \cdot [\mathbf{B} \times (\nabla \times \mathbf{B})] - \frac{\nu_\mathrm{m}}{\mu_0}(\nabla \times \mathbf{B})^2 + \nabla \cdot \boldsymbol{\Pi}^{(\mathrm{Maxw})} \cdot \nabla\nu_\mathrm{m}.$$

The last terms on the second and third rows cancel each other. Collecting the total differentials of the enthalpy per unit mass $h$ on the first row (first and last terms) and comparing with the energy conservation law in dissipative hydrodynamics Eq. (1.36), we write the MHD energy conservation law

$$\partial_t \left( \frac{1}{2}\rho v^2 + \rho\varepsilon + \frac{B^2}{2\mu_0} \right) = -\mathrm{div}\,\mathbf{q}, \tag{1.81}$$

$$\mathbf{q} = \rho\mathbf{v}\left( \frac{v^2}{2} + h \right) + \mathbf{v} \cdot \boldsymbol{\Pi}^{(\mathrm{visc})} - \varkappa\nabla T + \frac{1}{\mu_0}\mathbf{B} \times (\mathbf{v} \times \mathbf{B}) - \frac{\nu_\mathrm{m}}{\mu_0}\mathbf{B} \times \mathrm{rot}\,\mathbf{B}, \tag{1.82}$$

[Landau & Lifshitz (1957), Eqs.(51.9), (51.10)] and the equation of heat transfer in MHD

$$\rho T(\partial_t + \mathbf{v} \cdot \nabla)s = -\Pi_{ik}^{(\mathrm{visc})}\,\partial_k v_i + \nabla \cdot (\varkappa\nabla T) + \frac{\nu_\mathrm{m}}{\mu_0}(\mathrm{rot}\mathbf{B})^2 - (\partial_k\Pi_{ik}^{(\mathrm{Maxw})})(\partial_i\nu_\mathrm{m}). \tag{1.83}$$

For a constant magnetic diffusivity $\nabla\nu_\mathrm{m} = 0$ we obtain [Landau & Lifshitz (1957), Eq. (51.6)]

$$\rho T(\partial_t + \mathbf{v} \cdot \nabla)s = -\Pi_{ik}^{(\mathrm{visc})}\,\partial_k v_i + \nabla \cdot (\varkappa\nabla T) + \frac{\nu_\mathrm{m}}{\mu_0}(\mathrm{rot}\mathbf{B})^2. \tag{1.84}$$

The last two terms in the energy density flux Eq. (1.82) are the density of energy flux of the electromagnetic field, namely the Poynting vector [Landau & Lifshitz (1957), Sec. 51], while the last term in the heat transfer equation Eq. (1.83) can be a little modified by substitution of $\nu_\mathrm{m}$ from Eq. (1.61) and $\mathrm{rot}\mathbf{B}$ from Ampere's law Eq. (1.52)

$$\frac{\nu_\mathrm{m}}{\mu_0}(\mathrm{rot}\mathbf{B})^2 \equiv \frac{1}{\mu_0^2}\frac{j_e^2}{\sigma}, \tag{1.85}$$

to reveal that this is the Joule heating of the conducting fluid [Landau & Lifshitz (1957), Sec. 51].

Before moving on, let us analyse the last two terms of the energy flux density Eq. (1.82). This is simply the Poynting vector $\mathbf{S}$

$$\mathbf{S} = \frac{1}{\mu_0}\mathbf{B} \times (\mathbf{v} \times \mathbf{B}) - \frac{\nu_\mathrm{m}}{\mu_0}\mathbf{B} \times \mathrm{rot}\mathbf{B}. \tag{1.86}$$

We can easily show that by starting from the initial expression in SI units for the Poynting vector [Landau & Lifshitz (1971), Eq. (31.2)]

$$S = \frac{1}{\mu_0} \mathbf{E} \times \mathbf{B}, \tag{1.87}$$

and substituting $\mathbf{E}$ from the Lorentz transformations Eq. (1.54)

$$S = \frac{1}{\mu_0}[\mathbf{E}' \times \mathbf{B} - (\mathbf{v} \times \mathbf{B}) \times \mathbf{B}]. \tag{1.88}$$

We change the order of the vector product of the second term and substitute $\mathbf{E}'$ from Ohm's law Eq. (1.53)

$$S = \frac{1}{\mu_0}\left[\frac{\mathbf{j}_e}{\sigma} \times \mathbf{B} + \mathbf{B} \times (\mathbf{v} \times \mathbf{B})\right]. \tag{1.89}$$

We substitute the electric current $\mathbf{j}_e$ from Ampere's law Eq. (1.52), change the order of the vector product in it, too, and swap both terms to obtain

$$S = \frac{1}{\mu_0}\left[\mathbf{B} \times (\mathbf{v} \times \mathbf{B}) - \frac{1}{\mu_0\,\sigma}\mathbf{B} \times (\nabla \times \mathbf{B})\right]. \tag{1.90}$$

The multiplier in front of the second term is the magnetic diffusivity Eq. (1.61) and for the Poynting vector we obtain

$$S = \frac{1}{\mu_0}\mathbf{B} \times (\mathbf{v} \times \mathbf{B}) - \frac{\nu_{\mathrm{m}}}{\mu_0}\mathbf{B} \times (\nabla \times \mathbf{B}), \tag{1.91}$$

which is identical to Eq. (1.86).

In this section we have derived all the necessary MHD equations for solution of our problem, namely: the equation of motion Eq. (1.46), the equation for the evolution of the magnetic field Eq. (1.65), the equation for conservation of mass Eq. (1.2), the equation for the momentum flux density Eq. (1.48) and the equation for conservation of energy Eq. (1.81) and Eq. (1.82). In the next section we are going to derive the dependencies of all kinetic coefficients participating in our equations.

## 1.4 Elementary Statistics and Kinetics of Hydrogen Plasma

The Sun and its atmosphere consist of plasma, where the dominating chemical element is hydrogen with more than 90% abundance. At temperatures above 10 kK, which can be found in the solar chromosphere, the hydrogen is almost completely ionised and this is the reason to consider plasma properties and kinetic coefficients for electrons and protons only.

### 1.4.1 Debye radius (length)

A completely ionised gas as a whole is electrically neutral [Landau & Lifshitz (1980), Eq. (78.1)]

$$\sum_a z_a n_{a0} = 0, \tag{1.92}$$

where $n_{a0}$ is the mean density of ions of a-th kind and $z_a$ are positive or negative integers showing the multiple charge of these ions. In order the plasma to be in ideal state, the energy of the Coulomb interaction between its particles should be much smaller compared with their mean kinetic energy, which is given by the temperature $T = k_B T'$ ($T'$ is the temperature in degrees)

$$\frac{ze^2}{r} \ll T, \quad e^2 \equiv \frac{q_e^2}{4\pi\varepsilon_0}, \tag{1.93}$$

where $q_e$ is the electron charge. The potential energy of each ion is $\Phi_a = z_a q_e \varphi$, where $\varphi$ is the potential of the electric field given by the Poisson equation

$$\Delta\varphi = -\frac{q_e}{\varepsilon_0} \sum_a z_a n_a. \tag{1.94}$$

For an ideal plasma, we can use the Boltzmann distribution [Landau & Lifshitz (1980), Eq. (38.6)]

$$n_a = n_{a0} e^{-\Phi_a/T}, \tag{1.95}$$

which shows that at large distances from the ion, the density of other ions must be equal to the mean density $n_{a0}$. Expanding the last equation into Taylor series of $\varphi$

$$n_a \approx n_{a0} - n_{a0}\frac{z_a q_e}{T}\varphi + \dots, \tag{1.96}$$

and substituting only the linear term in the Poisson equation Eq. (1.94), we obtain

$$\Delta\varphi = \frac{q_e}{\varepsilon_0} \sum_a z_a^2 n_{a0} \frac{q_e}{T}\varphi. \tag{1.97}$$

The quantity in front of $\varphi$ in this differential equation has the dimension of reciprocal squared length and can be written

$$\frac{1}{r_D^2} = \frac{4\pi e^2}{T} \sum_a n_{a0} z_a^2, \tag{1.98}$$

and $r_D$ is called Debye radius (or length) [Landau & Lifshitz (1980), Eq. (78.8)]. In this way the differential equation becomes

$$\Delta\varphi = r_D^{-2}\varphi, \tag{1.99}$$

and for the spherically symmetric case

$$\Delta\varphi \equiv \frac{1}{r^2}\frac{d}{dr}\left(r^2\frac{d\varphi}{dr}\right) \equiv \frac{1}{r}\frac{d^2}{dr^2}(r\varphi) = r_D^{-2}\varphi, \tag{1.100}$$

with solution

$$\varphi = \frac{\text{const}}{r}e^{-r/r_D}. \tag{1.101}$$

At close distances to the ion, the field must be pure Coulomb field (which defines the constant) but for distances larger than $r_D$ the field becomes very small. In this way the ions around each ion create an ion cloud that effectively screens the electric field of the latter.

Lastly, we will write down the Debye radius for electrons and protons, for which $z = 1$

$$\frac{1}{r_D^2} = \frac{4\pi e^2 n}{T}. \tag{1.102}$$

### 1.4.2 Particle collisions in plasma

Another important property in plasma is the Coulomb logarithm $\Lambda$, which accounts for the Rutherford scattering between the plasma charged particles within a Debye radius. The values of $\Lambda$ for classical scattering, for which $ze^2/\hbar v_{T_a} \gg 1$ is

$$\Lambda = \ln\left(\frac{r_{\text{D}} T_a}{ze^2}\right), \tag{1.103}$$

and for the quantum mechanical scattering $ze^2/\hbar v_{T_a} \ll 1$ is

$$\Lambda = \ln\left(\frac{\sqrt{m_a T_a}\, r_{\text{D}}}{\hbar}\right), \tag{1.104}$$

where $T_a$ is the energy of the ion of a-th kind, $m_a$ its mass and $v_{T_a}$ its thermal velocity

$$m_a v_{T_a}^2 \sim T_a \tag{1.105}$$

[Lifshitz & Pitaevskii (1979), Eq. (42.6)]. For our classical treatment of the solar plasma ($z = 1$)

$$\Lambda = \ln\left(\frac{r_{\text{D}} T}{e^2}\right), \tag{1.106}$$

where $T$ is the energy of the particles in consideration.

### 1.4.3 Kinetic coefficients

In this section following the kinetics by Lifshitz and Pitaevskii [Lifshitz & Pitaevskii (1979)], we will repeat the gas kinetic formulae applied to the plasma. We introduce the variable $r_{\text{T}} = e^2/T$ with length dimension, which comes from the equality of the mean kinetic energy and the Coulomb potential energy

$$T = \frac{e^2}{r}, \tag{1.107}$$

without the Coulomb logarithm. The electron and proton transport sections for the Coulomb interaction have the form

$$\sigma_{ee} = \frac{\Lambda_e}{0.6} r_{\text{T}}^2, \quad \sigma_{pp} = \frac{\Lambda_p}{0.4} r_{\text{T}}^2, \tag{1.108}$$

where we have assumed that the temperature of electrons and protons have different temperature $T_e \neq T_p$ and the coefficients 0.6 and 0.4 are added to the final exact formulae for respectively the electric conductivity and viscosity coefficients of completely ionized hydrogen plasma.

In a gaseous approximation the reciprocal mean free path of one electron is additive

$$\frac{1}{l_e} = \sigma_{ee} n_e + \sigma_{ea} n_a \tag{1.109}$$

where $n_e$ is the density of the electrons, $\sigma_{ea}$ is the electron-atom cross section and $n_a$ is the atom density. The electron mean free path $l_e$

$$l_e = \left( n_e \frac{\Lambda_e}{0.6} r_{\mathrm{T}}^2 + \sigma_{ea} n_a \right)^{-1} \qquad (1.110)$$

is connected with the electron mean free time

$$\tau_e = \frac{l_e}{v_{T_e}} = \sqrt{\frac{m}{T_e}} \left( n_e \frac{\Lambda_e}{0.6} r_{\mathrm{T}}^2 + \sigma_{ea} n_a \right)^{-1}, \qquad (1.111)$$

where $m$ is the mass of the electron and $\nu_e = 1/\tau_e$ is the electron collision rate.

An elementary kinetic estimate of the electrical conductivity $\sigma$ is

$$\mathbf{j_e} = \sigma \mathbf{E} = \sigma \frac{\mathbf{F}_e}{q_e} = q_e n_e \mathbf{v}_{dr}, \qquad (1.112)$$

where $\mathbf{F}_e$ is the force of the electric field $\mathbf{E}$ and $\mathbf{v}_{dr}$ is the electron drift velocity. A simple dimensional analysis shows that the connection between the electric force and the drift velocity is

$$\mathbf{F}_e = m \frac{\mathbf{v}_{dr}}{\tau_e}, \qquad (1.113)$$

which substituted in Eq. (1.112) allows us to obtain

$$\sigma = \frac{q_e^2 n_e \tau_e}{m} \qquad (1.114)$$

and after substituting $\tau_e$ from Eq. (1.111)

$$\sigma = \frac{1}{\varrho} = \frac{q_e^2 n_e}{\sqrt{m T_e}} \left( n_e \frac{\Lambda_e}{0.6} r_{\mathrm{T}}^2 + \sigma_{ea} n_a \right)^{-1}. \qquad (1.115)$$

The magnetic diffusivity can be calculated from Eq. (1.61)

$$\nu_{\mathrm{m}} \equiv c^2 \varepsilon_0 \varrho = \frac{c^2}{4\pi} \frac{\sqrt{m T_e}}{e^2 n_e} \left( n_e \frac{\Lambda_e}{0.6} r_{\mathrm{T}}^2 + \sigma_{ea} n_a \right). \qquad (1.116)$$

For completely ionised plasma we have cf. [Lifshitz & Pitaevskii (1979), Eq. (43.8)]

$$\sigma = \frac{1}{\varrho} = 4\pi \varepsilon_0 \frac{0.6}{\Lambda_e} \frac{T_e^{3/2}}{e^2 m^{1/2}}, \qquad (1.117)$$

$$\nu_{\mathrm{m}} = \frac{c^2}{4\pi} \frac{\Lambda_e}{0.6} \frac{e^2 m^{1/2}}{T_e^{3/2}}. \qquad (1.118)$$

The heat flux is carried by the electrons from places with higher temperatures $T_{e2}$ to places with lower temperatures $T_{e1}$, meaning that it is dependent on the density, thermal velocity of the electrons and the temperature difference

$$q_h = -c_e n_e v_{T_e}(T_{e2} - T_{e1}) = -c_e n_e v_{T_e} l_e \frac{\Delta T_e}{l_e} = -\varkappa \nabla T_e, \tag{1.119}$$

where $c_e = 3/2$ is the heat capacity of one electron. An elementary dimensional analysis shows

$$\frac{\Delta T_e}{l_e} = \frac{\mathrm{d} T_e}{\mathrm{d} x} = \nabla T_e, \tag{1.120}$$

and therefore for the thermal conductivity we obtain

$$\varkappa = c_e n_e v_{T_e} l_e. \tag{1.121}$$

After substitution of $c_e = 3/2$, $v_{T_e}$ from Eq. (1.105) and $l_e$ from Eq. (1.110) we have

$$\varkappa = \frac{3}{2} n_e \sqrt{\frac{T_e}{m}} \left( n_e \frac{\Lambda_e}{0.6} r_{\mathrm{T}}^2 + n_a \sigma_{ea} \right)^{-1} \tag{1.122}$$

For completely ionised hydrogen plasma $n_a = 0$, the thermal conductivity becomes [Lifshitz & Pitaevskii (1979), Eq. (43.9)]

$$\varkappa = \frac{0.9}{\Lambda_p} \frac{T_e^{5/2}}{e^4 m^{1/2}}. \tag{1.123}$$

The last coefficient we are going to consider is the kinematic viscosity $\nu_{\mathrm{k}} = \eta/\rho$, which is determined by the plasma ion component. For hydrogen plasma, the proton mean free path is analogous to the same of electrons

$$\frac{1}{l_p} = \sigma_{pp} n_p + \sigma_{pa} n_a, \tag{1.124}$$

where $n_p$ is the density of the protons and $\sigma_{pa}$ is the proton-atom cross section. The mean free path of the protons expressed with the transport cross section is

$$l_p = \left( n_p \frac{\Lambda_p}{0.4} r_{\mathrm{T}}^2 + n_a \sigma_{pa} \right)^{-1}, \tag{1.125}$$

and the proton mean free times is

$$\tau_p = \frac{l_p}{v_{T_p}} = \sqrt{\frac{M}{T_p}} \left( n_p \frac{\Lambda_p}{0.4} r_{\mathrm{T}}^2 + \sigma_{pa} n_a \right)^{-1} \tag{1.126}$$

and $\nu_p = 1/\tau_p$ is the proton collision rate.

Dimensional analysis of the kinematic viscosity shows

$$\nu_{\mathrm{k}} = l_p v_{T_p},\tag{1.127}$$

and after substitution of $l_p$ from Eq. (1.125) and $v_{T_p}$ from Eq. (1.105) we obtain

$$\nu_{\mathrm{k}} = l_p v_{T_p} = \sqrt{\frac{T_p}{M}} \left( n_p \frac{\Lambda_p}{0.4} r_T^2 + n_a \sigma_{pa} \right)^{-1}\tag{1.128}$$

and for fully ionised plasma $n_a = 0$

$$\nu_{\mathrm{k}} = \frac{0.4}{\Lambda_p} \frac{T_p^{5/2}}{n_p e^4 M^{1/2}},\tag{1.129}$$

where $M$ is the proton mass. The dynamic viscosity for fully ionised plasma is [Lifshitz & Pitaevskii (1979), Eq. (43.10)]

$$\eta = \rho \nu_{\mathrm{k}} = \frac{0.4}{\Lambda_p} \frac{T_p^{5/2} M^{1/2}}{e^4},\tag{1.130}$$

where we have used $\rho = n_p M$.

### 1.4.4 Additional relations between kinetic coefficients and collision rates

Let us also mention the following relations between the kinetic coefficients

$$\varkappa \varrho = 1.5 \, T_e / q_e^2,\tag{1.131}$$

$$\eta / \varkappa \approx \frac{4}{9} \sqrt{mM}.\tag{1.132}$$

For equal electron and proton temperatures $T_e = T_p = T$ the temperature dependent magnetic Prandtl coefficient

$$\mathrm{Pr_m} \equiv \frac{\nu_{\mathrm{k}}}{\nu_{\mathrm{m}}} = \frac{4\pi e^2 n_e}{c^2 \sqrt{mM}} \left[ \left( n_e \frac{\Lambda r_T^2}{0.6} + n_a \sigma_{ea} \right) \left( n_p \frac{\Lambda r_T^2}{0.4} + n_a \sigma_{pa} \right) \right]^{-1},\tag{1.133}$$

and for high temperatures when $n_a \to 0$ we have

$$\mathrm{Pr_m} = \frac{0.96\pi}{\Lambda^2} \frac{T^4}{\sqrt{mM} c^2 e^6 n} \gg 1,\tag{1.134}$$

*i.e.* the electric resistance is negligible.

Let us give an elementary estimate for ratios of the corresponding times and collision rates for electron, proton and electron-proton collisions for energy exchange $\tau_{ep}^{\varepsilon}$ for completely ionized hydrogen plasma [Braginskii (1963), Lifshitz & Pitaevskii (1979)]

$$\left( \tau_e = \frac{1}{\nu_e} \right) : \left( \tau_p = \frac{1}{\nu_p} \right) : \left( \tau_{ep}^{\varepsilon} = \frac{1}{\nu_{ep}^{\varepsilon}} \right) = 1 : \sqrt{\frac{M}{m}} : \frac{M}{m}.\tag{1.135}$$

In the last proportion, it is taken into account that for elastic electron proton collision the exchange of energy is of order of $m/M$; it can be derived considering the scattering in the system of center of the mass. Our elementary consideration $\tau_{ep}^{\varepsilon} = \tau_e(M/m)$ gives

$$\tau_{ep}^{\varepsilon} = \frac{1}{8\sqrt{2\pi}} \frac{T^{3/2} M}{e^4 n\Lambda\, m^{1/2}}, \tag{1.136}$$

where the numerical coefficient $8\sqrt{2\pi} \approx 20$ in [Lifshitz & Pitaevskii (1979), Eq. (42.5)] requires state-of-the-art consideration. The kinetic of heat exchange between protons and electrons

$$\frac{dT_e}{dt} = -\frac{T_e - T_p}{\tau_{ep}^{\varepsilon}} \tag{1.137}$$

shows that in the transition region with width $\lambda$ we can consider equal temperatures $T_e = T_p = T$ if the flight time through the transition region with wind velocity $U$ is much longer than the time, for which the protons heat up the electrons

$$\frac{\lambda}{U} \gg \tau_{ep}^{\varepsilon}. \tag{1.138}$$

### 1.4.5 Influence of the magnetic field on the kinetic coefficients

The viscous momentum flux density tensor $\Pi_{ij}^{(\mathrm{visc})}$ in strong magnetic field is [Lifshitz & Pitaevskii (1979), Eq. (13.18)]

$$\begin{aligned}
\Pi_{ij}^{(\mathrm{visc})} =& 2\eta\left(v_{ij} - \frac{1}{3}\delta_{ij}v\right) + \zeta\delta_{ij}\mathrm{div}\mathbf{v} \\
&+ \eta_1(2v_{ij} - \delta_{ij}\mathrm{div}\mathbf{v} + \delta_{ij}v_{kl}b_k b_l - 2v_{ik}b_k b_j - 2v_{jk}b_k b_i + b_i b_j\mathrm{div}\mathbf{v} + b_i b_j v_{kl}b_k b_l) \\
&+ 2\eta_2(v_{ik}b_k b_j + v_{jk}b_k b_i - 2b_i b_j v_{kl}b_k b_l) + \eta_3(v_{ik}b_{jk} + v_{jk}b_{ik} - v_{kl}b_{ik}b_j b_l - v_{kl}b_{jk}b_i b_l) \\
&+ 2\eta_4 v_{kl}(b_{ik}b_j + b_{jk}b_i)b_l + \zeta_1(\delta_{ij}v_{kl}b_k b_l + b_i b_j\mathrm{div}\mathbf{v}), \tag{1.139}
\end{aligned}$$

$$v_{ij} = v_{ji} = \frac{1}{2}(\partial_i v_j + \partial_j v_i), \quad b_{ij} = -b_{ji} = \varepsilon_{ijk}b_k, \quad \mathbf{b} = \frac{\mathbf{B}}{B}, \tag{1.140}$$

where $\mathbf{B}$ is the external magnetic field, $B$ its magnitude and $\varepsilon_{ijk}$ is the Levi-Civita symbol. For $\mathbf{B} = 0$ it is easy to show that Eq. (1.139) is equal to Eq. (1.50).

The applicability of the usual MHD approach requires the inequality

$$\nu_p = \frac{e^2 n_p \Lambda_p}{T_p^{3/2}\sqrt{M}} \gg \omega_{Bp} = \frac{q_e B_0}{M}, \tag{1.141}$$

to be met, where we have assumed fully ionised plasma. For the Solar atmosphere this condition holds up to the lower layer of the transition region, where the proton density $n_p$ decreases steeply, while the proton temperature $T_p$ increases steeply. In order to account for the influence of magnetic field on viscosity for the waves, we use only the second viscosity coefficient [Braginskii (1963)]

$$\nu_{k,2} = \frac{\frac{6}{5}\tilde{x}^2 + \frac{7}{3}}{\tilde{x}^4 + 4.03\tilde{x}^2 + \frac{7}{3}}\,\nu_k, \tag{1.142}$$

where $\tilde{x} \equiv \omega_{Bp}\tau_p$. Instead of calculating the last expression, we use the Padé approximant [Mishonov & Varonov (2019a)]

$$\eta_2 = \frac{\eta}{1 + C(\omega_{Bp}\tau_p)^2}, \quad C \approx 0.9, \tag{1.143}$$

which has the same magnetic field dependence as longitudinal electric conductivity in perpendicular magnetic field in $\tau$-approximation. For strong magnetic fields case $\omega_{Bp} \gg \nu_p$ the suggested Padé approximant reproduces the well-known result [Lifshitz & Pitaevskii (1979), Eq. (59.38)]

$$\eta_2(B \to \infty) \approx \frac{8\pi^{1/2}e^4\Lambda n_p^2}{5(MT)^{1/2}\omega_{Bp}^2}. \tag{1.144}$$

For all other coefficients $\eta_1$-$\eta_4$ we can suggest analogous formulae but the simplest Padé approximation even quantitatively reproduces the main effect of the magnetic field at high temperatures and small densities – viscous heating stops in the hot corona.

CHAPTER **2**

# DERIVATION OF THE EQUATIONS

As it has already been shown from the observations, the narrow TR and the temperature of the plasma in the solar atmosphere are height dependent, therefore the problem for solar corona heating is one-dimensional (1D). Denoting height with $x$, AW generated in the photosphere travel upwards from $0 < x < l$ and their viscous damping heats the plasma in the solar atmosphere and accelerates the solar wind. From a given static spectral density of incoming AW the height profiles of the temperature $T(x)$ and solar wind velocity $U(x)$ are calculated in details within the frame of MHD. The calculation comprises only hydrogen, since it is the most abundant element in the solar atmosphere, more than 90% of its total chemical composition [Eddy (1979), Chap. 2]. The initial starting height is the upper chromosphere at $x = 0$, where the approximation for completely ionised hydrogen plasma, i.e. electrically neutral mixture of electrons and protons, can be used. For simplicity and illustration purpose an initial equal proton and electron temperatures $T_p = T_e = T$ is assumed. With these initial conditions set, a full theoretical derivation of all necessary equations for the MHD calculation of the self-induced opacity for AW follows in this chapter.

The starting point are the magnetic field evolution equation Eq. (1.65)

$$\mathrm{d}_t \mathbf{B} = (\mathbf{B} \cdot \nabla)\mathbf{v} - \mathbf{B}\,\mathrm{div}\,\mathbf{v} + \nu_\mathrm{m}\Delta\mathbf{B} - (\mathrm{grad}\,\nu_\mathrm{m}) \times (\mathrm{rot}\,\mathbf{B}), \tag{2.1}$$

the equation of motion Eq. (1.47)

$$\partial_t(\rho\mathbf{v}) + \nabla \cdot \mathbf{\Pi} = 0, \tag{2.2}$$

the laws for mass Eq. (1.2) and energy Eq. (1.81) flux conservation

$$\partial_t\rho + \mathrm{div}\,\mathbf{j} = 0, \quad \mathbf{j} = \rho\mathbf{v}, \tag{2.3}$$

$$\partial_t\left(\frac{1}{2}\rho\mathbf{v}^2 + \rho\varepsilon + \frac{\mathbf{B}^2}{2\mu_0}\right) + \mathrm{div}\,\mathbf{q} = 0, \tag{2.4}$$

where the energy flux density Eq. (1.82) is

$$\mathbf{q} = \rho\mathbf{v}\left(\frac{\mathbf{v}^2}{2} + h\right) + \mathbf{v}\cdot\mathbf{\Pi}^{(\text{visc})} - \varkappa\nabla T + \frac{1}{\mu_0}[\mathbf{B}\times(\mathbf{v}\times\mathbf{B}) - \nu_{\text{m}}\mathbf{B}\times\text{rot}\mathbf{B}], \qquad (2.5)$$

and the total momentum flux Eq. (1.51) is

$$\mathbf{\Pi} = \mathbf{\Pi}^{(\text{id})} + \mathbf{\Pi}^{(\text{visc})} + \mathbf{\Pi}^{(\text{Maxw})}, \qquad (2.6)$$

$$\Pi_{ik}^{(\text{id})} = \rho v_i v_k + P\,\mathbb{1}, \qquad (2.7)$$

$$\Pi_{ik}^{(\text{visc})} = -\eta\left(\partial_i v_k + \partial_k v_i - \frac{2}{3}\delta_{ik}\partial_l v_l\right) - \zeta\delta_{ik}\partial_l v_l, \qquad (2.8)$$

$$\Pi_{ik}^{(\text{Maxw})} = -\frac{1}{\mu_0}\left(B_i B_k - \frac{1}{2}\delta_{ik}B^2\right). \qquad (2.9)$$

These equations have already been derived in the previous chapter and are only repeated and summarized here. As mentioned before, the derivation of the equations for the self-induced opacity of AW starts from here.

## 2.1 Wave Equations

We analyse AW propagating along constant magnetic field lines $B_0$. For the velocity and magnetic field we assume

$$\mathbf{v}(t, x) = U(x)\,\mathbf{e}_x + u(t, x)\,\mathbf{e}_z,$$
$$\mathbf{B}(t, x) = B_0\,\mathbf{e}_x + B_0 b(t, x)\,\mathbf{e}_z, \qquad (2.10)$$

where $U(x)$ is the solar wind, $B_0\mathbf{e}_x$ is the homogeneous magnetic field perpendicular to the surface of the Sun, $u(t, x)$ is the amplitude of the AW velocity oscillation and $b(t, x)$ is the amplitude of the AW dimensionless magnetic field oscillation. Both AW oscillations are in the $z$-direction, perpendicular to the direction of propagation of the AW. These transverse wave amplitudes of the velocity $u(t, x)$ and dimensionless magnetic field $b(t, x)$ are represented with the Fourier integrals

$$u(t, x) = \int_{-\infty}^{\infty}\tilde{u}(\omega, x)\mathrm{e}^{-\mathrm{i}\omega t}\frac{\mathrm{d}\omega}{2\pi}, \qquad (2.11)$$

$$b(t, x) = \int_{-\infty}^{\infty}\tilde{b}(\omega, x)\mathrm{e}^{-\mathrm{i}\omega t}\frac{\mathrm{d}\omega}{2\pi}. \qquad (2.12)$$

For brevity it is convenient to consider monochromatic AW, for which

$$u(t, x) = \hat{u}(x)\mathrm{e}^{-\mathrm{i}\omega t},$$
$$b(t, x) = \hat{b}(x)\mathrm{e}^{-\mathrm{i}\omega t}. \qquad (2.13)$$

The summation on different frequencies, which finally gives Fourier integration, will be restored later on. After substitution of the Fourier transforms Eq. (2.13) of the AW oscillations in the anzatz Eq. (2.10), the Fourier representation for monochromatic AW of the anzatz is obtained

$$\mathbf{v}(x) = U(x)\,\mathbf{e}_x + \hat{u}(x)\mathrm{e}^{-\mathrm{i}\omega t}\mathbf{e}_z,$$
$$\mathbf{B}(x) = B_0\,\mathbf{e}_x + B_0\hat{b}(x)\mathrm{e}^{-\mathrm{i}\omega t}\mathbf{e}_z. \tag{2.14}$$

The Fourier representations of the $z$-component of the magnetic field evolution equation Eq. (2.1) is

$$-\mathrm{i}\omega\hat{b} = \mathrm{d}_x\hat{u} - \mathrm{d}_x(U\hat{b}) + \mathrm{d}_x(\nu_\mathrm{m}\mathrm{d}_x\hat{b}). \tag{2.15}$$

The Fourier representation of the $z$-component of the momentum flux conservation Eq. (2.2) is

$$\rho\partial_t v_z + \mathrm{d}_x\Pi_{zx} = 0. \tag{2.16}$$

A calculation of the divergence (which contains only derivatives in $x$) of the momentum flux tensors Eq. (2.7), Eq. (2.8), Eq. (2.9), the obtained equation is

$$-\mathrm{i}\omega\hat{u} + U\mathrm{d}_x\hat{u} - \frac{1}{\rho}\mathrm{d}_x(\eta_2\mathrm{d}_x\hat{u}) - V_\mathrm{A}^2\mathrm{d}_x\hat{b} = 0, \quad V_\mathrm{A}(x) = \frac{B_0}{\sqrt{\mu_0\rho(x)}} \tag{2.17}$$

where $V_\mathrm{A}$ is the Alfvén velocity and the shear viscosity in strong magnetic field $\eta_2$ is used in the viscous momentum flux tensor according to the analysis in Subsec. 1.4.5. The obtained equations Eq. (2.15) and Eq. (2.17) can be rearranged a little to be rewritten as

$$-\mathrm{i}\omega\hat{b} = \mathrm{d}_x\hat{u} - \mathrm{d}_x(U\hat{b}) + \mathrm{d}_x(\nu_\mathrm{m}\mathrm{d}_x\hat{b}), \tag{2.18}$$

$$-\mathrm{i}\omega\hat{u} = V_\mathrm{A}^2\mathrm{d}_x\hat{b} - U\mathrm{d}_x\hat{u} + \frac{1}{\rho}\mathrm{d}_x(\eta_2\mathrm{d}_x\hat{u}). \tag{2.19}$$

These equations form a complete linear system for $\hat{b}$ and $\hat{u}$. For a numerical solution, it is convenient these two second-order differential equations to be transformed to four first order differential equations, which can be represented in a matrix form

$$-\mathrm{id}_x\Psi = \mathsf{K}\Psi, \quad \Psi \equiv \begin{pmatrix} \hat{b} \\ \hat{u} \\ \hat{\gamma} \\ \hat{w} \end{pmatrix}, \quad \Psi^\dagger \equiv \left(\hat{b}^*, \hat{u}^*, \hat{\gamma}^*, \hat{w}^*\right), \quad \mathsf{K} \equiv \frac{\mathrm{i}\overline{\mathsf{M}}}{\nu_\mathrm{m}\nu_\mathrm{k,2}}, \tag{2.20}$$

where $\hat{\gamma} \equiv \mathrm{d}_x\hat{b}, \hat{w} \equiv \mathrm{d}_x\hat{u}$ and

$$\overline{\mathsf{M}} \equiv \begin{pmatrix} 0 & 0 & -\nu_\mathrm{m}\nu_\mathrm{k,2} & 0 \\ 0 & 0 & 0 & -\nu_\mathrm{m}\nu_\mathrm{k,2} \\ (\mathrm{i}\omega - \mathrm{d}_x U)\nu_\mathrm{k,2} & 0 & -(U + \mathrm{d}_x\nu_\mathrm{m})\nu_\mathrm{k,2} & \nu_\mathrm{k,2} \\ 0 & \mathrm{i}\omega\nu_\mathrm{m} & V_\mathrm{A}^2\nu_\mathrm{m} & -\left(U - \frac{\mathrm{d}_x(\eta_2)}{\rho}\right)\nu_\mathrm{m} \end{pmatrix}, \tag{2.21}$$

in some sense, $\mathsf{K}$ is the wave-vector operator.

For a homogeneous medium with constant $\eta_2$, $\nu_{\mathrm{m}}$, $\rho$, $V_{\mathrm{A}}$, and $U$, in short for constant wave-vector matrix $\mathsf{K}$, the exponential substitution $\Psi \propto \exp(\mathrm{i}kx)$ in Eq. (2.20) or equivalently Eq. (2.18) and Eq. (2.19) gives the secular equation $\det(\mathsf{K} - k\mathbb{1}) = 0$, which after some algebra gives the dispersion equation of AW

$$\left(\omega_{\mathrm{D}} + \mathrm{i}\nu_{\mathrm{m}}k^2\right)\left(\omega_{\mathrm{D}} + \mathrm{i}\nu_{\mathrm{k},2}k^2\right) = V_{\mathrm{A}}^2 k^2, \tag{2.22}$$

where $\omega_{\mathrm{D}} \equiv \omega - kU$ is the Doppler shifted frequency.

## 2.1.1 Reduced wave equations

For hot plasma $\nu_{\mathrm{k}} \gg \nu_{\mathrm{m}}$ Eq. (1.133) the magnetic diffusivity can be neglected and the matrix $\overline{\mathsf{M}}$ is reduced to 3×3. In this case the wave equations are

$$- \mathrm{i}\mathrm{d}_x\Psi_k = \mathsf{K}_\eta\Psi_\eta, \quad \Psi_\eta \equiv \begin{pmatrix} \hat{u} \\ \hat{b} \\ \hat{w} \end{pmatrix}, \quad \Psi_\eta^\dagger \equiv \left(\hat{u}^*, \hat{b}^*, \hat{w}^*\right), \quad \mathsf{K}_k \equiv \frac{\mathrm{i}\overline{\mathsf{M}}_\eta}{\nu_{\mathrm{k},2}U}, \tag{2.23}$$

$$\overline{\mathsf{M}}_\eta \equiv \begin{pmatrix} 0 & 0 & -\nu_{\mathrm{k},2}U \\ 0 & (-\mathrm{i}\omega + \mathrm{d}_x U)\nu_{\mathrm{k},2} & \nu_{\mathrm{k},2} \\ \mathrm{i}\omega U & -V_{\mathrm{A}}^2(-\mathrm{i}\omega + \mathrm{d}_x) & (V_{\mathrm{A}}^2 - U^2) + \frac{U}{\rho}\mathrm{d}_x\eta_2 \end{pmatrix}, \tag{2.24}$$

which has already been derived in [Mishonov et al. (2011)]. The secular equation for homogeneous medium in this case can be easily obtained by setting $\nu_{\mathrm{m}} = 0$ in Eq. (2.22)

$$\omega_{\mathrm{D}}(\omega_{\mathrm{D}} + \mathrm{i}\nu_{\mathrm{k},2}k^2) = V_{\mathrm{A}}^2 k^2. \tag{2.25}$$

For cold plasma $\nu_{\mathrm{k}} \ll \nu_{\mathrm{m}}$ and analogously the omission of the kinematic viscosity again reduces the matrix $\overline{\mathsf{M}}$ to 3×3. The wave equations now have the form

$$- \mathrm{i}\mathrm{d}_x\Psi_\nu = \mathsf{K}_\nu\Psi_\nu, \quad \Psi_\nu \equiv \begin{pmatrix} \hat{b} \\ \hat{u} \\ \hat{\gamma} \end{pmatrix}, \quad \Psi_\nu^\dagger \equiv \left(\hat{b}^*, \hat{u}^*, \hat{\gamma}^*\right), \quad \mathsf{K}_\nu \equiv \frac{\mathrm{i}\overline{\mathsf{M}}_\nu}{\nu_{\mathrm{m}}U}, \tag{2.26}$$

$$\overline{\mathsf{M}}_\nu \equiv \begin{pmatrix} 0 & 0 & -\nu_{\mathrm{m}}U \\ 0 & -\mathrm{i}\omega\nu_{\mathrm{m}} & -V_{\mathrm{A}}^2\nu_{\mathrm{m}} \\ -(-\mathrm{i}\omega + \mathrm{d}_x U)U & \mathrm{i}\omega & (V_{\mathrm{A}}^2 - U^2) + U\mathrm{d}_x\nu_{\mathrm{m}} \end{pmatrix}, \tag{2.27}$$

The secular equation for homogeneous medium in this case again can be easily obtained by setting $\nu_{\mathrm{k},2} = 0$ in Eq. (2.22)

$$\omega_{\mathrm{D}}(\omega_{\mathrm{D}} + \mathrm{i}\nu_{\mathrm{m}}k^2) = V_{\mathrm{A}}^2 k^2. \tag{2.28}$$

Numerically it is possible to solve the reduced wave equations: initially for cold plasma $\nu_{\mathrm{k}} < \nu_{\mathrm{m}}$ and therefore Eq. (2.26) is solved until the inequality is reversed $\nu_{\mathrm{k}} > \nu_{\mathrm{m}}$ and then a switch to Eq. (2.23) is performed.

## 2.2 Conservation Laws: the Non-linear Part of the Problem

The mass conservation law Eq. (2.3) for the initial $0$ and final $l$ ($0 < x < l$) calculation heights in 1D implies a constant flux

$$j = \rho(x)U(x) = \rho_0 U_0 = \rho_l U_l = \text{const},\qquad(2.29)$$

where $\rho_0 = \rho(0)$, $\rho_l = \rho(l)$, $U_0 = U(0)$, and $U_l = U(l)$. The energy conservation law Eq. (2.4) reduces to a constant energy flux density Eq. (2.5) along the $x$-axis

$$q = q_{\text{wind}}^{\text{ideal}}(x) + \tilde{q}(x) = \text{const},\qquad(2.30)$$

where the first term is the energy flux density of the ideal wind (wind from an ideal inviscid fluid) and the second term includes all other energy density fluxes. Using the mass density $\rho(x)$, the wind velocity $U(x)$ and the temperature $T(x)$ for the first two terms of Eq. (2.5), the energy of the ideal wind is

$$q_{\text{wind}}^{\text{ideal}} = \left(\frac{1}{2}U^2 + h\right)\rho U = q_{_U} + q_{_P}, \quad q_{_U} \equiv \left(\frac{1}{2}U^2\right)\rho U, \quad q_{_P} \equiv h\rho U,\qquad(2.31)$$

where the fully ionised hydrogen is an ideal in thermodynamic sense gas

$$h = \varepsilon + \frac{P}{\rho}, \quad P = n_{\text{tot}}T = \frac{\rho T}{\langle m \rangle} = \frac{j}{U}\frac{T}{\langle m \rangle}, \quad \varepsilon = \frac{3}{2}\frac{T}{\langle m \rangle},$$

$$\langle m \rangle = \frac{n_p M + n_e m}{n_p + n_e} \approx \frac{1}{2}M, \quad n_e = n_p = \frac{1}{2}n_{\text{tot}}, \quad \rho = \langle m \rangle\, n_{\text{tot}} = M n_p.$$

From these relations, the enthalpy per unit mass $h$ is

$$h = \frac{5}{2}\frac{T}{\langle m \rangle} = c_p \frac{T}{\langle m \rangle}, \quad \gamma = \frac{c_p}{c_v},$$

where $c_p = 5/2$ and $c_v = 3/2$ are the heat capacities per unit mass for constant pressure and volume, and $\gamma = 5/3$ is the adiabatic index. Although there are some hints for different values of the adiabatic index $\gamma$ [Doorsselaere et al. (2011)], the traditional value of 5/3 will be used since this choice will not change the essence of the presentation. Hydrogen plasma is well theoretically investigated and $\gamma = 5/3$ is used in thousands of works. There are no theoretical hints for significant deviations from this value obtained by an ideal gas approximation. Actually, a small variation of the value of gamma does not change qualitatively the temperature profile of the TR.

The second component $\tilde{q}$ of the energy flux density in Eq. (2.30) consists of a viscous wind $q_\xi$, head conductivity $q_\varkappa$ and AW $q_{\text{wave}}$ components

$$\tilde{q}(x) = q_\xi + q_\varkappa + q_{\text{wave}}.\qquad(2.32)$$

Substituting the wind velocity $U(x)$ into the viscous (the third) term and the temperature $T(x)$ into the heat conduction (the fourth) term of Eq. (2.5), the first two components of the non-ideal energy flux density are obtained

$$q_\xi \equiv q_{\text{wind}}^{\text{visc}} \equiv -\xi U \mathrm{d}_x U, \quad \xi = \frac{4}{3}\eta + \zeta,$$

$$q_\varkappa \equiv q_{\text{cond}}^{\text{heat}} \equiv -\varkappa \, \mathrm{d}_x T,$$

The second (bulk) viscosity for completely ionised plasma is zero $\zeta = 0$ [Lifshitz & Pitaevskii (1979), Sec. 8] and [Cramer (2012)] but nevertheless, the notation $\xi$ will be used for thoroughness.

The wave component of the non-ideal energy density flux is obtained by substituting the AW amplitudes of the monochromatic oscillations Eq. (2.14) into the total energy density flux Eq. (2.5). After the substitution, the wave energy density flux is time averaged

$$\left\langle \hat{u}^2 \right\rangle_t = \left\langle (\operatorname{Re}\hat{u})^2 \right\rangle_t = \left\langle \frac{1}{4}(\hat{u} + \hat{u}^*)^2 \right\rangle_t = \frac{1}{2}\left| \hat{u} \right|^2, \tag{2.33}$$

which is a standard procedure for alternating processes. The first and third terms of Eq. (2.5) give for the time averaged energy of a single wave

$$q_u \equiv \frac{j}{4}\left| \hat{u} \right|^2, \tag{2.34}$$

$$q_\eta \equiv -\frac{1}{4}\eta_2 \mathrm{d}_x \left| \hat{u} \right|^2. \tag{2.35}$$

The last two terms of Eq. (2.5) are actually the Poynting vector Eq. (1.86), which using Eq. (1.43) can be written in the form

$$\mathbf{S} = \frac{1}{\mu_0}\left[ \mathbf{v}B^2 - \mathbf{B}(\mathbf{v}\cdot\mathbf{B}) - \nu_{\text{m}}\left( \frac{\nabla B^2}{2} - (\mathbf{B}\cdot\nabla)\mathbf{B} \right) \right], \tag{2.36}$$

and its $x$-component is

$$S_x = \frac{1}{\mu_0}\left[ v_x B^2 - B_x(\mathbf{v}\cdot\mathbf{B}) - \nu_{\text{m}}\left( \frac{\mathrm{d}_x B^2}{2} - (B_x \mathrm{d}_x)B_x \right) \right]. \tag{2.37}$$

Substituting Eq. (2.14) into the last equation and performing the time averaging Eq. (2.33) for $\hat{b}^2$ and $\hat{b}\hat{u}$, three additional terms for a single wave are obtained

$$q_b \equiv \frac{B_0^2}{2\mu_0} U \left| \hat{b} \right|^2, \tag{2.38}$$

$$q_{ub} \equiv -\frac{B_0^2}{2\mu_0}\operatorname{Re}(\hat{b}^*\hat{u}), \tag{2.39}$$

$$q_\varrho \equiv -\frac{B_0^2}{2\mu_0}\nu_{\text{m}}\operatorname{Re}(\hat{b}^*\hat{\gamma}). \tag{2.40}$$

In total, the AW energy flux density consists of five terms, two of which Eq. (2.35) and Eq. (2.40) are dissipative, the rest can be considered as ideal. The total AW energy flux density can represented also as

$$q_{\text{wave}} = q_{\text{wave}}^{\text{ideal}} + q_{\text{wave}}^{\text{diss}},$$
$$q_{\text{wave}}^{\text{ideal}} = q_u + q_b + q_{ub}, \quad q_{\text{wave}}^{\text{diss}} = q_\eta + q_\varrho. \tag{2.41}$$

Combining all energy flux density ingredients and adding summation over multiple waves in the wave energy fluxes, the total energy flux density is

$$q = q_{\text{wind}}^{\text{ideal}}(x) + \tilde{q}(x) = \text{const}, \tag{2.42}$$

$$q_{\text{wind}}^{\text{ideal}} = q_U + q_P, \quad q_U \equiv \left(\frac{1}{2}U^2\right)\rho U, \quad q_P \equiv h\rho U, \tag{2.43}$$

$$\tilde{q}(x) = q_\xi + q_\varkappa + q_{\text{wave}}, \quad q_\xi \equiv -\xi U\mathrm{d}_x U, \quad q_\varkappa \equiv -\varkappa\mathrm{d}_x T, \tag{2.44}$$

$$q_{\text{wave}} = q_{\text{wave}}^{\text{ideal}} + q_{\text{wave}}^{\text{diss}}, \quad q_{\text{wave}}^{\text{ideal}} = q_u + q_b + q_{ub}, \quad q_{\text{wave}}^{\text{diss}} = q_\eta + q_\varrho, \tag{2.45}$$

$$q_u \equiv \frac{j}{4}\sum_{\text{waves}}|\hat{u}|^2, \quad q_b \equiv \frac{B_0^2}{2\mu_0}\sum_{\text{waves}}U\left|\hat{b}\right|^2, \quad q_{ub} \equiv -\frac{B_0^2}{2\mu_0}\sum_{\text{waves}}\text{Re}(\hat{b}^*\hat{u}),$$
$$q_\eta = -\frac{1}{4}\eta_2\sum_{\text{waves}}\mathrm{d}_x|\hat{u}|^2, \quad q_\varrho = -\frac{B_0^2}{2\mu_0}\nu_{\mathrm{m}}\sum_{\text{waves}}\text{Re}(\hat{b}^*\hat{\gamma}),$$

$$q_{\text{wave}}^{\text{ideal}} = \sum_{\text{waves}}\left\{\frac{j}{4}|\hat{u}|^2 + \frac{B_0^2}{2\mu_0}\left[U\left|\hat{b}\right|^2 - \text{Re}(\hat{b}^*\hat{u})\right]\right\}, \tag{2.46}$$

$$q_{\text{wave}}^{\text{diss}} = \sum_{\text{waves}}\left\{-\frac{1}{4}\eta_2\mathrm{d}_x|\hat{u}|^2 - \frac{B_0^2}{2\mu_0}\nu_{\mathrm{m}}\text{Re}(\hat{b}^*\hat{\gamma})\right\}, \tag{2.47}$$

$$q_{\text{wave}} = \sum_{\text{waves}}\left\{\frac{j}{4}|\hat{u}|^2 - \frac{1}{4}\eta_2\,\mathrm{d}_x|\hat{u}|^2 + \frac{B_0^2}{2\mu_0}\left(U\left|\hat{b}\right|^2 - \text{Re}(\hat{b}^*\hat{u}) - \nu_{\mathrm{m}}\text{Re}(\hat{b}^*\hat{\gamma})\right)\right\}. \tag{2.48}$$

Since the mass density flux is constant Eq. (2.29), the $xi$-components of the momentum flux density tensor Eq. (2.2) are also constant

$$\Pi = \Pi_{\text{wind}}^{\text{ideal}}(x) + \tilde{\Pi}(x) = \text{const}, \tag{2.49}$$

and analogously to the energy flux density, the first component is momentum flux density of the ideal wind and the second component includes all other momentum flux densities. A substitution of $\rho(x)$, $U(x)$ and $T(x)$ into the ideal momentum flux density tensor Eq. (2.7), gives

$$\Pi_{\text{wind}}^{\text{ideal}} = \rho UU + P = \Pi_U + \Pi_P, \quad \Pi_U \equiv \rho UU, \quad \Pi_P \equiv P.$$

The second component $\tilde{\Pi}(x)$ in Eq. (2.49) consists of

$$\tilde{\Pi}(x) = \Pi_\xi + \Pi_{\text{wave}}, \tag{2.50}$$

where the first term is the wind viscous component and the second term is the AW component. The viscous part is obtained by substituting $U(x)$ in the $xx$-component only (the other two $xy$ and $xz$ are evidently zero) of the viscous momentum tensor Eq. (2.8)

$$\Pi_\xi = \Pi_{xx}^{\text{visc}} = -\left(\frac{4}{3}\eta + \zeta\right)\mathrm{d}_x U = -\xi\mathrm{d}_x U, \tag{2.51}$$

where the minus sign in front of $\Pi_{xx}$ is due to the definition of $\Pi_\xi$ with a positive sign, i.e. the minus sign is moved into $\Pi_\xi$. Lastly, the wave term of the momentum flux density is obtained by substitution of the equation for the magnetic field from Eq. (2.14) into the $xx$, $xy$ and $xz$ components of the Maxwell stress tensor Eq. (2.9). Performing the subsequent time averaging according to Eq. (2.33), the only remaining term of the Maxwell stress tensor for a single wave is

$$\Pi_{\text{wave}} \equiv \frac{B_0^2}{4\mu_0}|\hat{b}|^2, \tag{2.52}$$

the others are either constant or $\propto \hat{b}$, which upon time averaging is zero (as any time averaged wave oscillation). Combining all the ingredients of the momentum flux density and adding summation over multiple waves to the wave component, the total momentum flux density is

$$\Pi = \Pi_{\text{wind}}^{\text{ideal}}(x) + \tilde{\Pi}(x) = \text{const}, \tag{2.53}$$

$$\Pi_{\text{wind}}^{\text{ideal}} = \Pi_U + \Pi_P, \quad \Pi_U \equiv \rho U U, \quad \Pi_P \equiv P, \tag{2.54}$$

$$\tilde{\Pi}(x) = \Pi_\xi + \Pi_{\text{wave}}, \quad \Pi_\xi \equiv -\xi\mathrm{d}_x U, \quad \Pi_{\text{wave}} \equiv \sum_{\text{waves}} \frac{B_0^2}{4\mu_0}\left|\hat{b}\right|^2. \tag{2.55}$$

A remark for the spectral density of AW should be made here. The summation over multiple waves is actually an integral over the spectral density $\mathcal{W}(\omega)$ of incoming AW

$$\sum_{\text{waves}} \mathcal{W}_{\text{wave}} \equiv \int_0^\infty \frac{\mathrm{d}\omega}{2\pi}\mathcal{W}(\omega). \tag{2.56}$$

In a numerical calculation, this integral is naturally converted into the sum to its left.

## 2.2.1 Dimensionless variables

At any given distance $x$ from the solar photosphere, the energy Eq. (2.42) and momentum fluxes Eq. (2.53) can be written as

$$q_{\text{wind}}^{\text{ideal}}(x) + \tilde{q}(x) = q_{\text{wind}}^{\text{ideal}}(0) + \tilde{q}(0), \tag{2.57}$$

$$\Pi_{\text{wind}}^{\text{ideal}}(x) + \tilde{\Pi}(x) = \Pi_{\text{wind}}^{\text{ideal}}(0) + \tilde{\Pi}(0). \tag{2.58}$$

Dimensionless variables

$$\chi(x) \equiv \left.\frac{\tilde{q}(x)}{\rho_0 U_0^3}\right|_x^0, \quad \tau(x) \equiv \left.\frac{\tilde{\Pi}(x)}{\rho_0 U_0^2}\right|_x^0, \tag{2.59}$$

representing the non-ideal parts of the energy $\chi$ and the momentum $\tau$ density fluxes are introduced. The solar wind $U(x)$ and temperature $T(x)$ are also expressed in dimensionless form

$$\overline{U}(x) \equiv \frac{U(x)}{U_0}, \quad U_0 = U(0),$$

$$\Theta(x) \equiv \frac{T(x)}{\langle m \rangle U_0^2}, \quad \Theta_0 = \Theta(0). \tag{2.60}$$

The energy Eq. (2.57) and momentum Eq. (2.58) fluxes in the newly introduced dimensionless notation take the form

$$\frac{1}{2}\overline{U}^2 + c_p\Theta - \chi = \frac{1}{2} + c_p\Theta_0, \tag{2.61}$$

$$\overline{U} + \Theta/\overline{U} - \tau = 1 + \Theta_0. \tag{2.62}$$

Expressing $\Theta$ from the second equation Eq. (2.62) and substituting it in the first one Eq. (2.61), a quadratic equation for $\overline{U}$ is obtained

$$\overline{U}^2(\gamma + 1) - 2\overline{U}(\gamma + \overline{s}^2 + \gamma\tau) + (2\chi + 1)(\gamma - 1) + 2\overline{s}^2 = 0 \tag{2.63}$$

with discriminant

$$\mathcal{D} = (\overline{s}^2 - 1)^2 - 2\chi(\gamma^2 - 1) + \gamma\tau\left[\gamma\tau + 2\left(\gamma + \overline{s}^2\right)\right], \tag{2.64}$$

$$\overline{s}^2 \equiv \frac{c_s^2(0)}{U_0^2} = \gamma\Theta_0, \quad c_s^2(x) = \left(\frac{\partial P}{\partial \rho}\right)_S = \frac{\gamma T(x)}{\langle m \rangle}, \tag{2.65}$$

and solution

$$\overline{U}(x) = \frac{1}{\gamma + 1}\left(\gamma + \overline{s}^2 + \gamma\tau(x) - \sqrt{\mathcal{D}(x)}\right), \quad U(x) = U_0\overline{U}(x). \tag{2.66}$$

For $\chi = \tau = 0$ there should be no heating nor acceleration and this condition determines the sign in front of the $\sqrt{\mathcal{D}}$. After substitution of the obtained expression for $\overline{U}$ Eq. (2.66) into Eq. (2.62), a solution for the dimensionless temperature is obtained

$$\Theta(x) = \overline{U}(x)\left[1 + \Theta_0 + \tau(x) - \overline{U}(x)\right], \quad \Theta_0 = \frac{\overline{s}^2}{\gamma}, \tag{2.67}$$

and the solution for the real temperature is

$$T(x) = \langle m \rangle U_0^2 \Theta(x). \tag{2.68}$$

The problem is formally reduced to analogous one for a jet engine, cf. Ref. [Feynmann, Leighton & Sands (1965), problems 40.4-5] and the non-linear part is solved with a quadratic equation. In one dimensional approximation the use of the three conservation lws at known mass, energy and momentum fluxes gives the analytical solution of the non-linear part of the MHD problem calculation of temperature and wind profiles. For an ideal gas the solution is given by the solution of the quadratic equation Eq. (2.63). Here we have to emphasize that propagation of AW is described by linear differential equations.

The energy Eq. (2.42) and momentum flux density Eq. (2.53) equations are non-linear differential equations, which have been reduced to the quadratic equation Eq. (2.63). Such an approach requires a specific numerical method for solution, which is given in the next chapter.

## 2.2.2   Energy-momentum equation analysis

The energy-momentum quadratic equation Eq. (2.63) and its solutions for the wind Eq. (2.66) and temperature Eq. (2.67) obtained in the last subsection definitely have physical meaning. In case of $\mathcal{D} = 0$ in Eq. (2.64), the solution for the dimensional wind velocity becomes

$$U(x) \equiv \overline{U} U_0 = \frac{U_0}{\gamma + 1} [\gamma + \overline{s}^2 + \gamma \tau(x)]. \tag{2.69}$$

Dividing this solution with the local sound velocity Eq. (2.65), a relation between the wind velocity and local sound velocity is obtained

$$\frac{U(x)}{c_{\mathrm{s}}(x)} = \frac{\overline{U} U_0}{\sqrt{\gamma \Theta} U_0} = \frac{\overline{U}}{\sqrt{\gamma \Theta}}. \tag{2.70}$$

After substitution the solution for the dimensionless temperature $\Theta$ from Eq. (2.67), a little arrangement of the terms, the velocities relation considered here becomes

$$\frac{U}{c_{\mathrm{s}}} = \sqrt{\frac{\overline{U}}{\gamma + \overline{s}^2 + \gamma \tau - \gamma \overline{U}}}, \tag{2.71}$$

where Eq. (2.65) for $\Theta_0$ has also been used. A final substitution for $\overline{U}$ from Eq. (2.69) gives

$$\frac{U}{c_{\mathrm{s}}} = \sqrt{\frac{\gamma + \overline{s}^2 + \gamma \tau}{(\gamma + 1)[\gamma + \overline{s}^2 + \gamma \tau - \gamma(\gamma + \overline{s}^2 + \gamma \tau)/(\gamma + 1)]}} = 1. \tag{2.72}$$

A zero discriminant means that the local sound velocity $c_{\mathrm{s}}(x)$ is reached and this is the maximal value $U(x)$ can have, which is evident from Eq. (2.66), and this result is in full agreement with [Landau & Lifshitz (1987), Sec. 83 and Sec. 97] that supersonic velocities can be achieved with a de Laval nozzle only. For $\mathcal{D} < 0$ the are no real solutions, while as $\mathcal{D}$ gets larger, $U(x)$ gets smaller. This behaviour certainly seems at least peculiar, since applying more momentum to the air flowing into our jet engine leads to a decrease of the velocity of the air flowing out. A plausible explanation for this phenomenon is that the smaller initial momentum allows more heating (more time spent in the combustion chamber), which results in larger acceleration (of the jet exhaust).

### 2.2.3   Additional energy and momentum density fluxes

Several physical effects in the energy-momentum equations derivation have been omitted so far: bremsstrahlung, solar gravity and radiative losses. Bremsstrahlung is negligible within the narrow TR compared to the energy necessary for the heating of the solar corona as free-free radiation may be neglected for temperatures below few million degrees [Landi & Landini (1999)] or even 10 MK [Gronenschild & Mewe (1978)]. It becomes significant in distances of the order of a solar radius from the solar photosphere and it is included in global solar atmosphere calculations [Suzuki & Inutsuka (2005)]. Since the theory is concerned with a region with dimensions comparable to the TR, rather to the solar radius, the influence of bremsstrahlung can be safely neglected.

Similarly, the energy loss due to the solar gravity within the TR is negligible compared to the heating of the corona

$$M g_\odot \lambda \ll k_{\mathrm{B}} \Delta T', \tag{2.73}$$

where $g_\odot$ is the solar surface gravity. Considering the width of the TR $\lambda \simeq 10$ km and the change of the temperature within the TR $\Delta T' = 10^5$ K, the energy loss due to the solar gravity is less than 1% than the obtained energy for the heating. Even before the observations from Skylab, [Parker (1958), Sec. 4] studied coronal heating and mass loss neglecting the gravitational potential energy.

Nevertheless, the introduction of the gravitational energy and momentum density fluxes is straightforward and can be performed with ease. Since the region of interest of our problem does not extend beyond 1 solar radius, the approximate form of the gravitation potential energy can be used and the gravitational energy density flux has the form

$$\mathbf{q}_g = \rho \mathbf{v} \Phi, \quad \Phi = g_\odot x + \mathrm{const}, \quad g_\odot > 0. \tag{2.74}$$

In 1D and in the used notations the gravitational energy density flux is

$$q_g \equiv \tilde{q}_{\mathrm{grav}} = j g_\odot x \tag{2.75}$$

and it is added to the non-ideal energy density flux Eq. (2.44)

$$\tilde{q}(x) = q_\xi + q_\varkappa + q_{\mathrm{wave}} + q_g. \tag{2.76}$$

According to Eq. (2.59), in dimensionless variables the gravitational energy flux is

$$\chi_{\mathrm{grav}} = -\frac{j g_\odot x}{\rho_0 U_0^3}, \tag{2.77}$$

and in this form it is simply added to the dimensionless energy $\chi$.

The gravitational force $\mathbf{f}_g$ per unit volume according to [Landau & Lifshitz (1957), Eq. (15.2)]

$$\mathbf{f}_g = -\nabla \cdot \mathbf{\Pi}^{(\mathrm{grav})}, \tag{2.78}$$

where $\mathbf{\Pi}^{(\mathrm{grav})}$ is the gravitational momentum flux density. The gravitational force has a $x$-component only

$$f_g(x) = -\rho(x) \nabla \Phi(x) = -\rho(x) \mathrm{d}_x \Phi(x) = -\rho(x) g_\odot, \tag{2.79}$$

and in 1D the gravitational momentum flux density has only $xx$-component, which from Eq. (2.78) is given by

$$\Pi_g \equiv \Pi_{xx}^{\text{grav}} = -\int_0^x f_g(x')\mathrm{d}x' = \int_0^x \rho(x')g_\odot \mathrm{d}x' \qquad (2.80)$$

and it is added to the non-ideal momentum density flux Eq. (2.55)

$$\tilde{\Pi}(x) = \Pi_\xi + \Pi_{\text{wave}} + \Pi_g. \qquad (2.81)$$

According to Eq. (2.59), in dimensionless variables the gravitational momentum flux is

$$\tau_{\text{grav}} = \frac{\Pi_g(x)}{\rho_0 U_0'^2}\Big|_x^0 = -\frac{g_\odot}{\rho_0 U_0'^2}\int_0^x \rho(x')\mathrm{d}x', \qquad (2.82)$$

and in this form it is added to the dimensionless momentum $\tau$ as well.

Finally radiative losses have to be considered in the energy balance of the TR. For this optically thin plasma we use volume density of radiated power

$$E_r = n_e n_p \mathcal{P}(T), \qquad (2.83)$$

where the radiative loss function function $\mathcal{P}(T)$ given in [Golub & Pasachoff (1997), Fig. 3.12] is presented in Fig. 2.1. In the MHD calculation by [Shoda et al. (2018)] a polynomial ap-proximation of radiative function $\log_{10}\mathcal{P}$ as function of $\log_{10}T$ is used from [Matsumoto & Suzuki (2014), Fig. 1, Eq. (14)], while here an interpolation from the digitized [Golub & Pasachoff (1997), Fig. 3.12] is used. These two functions defer hundred times in the maximum around 100 kK. Here the deviation from [Golub & Pasachoff (1997)] is in opposite direction $C_\lambda \sim 1/100$ but still in the framework of the acceptable for this topic disagreement.

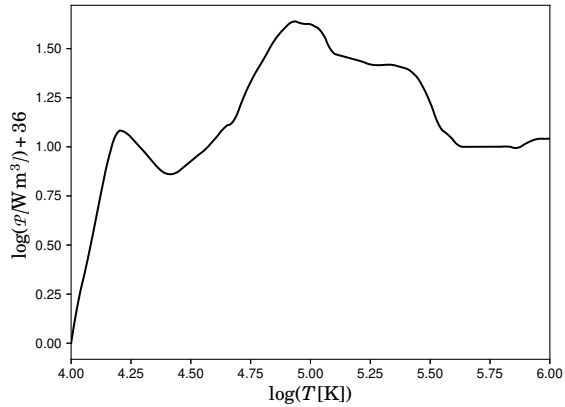

Figure 2.1: A segment of the Raymond radiative loss function [Raymond & Smith (1977)] from [Golub & Pasachoff (1997), Fig. 3.12]. In our MHD calculation we use digitalized data from this book. One order magnitude correction in $n_\text{H}$ in Table 2.1 will lead to two orders magnitude correction in $\mathcal{P}(T)$. That is why different estimations for $\mathcal{P}(T)$ differ by 2 orders of magnitude.

If a thin layer of plasma passing in altitude from 0 to $x$ is traced, for the total energy loss per unit volume

$$\int_0^x n_e n_p \mathcal{P}(T) \frac{\mathrm{d}x'}{U(x')} \tag{2.84}$$

the power with respect to proper time $\mathrm{d}t_{\mathrm{proper}} \equiv \mathrm{d}x/U(x)$ of this "liquid particle" has to be integrated. For the radiative losses of the energy flux we obtain

$$q_r(x) = -U(x) \int_0^x n_e(x') n_p(x') C_\lambda \mathcal{P}(T(x')) \frac{\mathrm{d}x'}{U(x')}, \tag{2.85}$$

where we have to take into account the charge neutrality of a completely ionized plasma. In this form Eq. (2.85) has to be added to non-ideal energy density flux Eq. (2.32). As expected, one order magnitude correction in the models for the hydrogen density $n_{\mathrm{H}}$, and different models for the radiative loss function $\mathcal{P}$ gives 2 orders of magnitude. Therefore a dimensionless factor $C_\lambda$ which absorbs the differences between different evaluation for $\mathcal{P}$ is introduced; i.e. $\mathcal{P} \rightarrow C_\lambda \mathcal{P}$. Dividing by $\rho_0 U_0^3$, we derive the dimensionless radiative energy transmission

$$\chi_r = -\frac{U(x)}{\rho_0 U_0^3} \int_0^x n_p^2 C_\lambda \mathcal{P}(T) \frac{\mathrm{d}x'}{U}, \tag{2.86}$$

where $n_e = n_p$ due to complete ionization of hydrogen plasma. This form is added to the non-ideal energy density flux Eq. (2.59). For the present spatial and temporal resolution for $n_{\mathrm{H}}$, all MHD calculations, for which $|\lg C_\lambda| \leq 2$ are realistic. The used value in the presented research is $C_\lambda = 10^{-2}$.

### 2.2.4 Wave power upward extrapolation

From the chromospheric side of the TR, the solar wind velocity is very small and there is hydrostatic balance between gravity and the gas pressure gradient. This balance is broken in the narrow TR because the acceleration forces there are significant. Temperature, density, small wind velocity and spectral density of AW are all fixed initially. In [Avrett & Loeser (2008)] a significant hint for existence of AW in the chromosphere can be found, the turbulent pressure velocity there, which is the same as non-thermal microturbulent velocity $V_{\mathrm{nt}}(x)$ is simply a mean square velocity of the plasma, through which AW propagate. In our notations

$$V_{\mathrm{nt}}(x) = \sqrt{\langle u^2(t,x) \rangle}. \tag{2.87}$$

If AW absorption in first approximation is neglected, their energy flux

$$q_{\mathrm{AW}} = V_{\mathrm{A}} \, \rho V_{\mathrm{nt}}^2. \tag{2.88}$$

Table 2.1 is [Avrett & Loeser (2008), Table I] with an extra column added $\log(q_{\mathrm{AW}})$, where for illustrative purposes a constant magnetic field $B_0 = 100$ G is assumed. One can see that while the density changes by 7 orders of magnitude, the variation of $\lg(q_\mathrm{AW})$ is moderate and $\lg(q_\mathrm{AW})$ is almost constant. This is an important hint that non-thermal broadening of the lines is due to propagation of the AW. These data give order evaluation of $q_\mathrm{AW} \simeq (10 \div 100)\ \mathrm{kW/m^2}$ [Avrett & Loeser (2008), Table I, analyzing SUMER observations]. According to other order estimations even energy flux flows of $1000\ \mathrm{kW/m^2}$ are not huge [Asgari et al. (2013)]; in this article the TR is treated as a discontinuity not as a result of MHD calculations and moreover the energy exchange between chromosphere and corona is not included. Other order evaluations give even higher energy flux densities. Authors (Avrett & Loeser) comment that these data need to be refined by comparison with observations that have higher temporal and spacial resolution. One order of magnitude correction in $n_{\mathrm{H}}$ could be not surprising but the qualitative picture has already been established.

| $\lg(n_\mathrm{H} \times \mathrm{cm}^3)$ | $V_\mathrm{nt}$ [km/s] | $\lg\left(\dfrac{q_\mathrm{AW}}{\mathrm{kW/m^2}}\right)$ |
|---|---|---|
| 16.0 | 0.34 | 0.62 |
| 15.5 | 0.37 | 0.45 |
| 15.0 | 0.4 | 0.27 |
| 14.5 | 0.5 | 0.21 |
| 14.0 | 0.6 | 0.12 |
| 13.5 | 0.7 | 0.00 |
| 13.0 | 1.0 | 0.06 |
| 12.5 | 1.5 | 0.16 |
| 12.0 | 2.5 | 0.36 |
| 11.8 | 3.1 | 0.44 |
| 11.5 | 3.8 | 0.47 |
| 11.3 | 4.7 | 0.56 |
| 11.0 | 6.0 | 0.62 |
| 10.8 | 7.6 | 0.72 |
| 10.5 | 9.3 | 0.75 |
| 10.3 | 11.4 | 0.83 |
| 10.0 | 15.0 | 0.91 |
| 9.8 | 19.8 | 1.06 |
| 9.5 | 24.5 | 1.09 |
| 9.3 | 29.3 | 1.15 |
| $\leq 9.0$ | 34.0 | 1.12 |

Table 2.1: Hydrogen density $\lg(n_\mathrm{H})$, non-thermal velocity $V_\mathrm{nt}$ after [Avrett & Loeser (2008), Table I] and AW energy density flux $\lg(q_\mathrm{AW})$. The change in the third column is 10 times smaller than the change in the first one therefore with 10% accuracy it can be stated that $\lg(q_\mathrm{AW}) \approx \mathrm{const}$. This constant AW energy flux density penetrating the transparent for them chromosphere gives an important estimate for the full power of the AW heating the solar corona. The non-thermal broadening of the chromospheric lines does not have an alternative explanation. Avrett and Loeser point out that data needs to be refined, but one order magnitude correction in $n_\mathrm{H}$ will not create qualitative changes.

For the frequency dependence of the spectral density of AW one can use the spectral density of the magnetic field obtained by satellite magnetometers. Analogously to Eq. (2.87) one can define the averaged square of the wave component of the magnetic field, in our notations

$$B_0^2 \left\langle b^2(t, x) \right\rangle = \int_0^\infty (B^2)_f \mathrm{d}f, \quad f = \frac{\omega}{2\pi}, \quad (B^2)_f = \frac{\mathcal{D}_b}{f^a}, \quad \omega_{\min} < \omega < \omega_{\max}. \quad (2.89)$$

The spectral density of the magnetic field fluctuations has dimension (in practical units) T$^2$/Hz, and the observations of Voyager 1 [Marsch (1991), Burlaga & Mish (1987)] give $a \approx 2$. The minimal frequency is limited with the variation of the conditions generating AW, while the maximal frequency of AW is determined by their Q-factor

$$\omega_{\mathrm{AW}}(k) = V_{\mathrm{A}}k - \mathrm{i}\nu_k k^2, \quad V_{\mathrm{A}}k_{\max} = \nu_k k_{\max}^2, \quad \omega_{\max}(T) = V_{\mathrm{A}}k_{\max} = 5\frac{e^4 \Lambda B_0^2/2\mu_0}{M^{1/2}T^{5/2}}. \quad (2.90)$$

For $T' = 1 \times 10^6$ K the almost density independent $f_{\max} \approx 300$ Hz and for $10 \times 10^3$ K is $10^5$ times higher. This difference of the frequencies shows the domain in which AW are absorbed heating the solar corona.

Qualitatively for an order evaluation one consider that in the TR all AW from infinitely to the $f_{\max}$ for the coronal side of the TR are completely absorbed and simple integration gives the evaluation

$$q_{\mathrm{AW}} = \frac{1}{a-1} V_{\mathrm{A}} \mu_0 f_{\max} (B^2)_f \big|_{f_{\max}} \quad (2.91)$$

$$(B^2)_f \big|_{f_{\max}} = \frac{\mathcal{D}_b}{f_{\max}^a}, \quad (2.92)$$

where all $V_{\mathrm{A}}$ is taken for the chromosperic side of the TR, and $f_{\max}$ from the coronal one. In the chromosperic side AW with the same constant $\mathcal{D}_b$ in front of the power like spectrum exists up to $(T/T_0)^{5/2}f_{\max}$. In the used logarithmic accuracy of this evaluation, the Fresnel reflection of low frequency waves from the jump of the mass density $\rho$ at the TR with reflection coefficient

$$R_{\mathrm{Fresnel}} = \left( \frac{\sqrt{\rho_1} - \sqrt{\rho_2}}{\sqrt{\rho_1} + \sqrt{\rho_2}} \right)^2 \quad (2.93)$$

is neglected. This Fresnel reflection implies much higher power of AW coming from the solar photosphere with a small part of it passing through the step-like TR to be measured by the spacecrafts scientific instruments. In other words, evaluation the Pointing flux coming from photosphere we have to add, say one order magnitude due to the long wavelength reflection from the step of the plasma density. Extrapolating to the transition region, the distancing of magnetic force lines and decreasing of the Pointing flux at approximately constant total energy flux have to be taken into account.

## 2.3   Wave Boundary Conditions

For known background solar wind $U(x)$ and temperature $T(x)$ for $0 < x < l$, the wave equations Eq. (2.20) can be solved for run-away AW at $x = l$. To obtain the run-away condition at $x = l$, a combination of left and right propagating waves at $x = 0$ has to be found.

First the AW energy density flux Eq. (2.48) for a single wave can be rewritten in the form

$$q_{\text{wave}}(\Psi(x)) \equiv \Psi^\dagger g(x) \Psi = \frac{j}{4} |\hat{u}|^2 - \frac{\eta_2}{2} \operatorname{Re}(\hat{u}^* \hat{w}) + \frac{B_0^2}{2\mu_0} \left( U|\hat{b}|^2 - \operatorname{Re}(\hat{b}^* \hat{u}) - \nu_{\text{m}} \operatorname{Re}(\hat{b}^* \hat{\gamma}) \right), \quad (2.94)$$

where

$$g(x) \equiv \begin{pmatrix} \frac{U B_0^2}{2\mu_0} & -\frac{B_0^2}{4\mu_0} & -\frac{B_0^2}{4\mu_0} \nu_{\text{m}} & 0 \\ -\frac{B_0^2}{4\mu_0} & \frac{1}{4} j & 0 & -\frac{1}{4} \eta_2 \\ -\frac{B_0^2}{4\mu_0} \nu_{\text{m}} & 0 & 0 & 0 \\ 0 & -\frac{1}{4} \eta_2 & 0 & 0 \end{pmatrix}.$$

Next the eigenvalues and eigenvectors of the matrix K are calculated. According to Eq. (2.20) the eigenvectors determine wave propagation in a homogeneous fluid with amplitude $\propto \exp(ikx)$ and the eigenvalues are complex wave vectors

$$k = k' + ik'' = \text{eigenvalue}(\mathsf{K}), \qquad \text{i.e.} \quad \det(\mathsf{K} - k\mathbb{1}) = 0. \quad (2.95)$$

In this way the characteristic dispersion equation Eq. (2.22) is obtained, whose solutions are the four eigenvectors

$$\mathrm{F} = \begin{pmatrix} \mathrm{F}_b(x) \\ \mathrm{F}_u(x) \\ \mathrm{F}_\gamma(x) \\ \mathrm{F}_w(x) \end{pmatrix}, \quad \mathrm{L} = \begin{pmatrix} \mathrm{L}_b(x) \\ \mathrm{L}_u(x) \\ \mathrm{L}_\gamma(x) \\ \mathrm{L}_w(x) \end{pmatrix}, \quad \mathrm{R} = \begin{pmatrix} \mathrm{R}_b(x) \\ \mathrm{R}_u(x) \\ \mathrm{R}_\gamma(x) \\ \mathrm{R}_w(x) \end{pmatrix}, \quad \mathrm{D} = \begin{pmatrix} \mathrm{D}_b(x) \\ \mathrm{D}_u(x) \\ \mathrm{D}_\gamma(x) \\ \mathrm{D}_w(x) \end{pmatrix} \quad (2.96)$$

ordered by the spatial decrements of their values

$$k_{\text{F}}'' < k_{\text{L}}'' < 0 < k_{\text{R}}'' < k_{\text{D}}'' \quad (2.97)$$

(the same applies for the real parts $k'$) and normalized by the conditions

$$- \mathrm{F}^\dagger g \mathrm{F} = -\mathrm{L}^\dagger g \mathrm{L} = \mathrm{R}^\dagger g \mathrm{R} = \mathrm{D}^\dagger g \mathrm{D} = 1, \quad (2.98)$$

where the sign corresponds to the wave propagation direction ($-$ left, $+$ right). Eigenvector F corresponds to left overdamped wave, L to left propagating wave, R to right propagating wave and D to right overdamped wave. For clarity let $\nu_{\text{k},2} \approx \nu_{\text{m}} \approx \nu$ and in this case Eq. (2.22) becomes

$$(\omega_{\text{D}} + i\nu k^2)^2 - V_{\text{A}}^2 k^2 = 0, \quad (2.99)$$

where remembering that $\omega_{\text{D}} \equiv \omega - kU$. This quartic equation consists of two quadratic equations

$$(\omega - kU + i\nu k^2 - V_{\text{A}} k)(\omega - kU + i\nu k^2 + V_{\text{A}} k) = 0, \quad (2.100)$$

or more conveniently written

$$i\nu k^2 \mp (V_A \pm U)k + \omega = 0. \tag{2.101}$$

The solutions for the wave vectors are

$$k_{1,2,3,4} = \pm \frac{(V_A \pm U) \pm \sqrt{\mathcal{D}_k}}{2i\nu}, \qquad \mathcal{D}_k = (V_A \pm U)^2 - 4i\omega\nu, \tag{2.102}$$

which can also be represented in the form

$$k_{1,2,3,4} = \pm \frac{(V_A \pm U)}{2i\nu} \left( 1 \pm \sqrt{1 - \frac{4i\nu\omega}{(V_A \pm U)^2}} \right). \tag{2.103}$$

Expanding the square root in Taylor series for low frequencies $\omega \to 0$ and tailing up to the quadratic term

$$\sqrt{1 - \frac{4i\nu\omega}{(V_A \pm U)^2}} \approx 1 - \frac{2i\nu\omega}{(V_A \pm U)^2} + \frac{2\nu^2\omega^2}{(V_A \pm U)^4}. \tag{2.104}$$

Substituting this expansion in Eq. (2.103), the obtained solutions for the wave vectors become

$$k_{1,2,3,4} \approx \pm \frac{(V_A \pm U)}{2i\nu} \left[ 1 \pm \left( 1 - \frac{2i\nu\omega}{(V_A \pm U)^2} + \frac{2\nu^2\omega^2}{(V_A \pm U)^4} \right) \right], \tag{2.105}$$

which written in details are

$$k_1 = k_1' + ik_1'' = -\frac{\omega}{V_A + U} - i \left( \frac{V_A + U}{\nu} + \frac{\nu\omega^2}{(V_A + U)^3} \right) \equiv k_F = k_F' + ik_F'', \tag{2.106}$$

$$k_2 = k_2' + ik_2'' = -\frac{\omega}{V_A - U} - i \frac{\nu\omega^2}{(V_A - U)^3} \equiv k_L = k_L' + ik_L'', \tag{2.107}$$

$$k_3 = k_3' + ik_3'' = +\frac{\omega}{V_A + U} + i \frac{\nu\omega^2}{(V_A + U)^3} \equiv k_R = k_R' + ik_R'', \tag{2.108}$$

$$k_4 = k_4' + ik_4'' = +\frac{\omega}{V_A - U} + i \left( \frac{V_A - U}{\nu} + \frac{\nu\omega^2}{(V_A - U)^3} \right) \equiv k_D = k_D' + ik_D''. \tag{2.109}$$

The stiffness ratio of the eigenvalues in this low frequency limit

$$r_{DR} = \frac{|k_D|}{|k_R|} \approx \frac{k_D''}{k_R'} \approx \frac{V_A^2 - U^2}{\nu\omega} \gg 1, \tag{2.110}$$

meaning that the wave equations Eq. (2.20) form a very stiff system and indispensably has to be solved using algorithms for stiff systems.

Let

$$\psi_L(x) = \begin{pmatrix} b_L(x) \\ u_L(x) \\ \gamma_L(x) \\ w_L(x) \end{pmatrix}, \qquad \psi_R(x) = \begin{pmatrix} b_R(x) \\ u_R(x) \\ \gamma_R(x) \\ w_R(x) \end{pmatrix}, \tag{2.111}$$

are the solutions of the wave equations Eq. (2.20) with boundary conditions at $x = 0$

$$\psi_{\mathrm{L}}(0) = \mathrm{L}(0), \qquad \psi_{\mathrm{R}} = \mathrm{R}(0). \tag{2.112}$$

Physically AW come from the Sun and some of them (L-modes) are reflected from the TR $\psi(0) = \mathrm{R}(0) + r\mathrm{L}(0)$ therefore a solution as a linear combination

$$\psi(x) = \psi_{\mathrm{R}}(x) + \tilde{r}\psi_{\mathrm{L}}(x) \tag{2.113}$$

is looked for. Assuming that from the low viscosity chromospheric plasma overdamped D modes do not come and the overdamped F modes have negligible amplitudes for smaller $x$ (at the start), the influence of both overdamped modes can be safely neglected. For the run-away boundary conditions at $x = l$ it is supposed that no waves come from infinity and therefore

$$\psi(l) = \tilde{t}\mathrm{R}(l) + \tilde{c}_{\mathrm{D}}\mathrm{D}(l) + \tilde{c}_{\mathrm{F}}\mathrm{F}(l). \tag{2.114}$$

Combining Eq. (2.113) and Eq. (2.114) the equations for calculating the reflection $\tilde{r}$, transmission $\tilde{t}$ and the two mode conversion coefficients $\tilde{c}_{\mathrm{D}}$ and $\tilde{c}_{\mathrm{F}}$ are obtained

$$\psi_{\mathrm{R}}(l) + \tilde{r}\psi_{\mathrm{L}} = \tilde{t}\mathrm{R}(l) + \tilde{c}_{\mathrm{D}}\mathrm{D}(l) + \tilde{c}_{\mathrm{F}}\mathrm{F}(l). \tag{2.115}$$

Written by components

$$\begin{pmatrix} b_{\mathrm{R}}(l) \\ u_{\mathrm{R}}(l) \\ \gamma_{\mathrm{R}}(l) \\ w_{\mathrm{R}}(l) \end{pmatrix} = -\tilde{r} \begin{pmatrix} b_{\mathrm{L}}(l) \\ u_{\mathrm{L}}(l) \\ \gamma_{\mathrm{L}}(l) \\ w_{\mathrm{L}}(l) \end{pmatrix} + \tilde{t} \begin{pmatrix} \mathrm{R}_b(l) \\ \mathrm{R}_u(l) \\ \mathrm{R}_\gamma(l) \\ \mathrm{R}_w(l) \end{pmatrix} + \tilde{c}_{\mathrm{D}} \begin{pmatrix} \mathrm{D}_b(l) \\ \mathrm{D}_u(l) \\ \mathrm{D}_\gamma(l) \\ \mathrm{D}_w(l) \end{pmatrix} + \tilde{c}_{\mathrm{F}} \begin{pmatrix} \mathrm{F}_b(l) \\ \mathrm{F}_u(l) \\ \mathrm{F}_\gamma(l) \\ \mathrm{F}_w(l) \end{pmatrix}, \tag{2.116}$$

from where the convenient for numerical calculation matrix form of these equations is easily visible

$$\begin{pmatrix} b_{\mathrm{R}}(l) \\ u_{\mathrm{R}}(l) \\ \gamma_{\mathrm{R}}(l) \\ w_{\mathrm{R}}(l) \end{pmatrix} = \begin{pmatrix} \tilde{r} \\ \tilde{t} \\ \tilde{c}_{\mathrm{D}} \\ \tilde{c}_{\mathrm{F}} \end{pmatrix} \begin{pmatrix} -b_{\mathrm{L}}(l) & \mathrm{R}_b(l) & \mathrm{D}_b(l) & \mathrm{F}_b(l) \\ -u_{\mathrm{L}}(l) & \mathrm{R}_u(l) & \mathrm{D}_u(l) & \mathrm{F}_u(l) \\ -\gamma_{\mathrm{L}}(l) & \mathrm{R}_\gamma(l) & \mathrm{D}_\gamma(l) & \mathrm{F}_\gamma(l) \\ -w_{\mathrm{L}}(l) & \mathrm{R}_w(l) & \mathrm{D}_w(l) & \mathrm{F}_w(l) \end{pmatrix}. \tag{2.117}$$

This concludes the derivation of the MHD equations for the problem of the self-induced opacity of Alfvén waves. In the next section the numerical method for the solution of these equations is presented.

# Numerical Methods for Solution of the Derived Equations

The derived wave Eq. (2.20), dimensionless solar wind Eq. (2.66) and temperature Eq. (2.67) equations have to be solved simultaneously to obtain the height profiles of the temperature $T(x)$ and solar wind velocity $U(x)$. In short, the temperature $T(x)$ and solar wind velocity $U(x) = j/\rho(x)$ are used to calculate the kinetic coefficients derived in Sec. 1.4, which determine the absorption of the AW and this absorption determines the temperature $T(x + h)$ and solar wind velocity $U(x + h)$ at the next step $x + h$ of the calculation.

The present chapter thoroughly describes the numerical methods used for the calculations of the solar wind temperature and velocity height profiles. The used programming language for performing the numerical calculations is Fortran.

## 3.1   WKB Method for the Wave Equations

For high frequency waves we can use WKB approximation to solve the wave equations. In this case our problem reads

$$\Psi(\omega, x) = \sqrt{\mathcal{W}(\omega)}\, \mathrm{R}(x) \exp\left(\mathrm{i} \int_0^x k_{\mathrm{R}}(x')\, \mathrm{d}x'\right), \quad k_{\mathrm{R}}(x) = k'_{\mathrm{R}}(x) + \mathrm{i}k''_{\mathrm{R}}(x), \tag{3.1}$$

where $\mathcal{W}(\omega)$ is the spectral density of the AW and $\mathrm{R}(x)$ is the eigenvector of the right propagating wave (in the positive $x$ direction), $(B^2)_f = 2\pi V_{\mathrm{A}} \mathcal{W}\, \delta(\omega)$. The wave energy and momentum fluxes for a single AW in the WKB approximation have the form

$$q_{\text{wave}}(x) = \mathcal{W} \, q_{\text{wave}}(x=0) \exp\left(-\int_0^x 2k_{\text{R}}''(x)\mathrm{d}x\right),$$

$$\Pi_{\text{wave}}(x) = \mathcal{W} \frac{B_0^2}{4\mu_0} \left|\hat{b}^2(x=0)\right| \exp\left(-\int_0^x 2k_{\text{R}}''(x)\mathrm{d}x\right).$$

The eigenvalues and eigenvectors are calculated with EISPACK subroutines, which has been superseded for the most part by LAPACK [Anderson et al. (1999)].

## 3.2 Numerical Method for the Energy-Momentum Equations

The energy-momentum equations Eq. (2.42) and Eq. (2.49) are also stiff differential equations as both kinetic coefficients $\varkappa$ and $\xi$ multiplying respectively the derivatives of the temperature $\mathrm{d}_x T$ and solar wind velocity $\mathrm{d}_x U$ are very small. In addition, these equations are non-linear and this greatly complicates the solution of the whole problem, which consists of four first order stiff linear differential equations coupled to two first order stiff non-linear differential equations.

As already mentioned in Subsec. 1.1.6, all available commercial computation software was able to reach 30-time temperature increase and beyond that it stopped working. Reaching the remaining multiplier of 2-3 for the solar temperature requires the development an own robust method for frequently encountered numerical problem (four first order linear differential equations coupled to four first order non-linear differential equations). This section describes the theory of the developed numerical method for the solution of the derived MHD equations for the solar corona heating by AW absorption in the solar TR and the next section includes numerical examples used for testing and getting to know the work of the numerical method.

Instead of solving the non-linear differential equations Eq. (2.42) and Eq. (2.49), we can solve their corresponding quadratic equation Eq. (2.63) with a specially developed predictor-corrector method. This method consists of an extrapolating part termed prediction, which enables calculating the derivatives and solving the quadratic equation and a correcting part serving to verify the obtained solutions.

### 3.2.1 Aitken's interpolation method

A robust interpolating method is necessary for performing extrapolation calculations. This is a starter for the predictor-corrector method for solving the energy-momentum equations Eq. (2.61) and Eq. (2.62). The chosen method for the numerical analysis in the current work is Aitken's interpolation method [Aitken (1932), Bronshtein & Semedyayev (1955), Abramowitz & Stegun (1972), Korn & Korn (1968)], which reduces Newton's fundamental formula for interpolation by divided differences to a simple procedure when the differences are not given [Aitken (1932), Whittaker & Bartlett (1968)]. This procedure is easily implemented with a small computing machine [Whittaker & Bartlett (1968)] providing excellent results on both interpolation and extrapolation, as it is shown in the test examples at the end of this section.

The following description of Aitken's method here is in the terms given in the mathematical handbooks [Bronshtein & Semedyayev (1955), Sec. 6] or [Abramowitz & Stegun (1972), Subsec. 25.2]. Let $x_0, x_1, x_2, \ldots, x_n$ is a set of points with corresponding values $f_0, f_1, f_2, \ldots, f_n$ and $x_0 < x_1 < x_2 < \cdots < x_n$. Using the first $x_0$ and any other point of the set $x_k, k = 1, 2, \ldots, n$, the value $f(x_0, x_k)$ at a point $x$, where $x_0 < x < x_n$ is given by

$$f(x_0, x_k) = \frac{1}{x_k - x_0} \begin{vmatrix} f_0 & x_0 - x \\ f_k & x_k - x \end{vmatrix}. \tag{3.2}$$

If $x \equiv x_k$, the respective value $f_k$ will be returned evidently. This is a second degree polynomial interpolation. For the next degree polynomial interpolation, a third point $x_1$ is included. Now the method uses the points $x_0, x_1, x_k$ with $k = 2, 3, \ldots, n$ and the obtained interpolated value is

$$f(x_0, x_1, x_k) = \frac{1}{x_k - x_1} \begin{vmatrix} f(x_0, x_1) & x_1 - x \\ f(x_0, x_k) & x_k - x \end{vmatrix}, \tag{3.3}$$

where the second degree polynomials $f(x_0, x_k)$ are calculated from Eq. (3.2). Analogously the fourth point $x_2$ is added, $k = 3, 4, \ldots, n$

$$f(x_0, x_1, x_2, x_k) = \frac{1}{x_k - x_2} \begin{vmatrix} f(x_0, x_1, x_2) & x_2 - x \\ f(x_0, x_1, x_k) & x_k - x \end{vmatrix}, \tag{3.4}$$

where the third degree polynomials are calculated from Eq. (3.3). This procedure continues until the last point $x_n$ is included. In this way, $f(x_0, x_1, x_2, \ldots, x_n)$ obtained from the $n$-th degree polynomial interpolation is the most accurate interpolated value at the point $x$.

Up to now it is assumed that $x_0 < x < x_n$. This is not a requirement and the method can be used when $x$ is smaller or larger than all the points from the set, i.e. $x < x_0 < \cdots < x_n$ or $x_0 < \cdots < x_n < x$. In both cases the procedure of finding the corresponding value $f(x_0, x_1, x_2, \ldots, x_n)$ for $x$ is exactly the same, with the difference that this time the Aitken method performs an extrapolation.

But what can one read about the process of extrapolation in the literature for numerical recipes [Press et al. (1992)]? *There, considerable care is taken with the monitoring of errors. Otherwise, the dangers of extrapolation cannot be overemphasized: An interpolating function, which is is perforce an extrapolating function, will typically go berserk when the argument $x$ is outside the range of tabulated values by more than the typical spacing of tabulated points.* By the very famous law: whatever can go wrong, will go wrong, in very interesting for the physics problems, the commercial software stops working and the research can continue only in co-authorship with the source code authors. Therefore the most appropriate approach in this situation is the researchers of the problem to develop and program own method for solution of the MHD equations. The next subsection thoroughly describes the developed numerical method.

### 3.2.2 Padé approximants and Wynn's epsilon algorithm

Having performed an extrapolation with the Aitken's interpolation method, a fast convergence algorithm is necessary for correcting and verifying the obtained extrapolated value. Padé approximants give fast convergence and are actually the best rational approximation, for a general introduction in the problem of Padé approximants, see the well-known monographs Refs. [Mishonov & Varonov (2019a), Brezinski and Zaglia (1991), Brezinski (1996)]. Since the problem was first systematically studied by Jacobi in the 1840s, later Frobenius [Wynn (1966), Brezinski (1991), Frobenius (1881)] and the Padé table introduced by Padé [Brezinski (1991), Padé (1892)] in the 19$^{\text{th}}$ century, the contemporary literature is enormous. However, Padé approximants are sensitive with respect to the noise of the discrete representation of real numbers [Press et al. (1992)]: *That is the downside of the Padé approximation: it is uncontrolled. There is, in general, no way to tell how accurate it is, or how far out in $x$ it can be usefully extended. It is powerful, but in the end still mysterious, technique.*

Peter Wynn's epsilon ($\varepsilon$) algorithm based on Shanks transformations [Shanks (1955), Wynn (1956)] is one of the most familiar and efficient acceleration convergent methods [Brezinski (1996)], which can be used for the calculation of the Padé table [Wynn (1966)]. This subsection follows Wynn's derivation from 1966 [Wynn (1966)] of the connection between the Padé table and the $\varepsilon$-algorithm, and of the Wynn identity.

A Padé approximant of a function $f(z)$, which can be expanded in the power series

$$f(z) = \sum_{k=0}^{\infty} \mathrm{u}_k z^k \tag{3.5}$$

is a rational fraction [Mishonov & Varonov (2019a)]

$$\mathcal{P}_{i,j}(z) = \frac{\sum_{k=0}^{j} \mathrm{d}_k z^k}{\sum_{k=0}^{i} \mathrm{q}_k z^k}, \tag{3.6}$$

where $\mathrm{d}_k$ and $\mathrm{q}_k$ are the numerator and denominator coefficients respectively. Let us mention that the convention for the indices $i, j$ here is according to [Wynn (1966)], unlike [Mishonov & Varonov (2019a)] where the first index is for the numerator and the second one is for the denominator. Taking the outer boundaries

$$\mathcal{P}_{-1,j} = \infty, \tag{3.7}$$
$$\mathcal{P}_{i,-1} = 0 \tag{3.8}$$

a complete Padé table can be constructed

$$\begin{array}{ccccc}
\infty & \infty & \ldots & \infty & \ldots \\
0 & \mathcal{P}_{0,0} & \mathcal{P}_{0,1} & \ldots & \mathcal{P}_{0,j} & \ldots \\
0 & \mathcal{P}_{1,0} & \mathcal{P}_{1,1} & \ldots & \mathcal{P}_{1,j} & \ldots \\
\vdots & \vdots & \vdots & & \vdots & \\
0 & \mathcal{P}_{i,0} & \mathcal{P}_{i,1} & \ldots & \mathcal{P}_{i,j} & \ldots \\
\vdots & \vdots & \vdots & \ldots & \vdots &
\end{array} \tag{3.9}$$

The fundamental relationships of the $\varepsilon$-algorithm are [Wynn (1956), Wynn (1966)]

$$(\epsilon_{s+1}^{(m)} - \epsilon_{s-1}^{(m+1)})(\epsilon_s^{(m+1)} - \epsilon_s^{(m)}) = 1 \tag{3.10}$$

or else written

$$\epsilon_{s+1}^{(m)} = \epsilon_{s-1}^{(m+1)} + \frac{1}{\epsilon_s^{(m+1)} - \epsilon_s^{(m)}} \tag{3.11}$$

and they relate the functions $\epsilon_s^{(m)}$, which may be placed in a table, where the subscript indicates a column and the superscript indicates a diagonal. Such a table (or an extract from a table more precisely) looks like

$$\begin{array}{ccc} & \epsilon_s^{(m)} & \\ \epsilon_{s-1}^{(m+1)} & & \epsilon_{s+1}^{(m)} \\ & \epsilon_s^{(m+1)} & \end{array} \tag{3.12}$$

and if we make a comparison with a table containing the world direction $\mathcal{E}$ (East), $\mathcal{W}$ (West), $\mathcal{S}$ (South) and $\mathcal{N}$ (North)

$$\begin{array}{ccc} & \mathcal{N} & \\ \mathcal{W} & & \mathcal{E}, \\ & \mathcal{S} & \end{array} \tag{3.13}$$

then the relationships of the $\varepsilon$-algorithm Eq. (3.10) and Eq. (3.11) can be simply written

$$(\mathcal{E} - \mathcal{W})(\mathcal{S} - \mathcal{N}) = 1 \quad \text{and} \quad \mathcal{E} = \mathcal{W} + \frac{1}{\mathcal{S} - \mathcal{N}}. \tag{3.14}$$

Applying the initial conditions Eq. (3.5), Eq. (3.7) and Eq. (3.8) to the $\varepsilon$-algorithm relationships Eq. (3.10) (or Eq. (3.11)) we have

$$\epsilon_{-1}^{(m)} = 0, \qquad m = 1, 2, \ldots, \tag{3.15}$$

$$\epsilon_{2r}^{(-r-1)} = 0, \qquad r = 0, 1, \ldots, \tag{3.16}$$

$$\epsilon_0^{(m)} = \sum_{k=0}^{m+l} \mathrm{u}_k z^k, \qquad m = 0, 1 \ldots \tag{3.17}$$

and the following connection between the $\varepsilon$-algorithm and the Padé table is obtained

$$\epsilon_{2r}^{(m)} = \mathcal{P}_{r,m+r}, \qquad r = 0, 1, \ldots, \qquad m = -r, -r+1, \ldots \quad , \tag{3.18}$$

cf. [Wynn (1966), Eq. (10)]. It is important to note that the functions of the even order $\epsilon_{2r}^{(m)}$ occur in the transpose of the Padé table, i.e. $\epsilon_{2r}^{(m)}$ lie in a column, while $\mathcal{P}_{r,m+r}$ lie in a row of the Padé table.

Let us now consider the following array of $\epsilon$ functions

$$\begin{array}{ccccc} & & \epsilon_{2r}^{(m-1)} & & \\ & \epsilon_{2r-1}^{(m)} & & \epsilon_{2r+1}^{(m-1)} & \\ \epsilon_{2r-2}^{(m+1)} & & \epsilon_{2r}^{(m)} & & \epsilon_{2r+2}^{(m-1)} \\ & \epsilon_{2r-1}^{(m+1)} & & \epsilon_{2r+1}^{(m)} & \\ & & \epsilon_{2r}^{(m+1)} & & \end{array} \quad . \tag{3.19}$$

Applying Eq. (3.10) for the two diamonds above and below with $\epsilon_{2r}^{(m)}$ a common vertex, we obtain

$$\epsilon_{2r+1}^{(m-1)} - \epsilon_{2r-1}^{(m)} = (\epsilon_{2r}^{(m)} - \epsilon_{2r}^{(m-1)})^{-1}, \tag{3.20}$$

$$\epsilon_{2r+1}^{(m)} - \epsilon_{2r-1}^{(m+1)} = (\epsilon_{2r}^{(m+1)} - \epsilon_{2r}^{(m)})^{-1}. \tag{3.21}$$

A subtraction of these equations results in

$$(\epsilon_{2r+1}^{(m-1)} - \epsilon_{2r+1}^{(m)}) - (\epsilon_{2r-1}^{(m)} - \epsilon_{2r-1}^{(m+1)}) = (\epsilon_{2r}^{(m+1)} - \epsilon_{2r}^{(m)})^{-1} - (\epsilon_{2r}^{(m)} - \epsilon_{2r}^{(m-1)})^{-1}. \tag{3.22}$$

Analogously applying Eq. (3.10) for the two diamonds from the right and left with $\epsilon_{2r}^{(m)}$ a common vertex, we obtain

$$\epsilon_{2r+1}^{(m)} - \epsilon_{2r+1}^{(m-1)} = (\epsilon_{2r+2}^{(m-1)} - \epsilon_{2r}^{(m)})^{-1}, \tag{3.23}$$

$$\epsilon_{2r-1}^{(m+1)} - \epsilon_{2r-1}^{(m)} = (\epsilon_{2r}^{(m)} - \epsilon_{2r-2}^{(m+1)})^{-1}. \tag{3.24}$$

The left terms of both equations are the same as respectively the first and second left hand side terms in Eq. (3.22). The substitution of the right hand terms into Eq. (3.22) yields the relationship

$$(\epsilon_{2r+2}^{(m-1)} - \epsilon_{2r}^{(m)})^{-1} - (\epsilon_{2r}^{(m)} - \epsilon_{2r-2}^{(m+1)})^{-1} = (\epsilon_{2r}^{(m+1)} - \epsilon_{2r}^{(m)})^{-1} - (\epsilon_{2r}^{(m)} - \epsilon_{2r}^{(m-1)})^{-1}. \tag{3.25}$$

cf. [Wynn (1966), Eq. (13)], which involves only even suffix functions. This equation is called the Wynn identity, which by using $r = i$, $m + r = j$, $i, j = 0, 1, \ldots$ and Eq. (3.18) can also be written with Padé approximants

$$(\mathcal{P}_{i+1,j} - \mathcal{P}_{i,j})^{-1} - (\mathcal{P}_{i,j} - \mathcal{P}_{i-1,j})^{-1} = (\mathcal{P}_{i,j+1} - \mathcal{P}_{i,j})^{-1} - (\mathcal{P}_{i,j} - \mathcal{P}_{i,j-1})^{-1}, \tag{3.26}$$

cf. [Wynn (1966), Eq. (15)]. This equation can also be written in the notations of the world directions [Wynn (1966)] and Gragg [Gragg (1972), Theorem 5.5], who even baptized this equation as *missing identity of Frobenius* (discrediting P. Wynn most probably because of inability to produce that kind of a result). Writing the Padé approximants from Eq. (3.26) in a table

$$\begin{array}{ccc} & \mathcal{P}_{i-1,j} & \\ \mathcal{P}_{i,j-1} & \mathcal{P}_{i,j} & \mathcal{P}_{i,j+1} \\ & \mathcal{P}_{i+1,j} & \end{array} \tag{3.27}$$

and comparing them with the table of the world directions

$$\begin{array}{ccc} & \mathcal{N} & \\ \mathcal{W} & \mathcal{C} & \mathcal{E} \\ & \mathcal{S} & \end{array}, \tag{3.28}$$

where $\mathcal{C}$ stands for Center, after substitution and sign changes of the terms, the Wynn identity can be written also as

$$(\mathcal{C} - \mathcal{S})^{-1} + (\mathcal{C} - \mathcal{N})^{-1} = (\mathcal{C} - \mathcal{E})^{-1} + (\mathcal{C} - \mathcal{W})^{-1}, \tag{3.29}$$

cf. [Wynn (1966), Eq. (16)].

Together with the Aitken interpolation method described in Subsec. 3.2.1, Wynn epsilon algorithm and identity are the basis of the prediction part of the developed numerical method for solution of the derived in Sec. 2.2 energy and momentum equations Eq. (2.61) and Eq. (2.62). The technical details of the developed prediction algorithm together with its performance are given in the next section, while the next subsection contains the corrector part of our numerical method.

### 3.2.3 Quasi-Newton method

Let us have the following recursive functions

$$x = f(x, y), \tag{3.30}$$

$$y = g(x, y). \tag{3.31}$$

An expansion in their Taylor series for $x$ and $y$ gives

$$x - f = x_0 - f_0 + \partial_x f_0 (x - x_0) + \partial_y f_0 (y - y_0), \tag{3.32}$$

$$y - g = y_0 - g_0 + \partial_x g_0 (x - x_0) + \partial_y g_0 (y - y_0), \tag{3.33}$$

where

$$\partial_x \equiv \frac{\partial}{\partial x}, \qquad f_0 \equiv f(x_0, y_0), \qquad \partial_y g_0 \equiv \frac{\partial g}{\partial y}\bigg|_{(x_0, y_0)}.$$

Supposing a solution for Eq. (3.30) and Eq. (3.31) has not been reached, therefore the Taylor expansions are different from 0. Therefore we seek

$$x_0 - f_0 + \partial_x f_0 (x - x_0) + \partial_y f_0 (y - y_0) = 0, \tag{3.34}$$

$$y_0 - g_0 + \partial_x g_0 (x - x_0) + \partial_y g_0 (y - y_0) = 0. \tag{3.35}$$

Both equations can be written in the matrix representation

$$\begin{pmatrix} f_0 - x_0 \\ g_0 - y_0 \end{pmatrix} = \mathcal{Q} \begin{pmatrix} x - x_0 \\ y - y_0 \end{pmatrix}, \qquad \mathcal{Q} \equiv \begin{pmatrix} \partial_x f_0 & \partial_y f_0 \\ \partial_x g_0 & \partial_y g_0 \end{pmatrix}, \tag{3.36}$$

where the solution for $x$ and $y$ is

$$\begin{pmatrix} x \\ y \end{pmatrix} = \begin{pmatrix} x_0 \\ y_0 \end{pmatrix} + \mathcal{Q}^{-1} \begin{pmatrix} f_0 - x_0 \\ g_0 - y_0 \end{pmatrix}. \tag{3.37}$$

By differentiating the equations Eq. (3.32) and Eq. (3.33) the derivatives are expressed as

$$\partial_x f_0 = 1 - \partial_x f, \qquad \partial_y f_0 = -\partial_y f, \tag{3.38}$$

$$\partial_y g_0 = 1 - \partial_y g, \qquad \partial_x g_0 = -\partial_x g. \tag{3.39}$$

Substituting these in the matrix $\mathcal{Q}$, the inverse matrix

$$\mathcal{Q}^{-1} \equiv \frac{1}{\mathrm{Det}(\mathcal{Q})} \begin{pmatrix} \partial_y g_0 & -\partial_y f_0 \\ -\partial_x g_0 & \partial_x f_0 \end{pmatrix} = \frac{1}{\mathrm{Det}(\mathcal{Q})} \begin{pmatrix} 1 - \partial_y g & \partial_y f \\ \partial_x g & 1 - \partial_x f \end{pmatrix}, \tag{3.40}$$

where

$$\mathrm{Det}(\mathcal{Q}) \equiv (1 - \partial_x f)(1 - \partial_y g) - (\partial_x g)(\partial_y f). \tag{3.41}$$

Finally substituting the derived expression for the inverse matrix into Eq. (3.37) the solutions for $x$ and $y$ are

$$\begin{pmatrix} x \\ y \end{pmatrix} = \begin{pmatrix} x_0 \\ y_0 \end{pmatrix} + \frac{1}{\mathrm{Det}(\mathcal{Q})} \begin{pmatrix} 1 - \partial_y g & \partial_y f \\ \partial_x g & 1 - \partial_x f \end{pmatrix} \begin{pmatrix} f_0 - x_0 \\ g_0 - y_0 \end{pmatrix}. \tag{3.42}$$

This is a method for calculation of consecutive approximations and a general method for finding of zeros of functions. Having approximate solutions $f_0$ and $g_0$ with the arguments $x_0$ and $y_0$, the new arguments $x$ and $y$ can be found, giving closer approximate solutions. This procedure can be implemented numerous times up to the required numerical precision.

## 3.3 Wynn-Epsilon Algorithm for Choice of the Optimal Padé Approximant

Often the difference between sequential Padé approximants gives a reasonable evaluation of the error [Mishonov & Varonov (2019a)]. In case of convergence, we have vanishing differences between the values of different cells of the Padé table $(\mathcal{P}_{i,j+1} - \mathcal{P}_{i,j}) \to 0$ for $i, j \to \infty$ and $i/j = \mathrm{const}$. Therefore the used criterion for choosing the optimal Padé approximant is based on the Wynn identity Eq. (3.26)

$$\frac{1}{\eta_{i,j}} \equiv \frac{1}{\mathcal{P}_{i+1,j} - \mathcal{P}_{i,j}} + \frac{1}{\mathcal{P}_{i-1,j} - \mathcal{P}_{i,j}} = \frac{1}{\mathcal{P}_{i,j+1} - \mathcal{P}_{i,j}} + \frac{1}{\mathcal{P}_{i,j-1} - \mathcal{P}_{i,j}}. \tag{3.43}$$

In this case $\eta_{i,j} \to 0$ also and the minimal value of $\eta_{i,j}$ gives a reasonable criterion for the minimal error and the choice of optimal Padé approximant. This theoretical hint has been proved on many examples of calculation of Padé approximations and has been arrived at the conclusion that the long sought criterion for the choice of the optimal Padé approximant is simply a search for the minimal value $|\eta_{i,j}|$, i.e.

$$\eta_{\min} \equiv |\eta_{I,J}| = \min_{i,j} |\eta_{i,j}| \tag{3.44}$$

and $f(z) \approx \mathcal{P}_{I,J}$. It is technologically and aesthetically attracting that the Wynn identity gives simultaneously a method for the calculation of Padé approximants and a method for the evaluation of the error.

### 3.3.1 Technical implementation of the algorithm

Having a numerical sequence $\{f_0, f_1, f_2, \ldots, f_n\}$ that has already been extrapolated with the Aitken's interpolation method, we need to calculate the limit $f = \lim_{l \to \infty} f_l$. The well-known method is to initialize $\mathcal{P}_{0,j} = f_l$, for $j = 0, 1, \ldots, n$ and $\mathcal{P}_{-1,j} = \infty$ according to Eq. (3.7). Then we calculate in the south direction the corresponding Padé approximants

$$\mathcal{P}_{i+1,j} = \mathcal{P}_{i,j} + \cfrac{1}{\cfrac{1}{\mathcal{P}_{i,j+1} - \mathcal{P}_{i,j}} + \cfrac{1}{\mathcal{P}_{i,j-1} - \mathcal{P}_{i,j}} - \cfrac{1}{\mathcal{P}_{i-1,j} - \mathcal{P}_{i,j}}} \qquad (3.45)$$

and simultaneously calculate the empirical error

$$\eta_{i,j} = \cfrac{1}{\cfrac{1}{\mathcal{P}_{i,j+1} - \mathcal{P}_{i,j}} + \cfrac{1}{\mathcal{P}_{i,j-1} - \mathcal{P}_{i,j}}}. \qquad (3.46)$$

The minimal absolute value in the η-table $\eta_{\min}$ is the criterion for the determination of the optimal Padé approximant. According to the best we know, this criterion has never been implemented in the numerical recipes so far.

For the programming task, we have to calculate all the values, for which division is possible. In the next subsection several technical examples are thoroughly described.

### 3.3.2 Performance of the algorithm

This subsection contains several well-known examples for calculation of divergent series and function extrapolation. These practical tests are crucial for the understanding of the Wynn-Epsilon algorithm for the determination of the optimal Padé approximant since its final result can be compared to the real results. In this way the applicability and limitations of the algorithm become known, which must be accounted for before using the algorithm for real problems calculation.

Before starting with the examples, the key technical notions of the algorithm should be introduced. As we have already seen in the beginning of the current section, $|\eta_{i,j}|$ gives the criterion for minimal error. Since the algorithm as a final result chooses the Padé approximant with smallest error $\eta_{\min}$, this error is the empirical error, which from now on is denoted as $\varepsilon_{\mathrm{emp}}$. The real error is denoted as $\varepsilon_{\mathrm{real}}$, the machine error (also called machine epsilon) is $\varepsilon_{\mathrm{mach}}$ and both empirical and real errors are given in machine error units. This ensures that the results from the performed tests are machine, platform and operating system independent, while for thoroughness let us note that double precision floats have been used. Alongside the calculation errors, the indices of the Padé approximants are used also. However, the implementation of the algorithm with respect to the indices is reversed in comparison with the Padé convention used in Subsec. 3.2.2, Eq. (3.6) and Eq. (3.9), meaning that the first index denoted by $i_{\mathrm{Pade}}$ from now on shows the numerator and the second one denoted by $k_{\mathrm{Pade}}$ from now on shows the denominator. This also means that the $\varepsilon$-table and the Padé table are not transposed, which is a little bit easier for analysis. Finally, an optimal number $N_{\mathrm{opt}}$ showing the number of the used input elements is necessary because not always all elements from the input sequence $\{f_0, f_1, f_2, \ldots, f_n\}$ will be used, i.e. $N_{\mathrm{opt}} \leq n + 1$. And since the programmed arrays and cycles of the algorithm start from 0, $N_{\mathrm{opt}} = i_{\mathrm{Pade}} + k_{\mathrm{Pade}} + 1$.

The first test example is the infinite series with a special sign pattern ($[k/2]$ means that the first 2 terms are positive, the next 2 are negative, the next 2 are positive and so on)

$$S_p = \sum_{k=0}^{\infty} \frac{(-1)^{[k/2]}}{k+1} = 1 + \frac{1}{2} - \frac{1}{3} - \frac{1}{4} + \frac{1}{5} + \frac{1}{6} - \cdots = \frac{\pi}{4} + \frac{1}{2}\ln 2, \qquad (3.47)$$

which poses a real challenge for most of the known convergence acceleration methods [Sidi (2006)]. Without the usage of Aitken's iteration method and varying the number of input terms $n$ from the minimum number 3 up to 100, the solutions for each $n$ the algorithm returns are shown in Fig. 3.1. Both real and empirical errors Fig. 3.2, $N_{opt}$ Fig. 3.3 and indices $i_{Pade}$ and $k_{Pade}$ Fig. 3.4 depending on $n$ are shown. Pixel accuracy is quickly reached for $n = 8$ in Fig. 3.1 and after that this figure is no more informative. The errors in Fig. 3.2 give a detailed idea of the algorithm precision. Both $\varepsilon_{emp}$ and $\varepsilon_{real}$ decrease linearly with the former being slightly larger, which is to be expected. This behaviour continues to $n = 40$, where both errors have minimum values with $\varepsilon_{emp} = 0$ and therefore disappears from the log graph. For $n > 40$ $\varepsilon_{real}$ slightly increases in steps, while $\varepsilon_{emp} = 0$ except for 4 occasions, where it is around 1. Analogous linear behaviour, albeit increase, up to $n = 40$ of $N_{opt}$ in Fig. 3.3 and the Padé indices in Fig. 3.4 is seen. For $n > 40$ $N_{opt}$ continues increasing in small steps until $n = 57$ and $n = 73$, where the steps are a lot larger, meaning that the algorithm does not need the new series terms. This leads to a decreased efficiency since the additional terms are consumed but turn out to yield Padé approximants with larger empirical errors. Looking back to $\varepsilon_{real}$ in Fig. 3.2, it is obvious that the steps slightly worsen the final result, as $\varepsilon_{real}$ slightly increases with almost the same steps, although there is a barely visible decrease. This strengthens the statement about the decreased efficiency, since with the increased consumption, the precision is getting lower, albeit a little bit, but the focus remains on the increased calculation time, which is much larger for larger $n$. The indices in Fig. 3.4 also show when the efficiency starts decreasing, for $n \leq 40$ $i_{Pade} \approx k_{Pade}$, meaning that the solution is a diagonal (in the case of $i_{Pade} = k_{Pade}$) or close to the diagonal Padé approximant. For $n > 40$ $i_{Pade}$ and $k_{Pade}$ diverge with the former increasing and the latter decreasing in steps. The large steps $57 \leq n < 73$ and $73 \leq n \leq 100$ in Fig. 3.3 are also present in Fig. 3.4, which is to be expected, albeit slight perturbations of $i_{Pade}$ and $k_{Pade}$ should not be excluded as a possible outcome.

Perhaps the most famous example summation of divergent series by Padé approximants is the calculation of the divergent Taylor series $\ln(1+x)$ beyond the radius of convergence $|x| < 1$

$$\ln(1+x) = \lim_{n \to \infty} S_n, \qquad S_n(x) \equiv \sum_{k=1}^{n} (-1)^{(k+1)} \frac{x^k}{k}. \qquad (3.48)$$

The optimal Padé approximant for every positive $x$ shown in Fig. 3.5, the errors of the calculation are shown in Fig. 3.6, the optimal number of used terms $N_{opt}$ is shown in Fig. 3.7 and the Padé approximant indices are shown in Fig. 3.8. Even for $x \approx 20$ for the calculation of the series $\ln(1+x)$ a pixel accuracy is present. For larger values of $x$ the dots representing the calculation *evaporate* from the line representing the exact value, meaning that the pixel accuracy has just been lost. The empirical error $\varepsilon_{emp}$ shows a saturation and is a reliable indicator for the calculation accuracy.

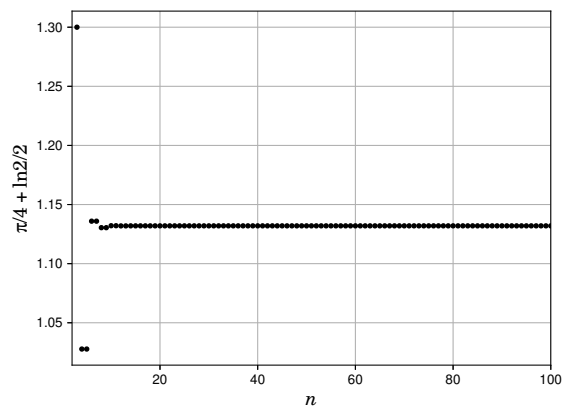

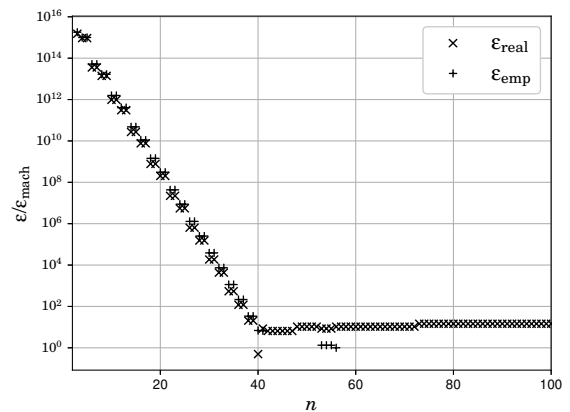

Figure 3.1: The results of the series $S_p$ Eq. (3.47) calculation depending on the number of input terms $n$. After $n = 8$ pixel (below percent) accuracy is reached, which can also be seen in the errors in Fig. 3.2. For $n > 8$ this figure is useless, unless rescaled for $9 \leq n \leq 100$, of course. However, much more information is present in the errors Fig. 3.2, $N_{opt}$ Fig. 3.3 and indices Fig. 3.4 dependencies corresponding to the calculation shown here.

Figure 3.2: The errors in machine epsilon $\varepsilon_{mach}$ units of the series $S_p$ Eq. (3.47) calculation in Fig. 3.1 depending on the number of input terms $n$. Both real $\varepsilon_{real}$ and calculated $\varepsilon_{emp}$ errors have the same behaviour and are close in values until $n = 40$. For $n = 40$ $\varepsilon_{real}$ is minimal, while $\varepsilon_{emp} = 0$ and therefore disappears from the log axis. For $n > 40$ $\varepsilon_{real}$ slightly increases in long step-like manner, while $\varepsilon_{emp}$ gets slightly larger or equal than 1 in 4 occurrences.

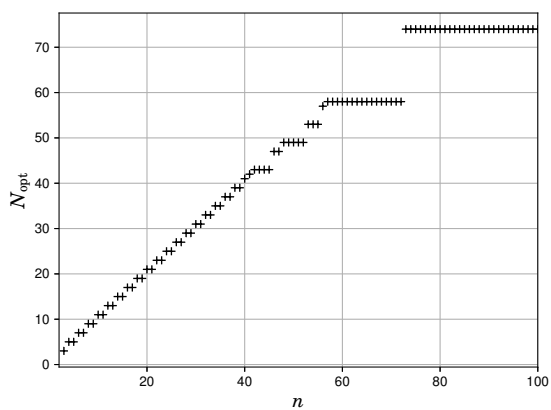

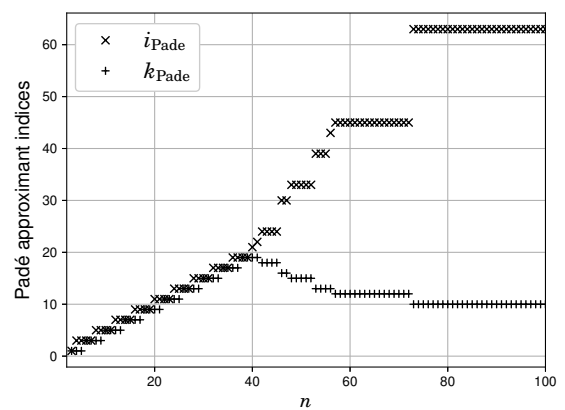

Figure 3.3: The optimal number of the used input terms $N_{opt}$ for the series $S_p$ Eq. (3.47) calculation in Fig. 3.1 depending on the number of input terms $n$. A steady linear increase of $N_{opt}$ is present up to $n = 40$, while for larger values of $n$ the algorithm starts not using all the elements. For $n > 40$ $N_{opt}$ is kept constant for a few $n$ and then it is increased but after $n = 57$ and the jump in $n = 73$ there are large plateaus.

Figure 3.4: The indices of the optimal Padé approximants of the series $S_p$ Eq. (3.47) calculation in Fig. 3.1 depending on the number of input terms $n$. The behaviour here is analogous to that of $N_{opt}$ in Fig. 3.3 for $n \leq 40$. For $n > 40$ the numerator $i_{Pade}$ increases in steps, while the denominator $k_{Pade}$ decreases also in steps. After $n = 57$ and the jump at $n = 73$ both indices remain constant as it is in Fig. 3.3.

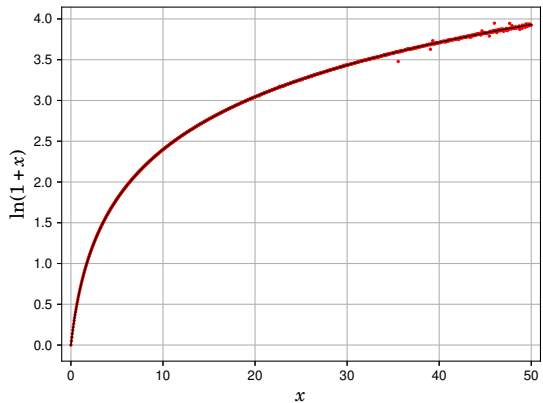

Figure 3.5: The continuous line is the logarithmic function and the dots represent the series summation with $\varepsilon_{\text{mach}}$ criterion for the optimal choice. One can see that up to $x \approx 20$ for $n = 101$ the pixel accuracy is conserved. The most important detail is how and when the method stops working and when the resources of the numerical accuracy are exhausted. The calculated values are *evaporated* from the analytical curve, but $\varepsilon_{\text{emp}}$ criterion gives reliable warning depicted in Fig. 3.6.

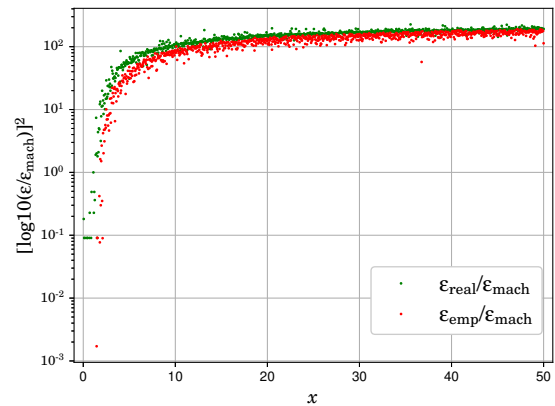

Figure 3.6: Square of decimal logarithms of empirical $\varepsilon_{\text{emp}}$ and real $\varepsilon_{\text{real}}$ errors in $\varepsilon_{\text{mach}}$ units versus the argument of the function for the calculation of the $\ln(1 + x)$ series in Fig. 3.5. It is evident that close to the convergence radius both errors are small and almost linearly increase. Then the errors reach saturation when the numerical resources of the fixed accuracy are exhausted, again just like Fig. 3.2 both errors have the same behaviour.

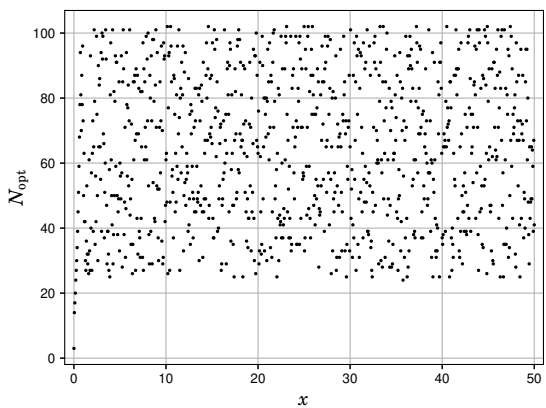

Figure 3.7: The optimal number of used input terms $N_{\text{opt}}$ for the calculation of the series Eq. (3.48) shown in Fig. 3.5. Inside the radius of convergence $|x| < 1$, where $N_{\text{opt}}$ is small with the tendency of a linear increase, a result is quickly and very precisely obtained seen in the errors in Fig. 3.6. For $|x| > 1$ $N_{\text{opt}}$ is evenly dispersed giving no hints of the accuracy behaviour beyond the convergence radius.

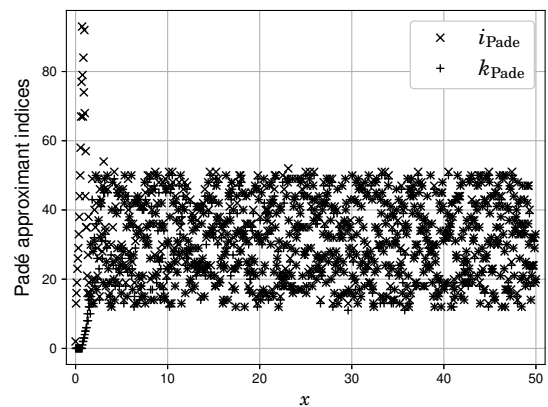

Figure 3.8: The indices of the optimal Padé approximants for the calculation of the series Eq. (3.48) shown in Fig. 3.5. In the radius of convergence $|x| < 1$ a very precise result is quickly obtained, which is seen in the large $i_{\text{Pade}}$ and small (even 0 at the beginning) $k_{\text{Pade}}$. For $|x| > 1$ diagonal or close to diagonal approximants are chosen, again giving no hints of the accuracy behaviour.

Moreover, the real error $\varepsilon_{\text{real}}$ shows the same and it is evident that both errors again have the same behaviour, just like in the previous example in Fig. 3.2. This saturation is the most important detail in the analysis of this example, since it indicates how and when the numerical method stops working, and when the resources of the numerical accuracy have been exhausted. Looking at the other two figures, however, $N_{\text{opt}}$ in Fig. 3.7 and the indices of the optimal Padé approximants in Fig. 3.8 give no hints about the accuracy behaviour beyond the radius of convergence, since $N_{\text{opt}}$, $i_{\text{Pade}}$ and $k_{\text{Pade}}$ are all evenly dispersed beyond the convergence radius $|x| > 1$. The only information shown in these two figures is that inside the convergence radius $|x| < 1$ a very precise result is quickly obtained, since $N_{\text{opt}}$ is small with the tendency of linear increase, $i_{\text{Pade}}$ is large and $k_{\text{Pade}}$ is small, even 0 at the beginning. Finally a comparison between the empirical $\varepsilon_{\text{emp}}$ and real $\varepsilon_{\text{real}}$ is shown in Fig. 3.9. A linear regression performed on the logarithms of these errors reveals a very high correlation coefficient 0.961, which qualitatively confirms the statement that $\varepsilon_{\text{emp}} \equiv \eta_{\text{min}}$ is a reliable indicator for the calculation accuracy. Now it can be safely concluded that the long sought criterion for empirical evaluation of the accuracy of the Padé approximants calculated by $\varepsilon$-algorithm has already been found.

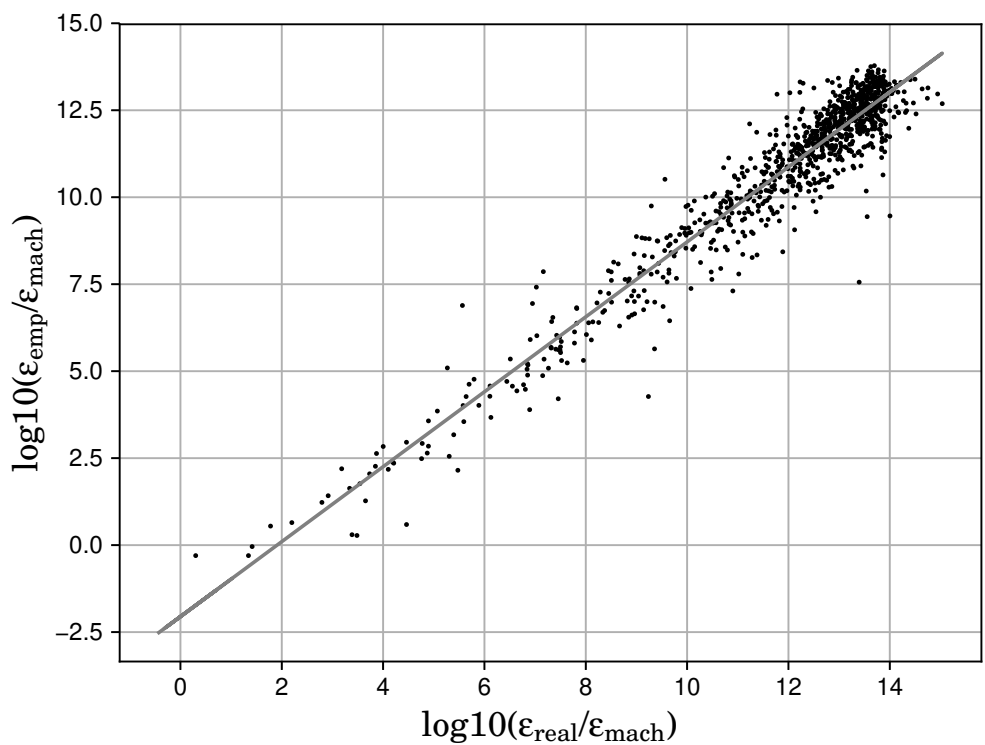

Figure 3.9: The logarithm of the empirical $\varepsilon_{\text{emp}}$ versus logarithm of the real $\varepsilon_{\text{real}}$ error both in $\varepsilon_{\text{mach}}$ units for the calculation of the $\ln(1 + x)$ series in Fig. 3.5. The high correlation coefficient 0.961 of the linear regression reveals that the long sought criterion for empirical evaluation of the accuracy of the Padé approximants calculated by $\varepsilon$-algorithm has already been found.

The last technical example to be considered is the problem of extrapolation of functions, which is illustrated in the case of the $\sin(x)$ function. $N$ equidistant interpolation points from one arch of the $\sin(x)$ function are taken and the next arch and even beyond is extrapolated. Aitken's interpolation method is used here to order the interpolation points within the arch and the point to be extrapolated. In this manner a numerical sequence is given to the Wynn-Epsilon algorithm to calculate its limit. The described calculation for the $\sin(x)$ function is shown in Fig. 3.10, together with its errors Fig. 3.11, optimal number of used interpolation points $N_{opt}$ Fig. 3.12 and indices of the optimal Padé approximants Fig. 3.13. The preceding arch in the interval $[-\pi, 0]$ contains 21 interpolation points. Using these points, one arch with 2000 points in the interval $(0, \pi]$ is extrapolated and continuation of the extrapolation in an attempt to obtain a second arch with the same number of points in the next interval $(\pi, 2\pi]$ is performed. The deviation of the points from the real function represented with the line in Fig. 3.10 shows the limit of applicability of the Aitken-Wynn extrapolation algorithm. Detailed error estimates of the extrapolation are shown in Fig. 3.11. Again, both $\varepsilon_{emp}$ and $\varepsilon_{real}$ errors have the same behaviour, meaning that the $\varepsilon_{emp}$ criterion gives reliable order estimation of the error. A comparison of both errors is given in Fig. 3.14, where correlation coefficient of the linear regression is 0.979. Finally to mention the number of the used interpolation points $N_{opt}$ only reveals that an extrapolation close to the interpolating points needs fewer interpolation points and is more accurate as it can be seen in Fig. 3.11, which is of course to be expected as already discussed at the end of Subsec. 3.2.1 based on [Press et al. (1992)]. The indices of the optimal Radé approximant reveal nothing about performance, nor accuracy as they are evenly distributed in Fig. 3.13. Maybe it's worth noting that usually $i_{Pade} > k_{Pade}$, although there are occasions where both are equal and the diagonal approximant has been chosen.

### 3.3.3 Concluding remarks

These new results of the implementation of calculation of Padé approximants and their application is the modulus minimization $\eta_{min}$ of the Wynn identity as a reliable empirical criterion of the error [Mishonov & Varonov (2019a), Mishonov & Varonov (2019b)].

The preformed analysis of several simple examples has revealed that for practical implementation of Padé approximants the empirical $\eta_{min}$ error extracted from the Wynn's identity ca be reliably used. In the agenda the statistical problem of calculation of probability distribution function (PDF) of the Padé approximants has already been set. The comparison of descriptive statistics data for the PDF of errors of calculation of Padé approximants by different criteria will give what the answer what general recommendation as a numerical recipe have to be given to users not willing to understand how.

In short, the practical implementation of Padé approximants can reach one order of magnitude more applications in theoretical physics and applied mathematics. More than half a century after its discovery, the $\varepsilon$-algorithm has not yet been included for calculation of divergent series with convergent Padé approximants and for extrapolation of functions in commercial software. Now the time for this inclusion has come, the herein implemented control mechanism has rendered this mission possible. Last but not least, the suggested criterion Eq. (3.46) is applicable in solution of differential equations, numerical analytical continuation, perturbation, series summation and other analogous problems of theoretical physics.

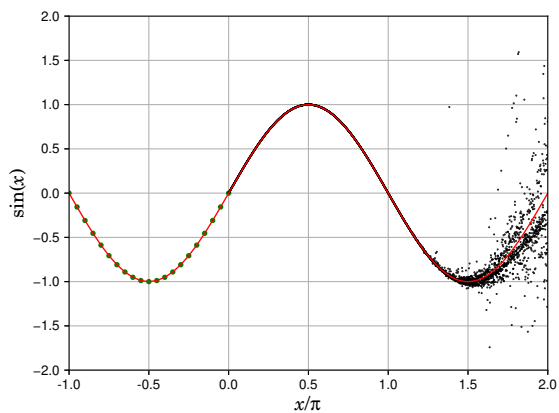

Figure 3.10: Extrapolation of the function $\sin(x)$ represented with the line: in the interval $[-\pi, 0]$ there are 21 interpolation points and in the interval $(0, \pi]$ the function is reliably extrapolated using these interpolation points. The limit of the numerical implementation of the Aitken-Wynn extrapolation is evident as in the interval $(\pi, 2\pi]$ a *gas* of extrapolated points is *evaporated* from the analytical function.

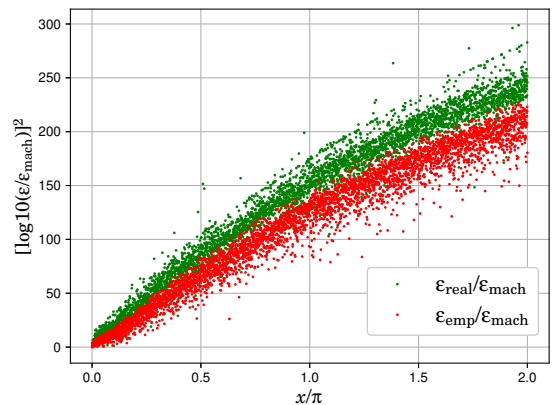

Figure 3.11: Squared logarithm of the error estimates of the $\sin(x)$ extrapolation shown in Fig. 3.10. The similar behaviour of both errors shows that our criterion is a reliable method for error estimation, which is also evident in Fig. 3.14. The important problem in front of the applied mathematics is to research real extrapolation error beyond the extrapolation interval and to compare the results by many other algorithms.

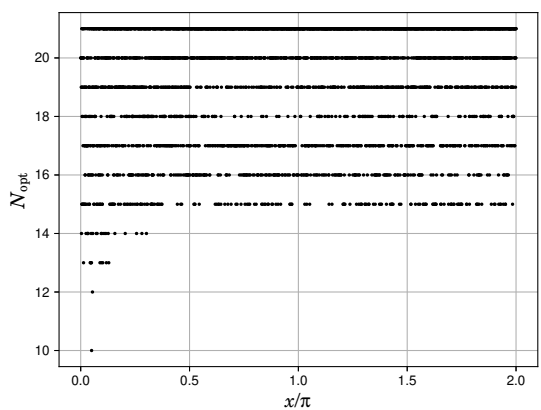

Figure 3.12: The optimal number of used input terms $N_{opt}$ for for the extrapolation of $\sin(x)$ shown in Fig. 3.10. For $x \lesssim 0.4\pi$ fewer interpolation points are used since the extrapolation is closer to the interpolating points, while for $x \gtrsim 0.4\pi$ $N_{opt}$ is evenly distributed revealing nothing about performance nor accuracy.

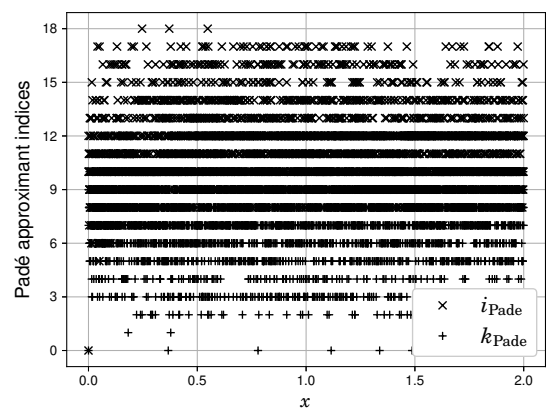

Figure 3.13: The indices of the optimal Padé approximants for the extrapolation of $\sin(x)$ shown in Fig. 3.10. Generally $i_{Pade} > k_{Pade}$, although there are occasions where both are equal and a diagonal approximant has been chosen. The even distributions of both indices reveal nothing about the performance and accuracy.

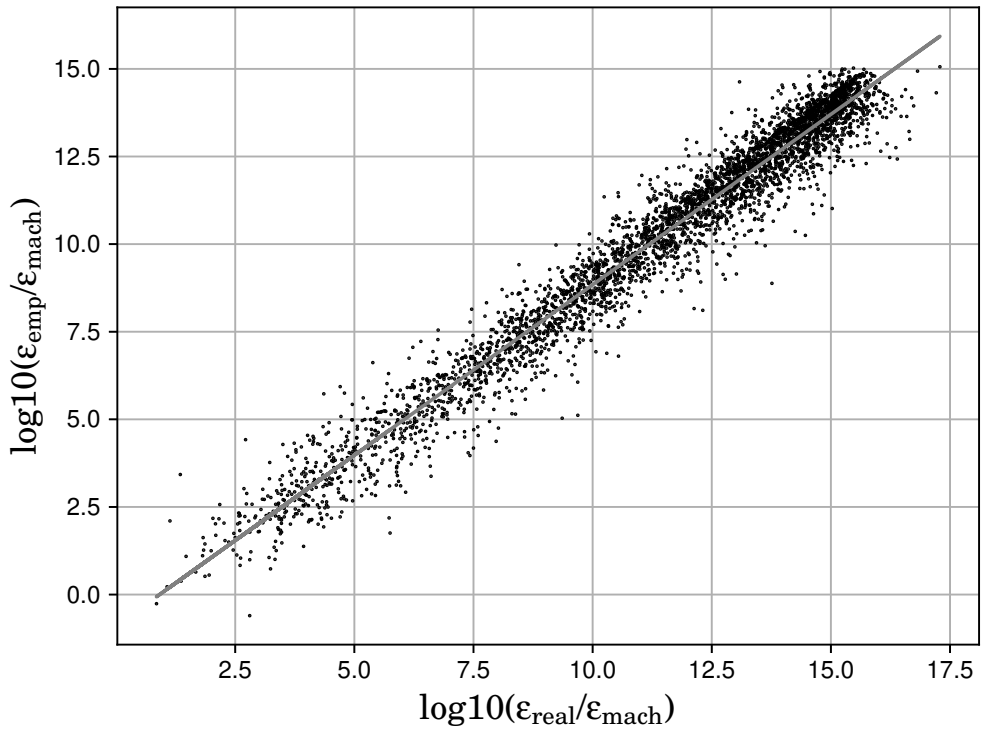

Figure 3.14: Logarithmic dependence of $\varepsilon_{emp}$ as a function of the real error $\varepsilon_{real}$ for the extrapolation shown in Fig. 3.10. The slope of the linear regression is 0.973, the intercept is -0.887 with a correlation coefficient 0.979. The use of $\varepsilon_{mach}$ units gives an invariant problem and system independent results. This strong correlation sets in the agenda the problem of the statistical properties of the Padé approximants and the investigation of the corresponding probability distribution functions (PDF).

# RESULTS FROM THE MHD CALCULATION

The first section of this chapter includes the calculated height dependent solar temperature and wind profiles, while the second section includes the calculated energy and momentum density fluxes. The third section discusses some additional parameters and the forth presents analysis of the performance of the developed numerical method from the performed MHD calculation.

The initial values to start the MHD calculation are: density $n_0 = 10^{19}$ m$^{-3}$, solar wind $U_0 = 1.1$ km/s, solar magnetic field $B_0 = 30$ G, initial temperature $T_0' = 7.7$ kK, a single AW with frequency of 96 Hz and energy flux density $\approx 628$ kW/m$^2$, which according to some order estimations energy flux flows of $1000$ kW/m$^2$ are not huge [Asgari et al. (2013)].

## 4.1 Solar Temperature and Wind Profiles

The performed MHD calculation results in the temperature profile $T(x)$ shown in Fig. 4.1 and solar wind profile $U(x)$ in Fig. 4.2. Both temperature Fig. 4.1 and solar wind Fig. 4.2 height profiles exhibit step-like behavior and the former agrees with the temperature observations of the TR [Withbroe & Noyes(1977), Peter (2004), Avrett & Loeser (2008)], confirming that AW absorption in the TR heats the solar corona and accelerates the solar wind. The width of the TR in this MHD calculation $\lambda \approx 2$ km. A lower frequency AW with the same initial conditions will result in a slower absorption and therefore a larger width of the TR. In this sense, the width of the TR could vary from even half a kilometer up to more than 50 km depending on the frequency of the AW.

A comparison between Avrett-Loeser model C7 [Avrett & Loeser (2008), Fig. 2] and the current MHD calculation for a single AW is given in Fig. 4.3.

The high AW frequency of 96 Hz is chosen for an illustration in order to obtain a steeper step-like height dependent temperature profile that can be at least qualitatively comparable to the semi-empirical observational profiles in [Avrett & Loeser (2008)]. The existence of high frequency AW in the inner solar corona far beyond the cadence of any existing high-time resolution instrument is a prediction of this MHD research. Such high frequency waves may never be observed in the solar corona because they should be detectable close to the TR, where they have not yet been completely absorbed. The argument against their existence because of the absence of observation is not appropriate, it is equivalent to the argument that a thunderstorm generates only low-frequency sound waves, because nobody has gotten close enough to hear the higher pitches yet. The first results from the Parker Solar Probe will definitely show higher frequency AW waves since no other spacecraft has ever gotten so close to the Sun and this statement can be considered as a prediction of MHD. Moreover, the existence of high frequency AW can be extrapolated from the low frequency power like spectrum of coronal AW and their total energy flux can be evaluated by the non-thermal broadening of the spectral lines in chromosphere and lower TR.

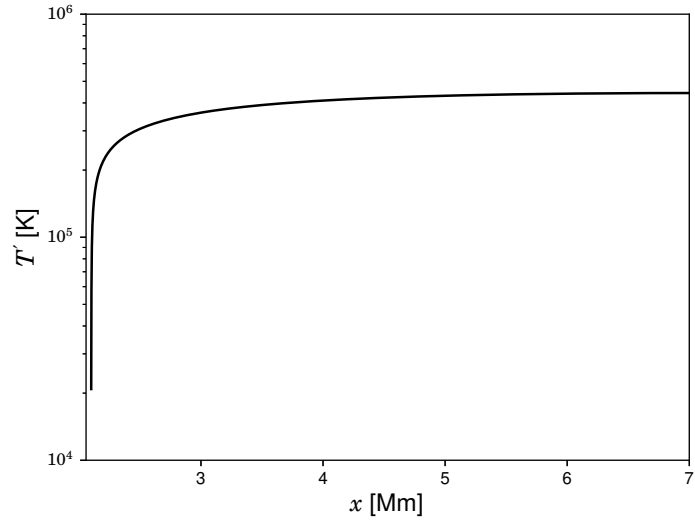

Figure 4.1: Height dependent temperature profile $T(x)$ calculated within MHD. The step-like behavior confirms that AW absorption heats the solar corona.

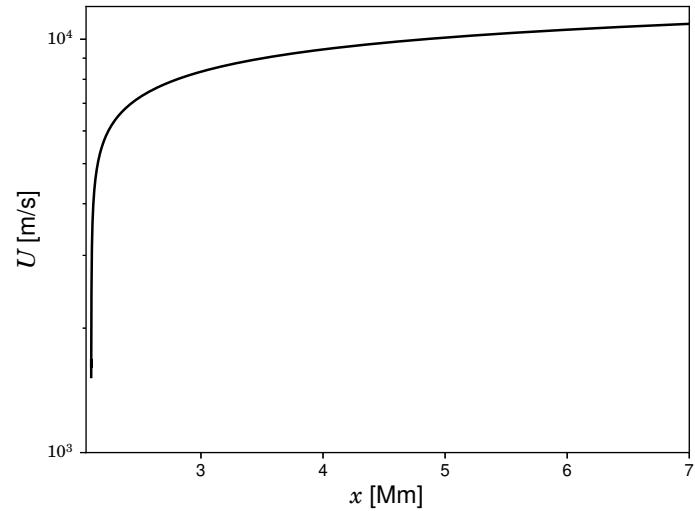

Figure 4.2: Height dependent solar wind velocity profile $U(x)$ calculated within MHD. Alongside heating of the solar corona, AW absorption also accelerates the solar wind.

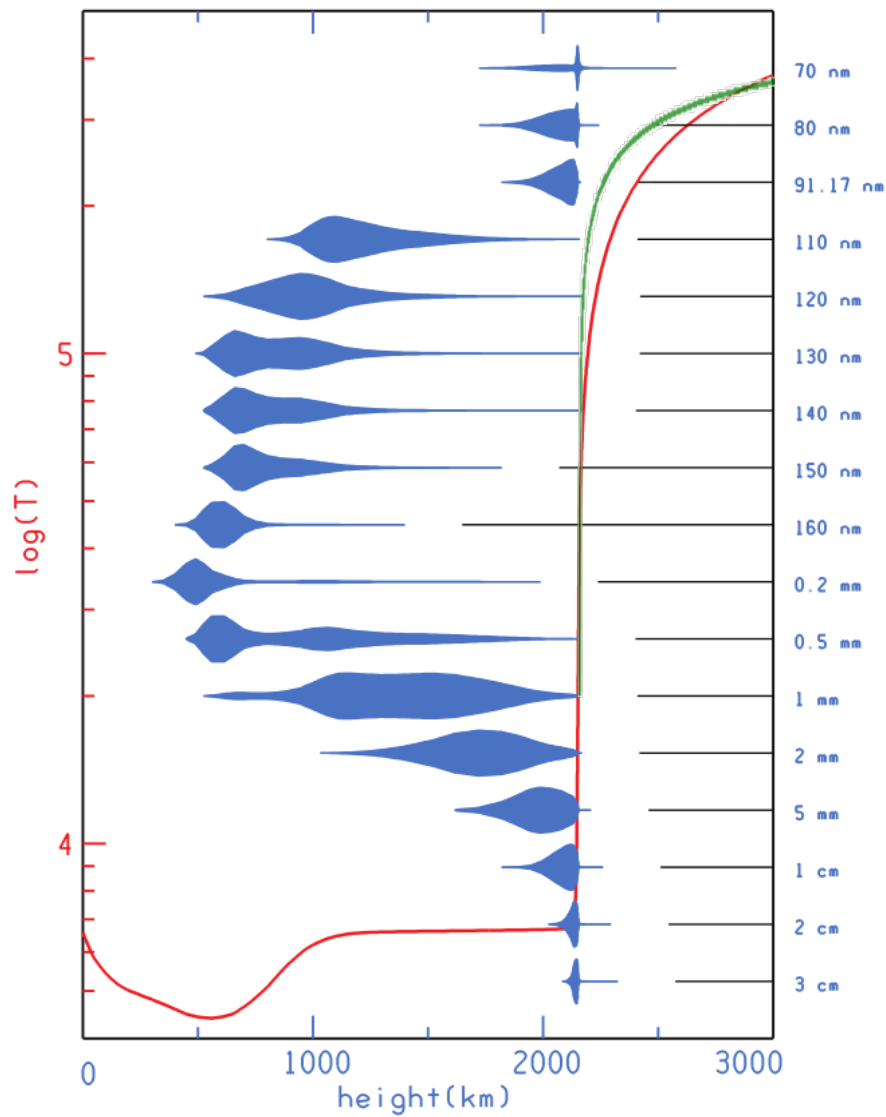

Figure 4.3: Comparison between a semi-empirical model of TR [Avrett & Loeser (2008), Fig. 2] (continuous curve in the interval 0–3 Mm) and our MHD theory (continuous line starting from the TR). One can see the qualitative agreement, which hopefully can be improved by a realistic frequency spectrum of AW.

Introducing few more AW with lower frequency compared to the frequency of the AW from this calculation will flatten a little bit the calculated theoretical profile in Fig. 4.1 so that it can almost perfectly fit the with the semi-observational one Fig. 4.3. Nevertheless, this theoretical profile is sufficient enough to explain the semi-empirical observation models and the mechanism for the solar corona heating by self-induced opacity of AW.

## 4.2   Energy and Momentum Density Fluxes

The energy and momentum density fluxes obtained from the MHD calculation described in this chapter are shown respectively in Fig. 4.4 and Fig. 4.5. Initially the AW energy flux is dominant and as the wave is being absorbed its energy quickly decreases, while the pressure and ideal wind energy fluxes quickly increase in Fig. 4.4. In this manner the absorption of the AW heats the solar corona and accelerates the solar wind. The AW, pressure and ideal wind momentum fluxes in Fig. 4.5 have identical behavior, with the difference being that the pressure momentum flux is dominant from the beginning. The identical rate of damping of both AW dominant energy and momentum fluxes is obvious and of course, this is what to be expected since these fluxes belong to the very same AW that is being absorbed. In both energy and momentum analyses both wave and pressure terms dominate in the TR, while the other terms have negligible influence. However, the absorption of AW is through $\eta_2$ and the neglection of this term will not produce a TR at all. This at first neglectable viscous term responsible for the whole absorption of AW excellently illustrates the stiffness of the derived and numerically solved MHD equations.

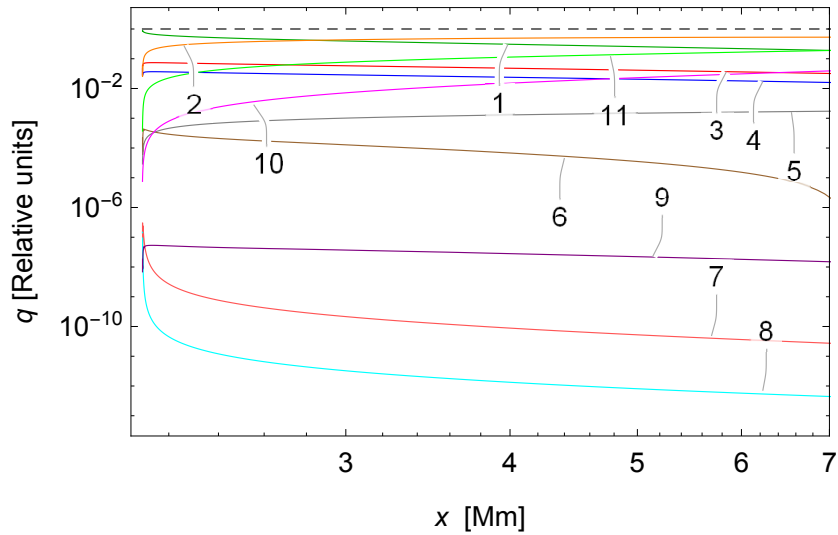

Figure 4.4: Detailed analysis of the energy flux density Eq. (2.42) in units of total energy density flux $q = \mathrm{const}$ represented by the horizontal dashed line – this is the energy conservation law. The energy density fluxes in the graph are: (1) $q_{ub}$, (2) $q_P$, (3) $q_b$, (4) $q_u$, (5) $q_U$, (6) $-q_\varkappa$, (7) $q_\eta$, (8) $q_\varrho$, (9) $-q_\xi$, (10) $-q_g$, (11) $-q_r$, where the definitions of the separate terms are given in Eq. (2.43), Eq. (2.44), Eq. (2.45), Eq. (2.75) and Eq. (2.85). One can easily see the decreasing AW energy flux of the wave $q_{ub}$ (1), which contains magnetic and kinetic energy, and the increasing energy flux of the wind $q_P$ (2). Those two terms approximately give the energy balance. The rest of the fluxes are programmable but have negligible part of the energy flux and conservation. The gravitational energy density flux (10) has less than 10% influence at height 5 Mm above the TR, while the width of the TR is of the order of several km.

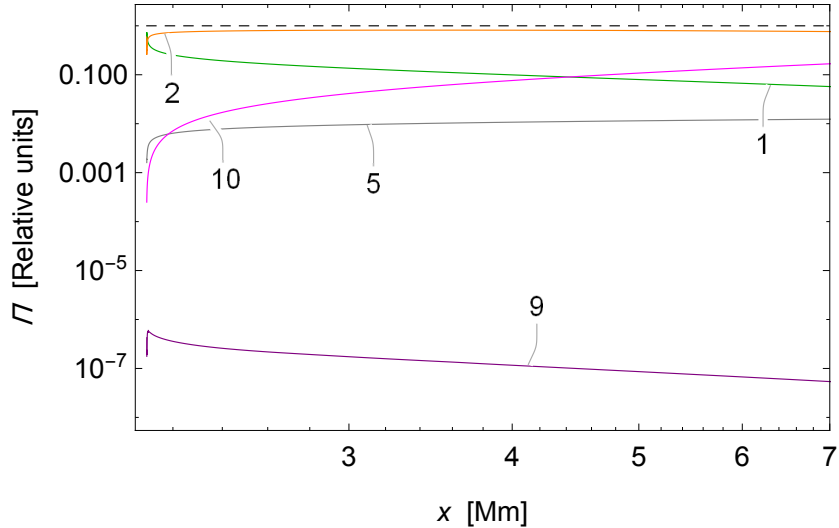

Figure 4.5: Detailed analysis of the momentum flux density Eq. (2.49) in units of total momentum density flux $\Pi = \mathrm{const}$ represented by the horizontal dashed line – this is the momentum conservation law. The momentum density fluxes are: (1) $\Pi_{\mathrm{wave}}$, (2) $\Pi_P$, (5) $\Pi_U$, (9) $\Pi_\xi$, (10) $\Pi_g$, where the definitions of the separate terms are given in Eq. (2.54), Eq. (2.55) and Eq. (2.80). One can easily see the dominating pressure term (2), the decreasing momentum flux of the AW (1) and the increasing momentum flux of the wind (5). The gravitational momentum density flux (10) becomes significant at heights larger than 2 Mm above the TR, where it is visible from the decreasing pressure and acceleration terms.

Finally it's worth noting the gravitational effects both in energy and momentum. As already stated for the thin TR the gravitational energy and momentum density are both too small to be a factor. With the increasing distance, their influence increases and within approximately a solar radius from the TR, the energy has risen to a few percent of the total energy flux density, while the momentum to slightly more than 10% of the total momentum flux density.

The preliminary results of the presented study were first presented and published in [Mishonov et al. (2018)], followed by [Mishonov & Varonov (2019c)] and the final results here are published in [Mishonov et al. (2019)].

## 4.3 Additional Calculated Parameters

This section contains results of several additional parameters from the numerical MHD calculation presented in this chapter.

The volume density of the power of some of the terms is shown in Fig. 4.6. The volume density power of the radiative cooling is dominant in the TR and slightly above it, corresponding to temperatures of around $2 \times 10^5$ K. For larger temperature the radiative cooling decreases and the constant volume density power of the solar gravity becomes dominant in the lower corona, as both energy Fig. 4.4 and momentum Fig. 4.5 flux densities show that. Of course, it should be remembered that bremsstrahlung has not been included in the MHD calculation (Subsec. 2.2.3).

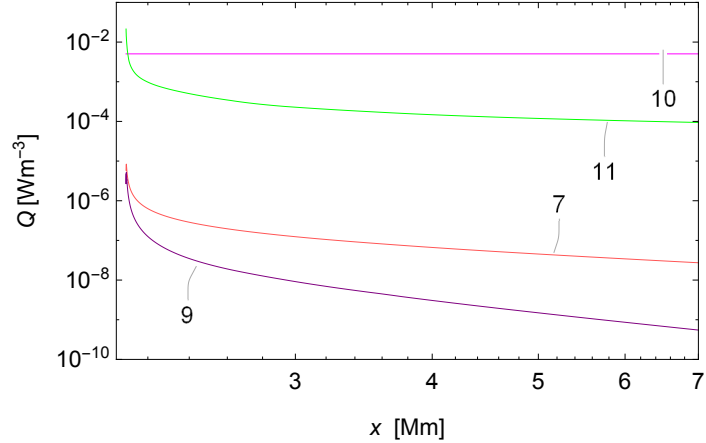

Figure 4.6: Volume densities of the power as functions of height. The biggest power at the TR (11) is the radiative cooling $Q_r = n_e n_p \mathcal{P}$, which quickly decreases simultaneously with the density decrease (and solar wind and temperature increase). The power $Q_g = jg$ related to the loss of kinetic energy due to the solar gravity is constant (10), the dissipative friction $Q_\eta = \frac{1}{2}\eta|w|^2$ of the wave is related to the plasma heating (7), and the smallest power (9) is related to heating by the fluid expansion $Q_\xi = 2\eta(\mathrm{d}_x U)^2$.

The numerical MHD analysis can easily separate the influence of radiative loss and the viscosity in the step like temperature distribution, which is a smooth function only in the scale of Mm and less. The same can be said for the coronal loops, for which the temperature is smooth in small scale. As the heat conductivity of the plasma is well-known, one can easily calculate the heat flux and even corresponding semi-empirical heating function for every model of interpretation of spectroscopic data.

The next parameter worth analysing is the magnetic Prandtl number Eq. (1.134) in Fig. 4.7. As $\mathrm{Pr_m} \equiv \nu_k/\nu_m \propto T^4/n$, it directly shows the ratio between the kinematic viscosity $\nu_k$ and the magnetic diffusivity $\nu_m$. Its profile resembles very much the temperature one in Fig. 4.1 with the difference being in the almost 6 orders of magnitude change, which shows the quickly overwhelming dominance of $\nu_k$. But initially $\mathrm{Pr_m} < 1$, meaning that $\nu_m > \nu_k$ for cooler and denser plasma. As the ratio between $\nu_k$ and $\nu_m$ changes 6 orders of magnitude, plotting them in a single log graph would not be more informative than $\mathrm{Pr_m}$ in Fig. 4.7.

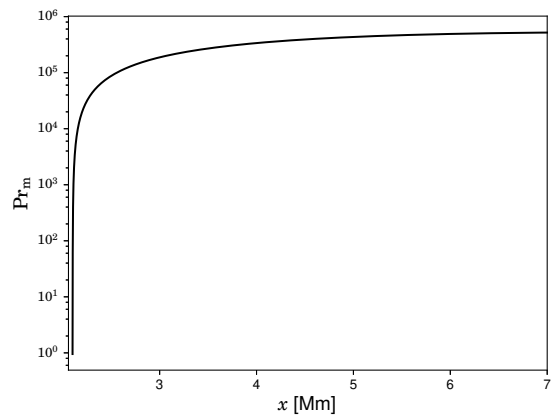

Figure 4.7: The magnetic Prandtl number $\mathrm{Pr_m}$ as a function of the height. For cool and dense plasma $\mathrm{Pr_m} < 1$, while in the TR and corona $\mathrm{Pr_m} \gg 1$ since $\mathrm{Pr_m} \propto T^4/n$.

## 4.4 Numerical Method Performance

As already stated, the developed numerical method in Subsec. 3.2.2 is specifically designed for the current MHD problem. This MHD calculation is its first real usage and its performance has to be discussed. The analysis refers only to the predictor part, including Aitken's interpolation method and the Wynn-Epsilon algorithm and it is almost the same as the analyses of the last two test examples in Subsec. 3.3.2.

The empirical error $\varepsilon_{\mathrm{emp}}$ is the only reliable criterion of the real error as already shown and discussed in Sec. 3.3. In Fig. 4.8 the squared logarithm of the dimensionless empirical errors in $\varepsilon_{\mathrm{mach}}$ units of the extrapolation of the solar temperature $\varepsilon_T/T$ and wind velocity $\varepsilon_U/U$ are shown, where $T$ and $U$ are the corresponding extrapolated values or simply the values of the optimal Padé approximants. The extrapolation procedure in the TR is much more inaccurate, which is to be expected since both height dependent profiles $T(x)$ in Fig. 4.1 and $U(x)$ in Fig. 4.2 are in practice vertical. Also, since the temperature profile is steeper than the solar wind velocity one, $\varepsilon_T/T$ is larger than $\varepsilon_U/U$ in the TR but apart from this, the errors have almost identical behaviour as shown in Fig. 4.9. Beyond the TR both errors decrease substantially since both profiles remain almost constant. However, the errors locally increase several times, which is contained by a correction of the calculation step h, including both minimum and maximum allowed steps, and

$$\mathrm{h} = \mathrm{h}_\varepsilon \frac{T(x)}{|\mathrm{d}T(x)|}, \qquad \mathrm{h}_\varepsilon = 10^{-2}, \tag{4.1}$$

which is calculated for each next iteration step. This correction is also easily visible in the TR, where $|\mathrm{d}T(x)|$ is very large, meaning that the step is very small and therefore the points density is higher. The step correction method also confines the extrapolation errors, since an extrapolation of an almost vertical function to a larger step is quite inaccurate if it does not fail.

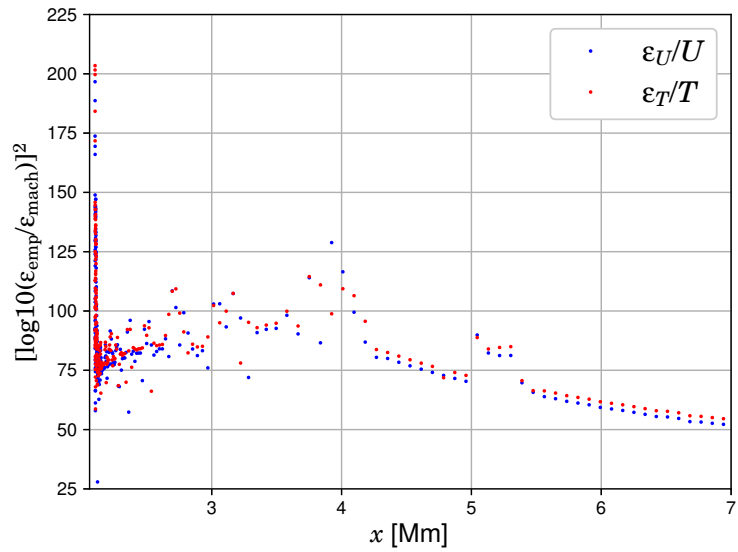

Figure 4.8: Square logarithm of the dimensionless empirical errors $\varepsilon_{\mathrm{emp}}$ in $\varepsilon_{\mathrm{mach}}$ units of the extrapolation of the solar temperature $\varepsilon_T/T$ and wind velocity $\varepsilon_U/U$, $T$ and $U$ are the corresponding extrapolated values. The extrapolation errors in the TR are larger but the smaller step there allows their confinement.

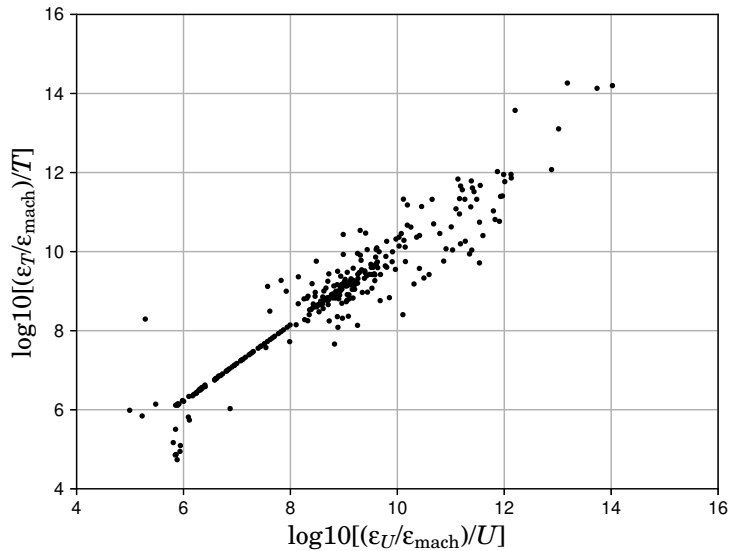

Figure 4.9: Dependency between logarithm of solar wind velocity $\varepsilon_U/U$ and temperature $\varepsilon_T/T$ extrapolation errors in $\varepsilon_{\mathrm{mach}}$ units. The linear dependence is obvious with little deviations at both minimal and maximal values of both errors.

## 4.5 Solar Temperature and Wind Profiles Comparison

Varying the initial values of the MHD calculation, the calculated temperature and solar wind profiles will be different. It is worth illustrating this difference between 3 separate calculations with solar temperature and wind profiles denoted respectively by T1 and U1, T2 and U2, T3 and U3. The profiles T1 and U1 of the first calculation are shown in Fig. 4.1 and Fig. 4.2 and its initial values are presented in the beginning of the current chapter. The initial values of the second calculation with profiles T2 and U2 are: $n_0 = 10^{19}$ m$^{-3}$, solar wind $U_0 = 0.5$ km/s, solar magnetic field $B_0 = 25$ G, initial temperature $T'_0 = 7.7$ kK, a single AW with frequency of 60 Hz and energy flux density $\approx 285$ kW/m$^2$. And the initial values of the third calculation with profiles T3 and U3 are: $n_0 = 5 \times 10^{18}$ m$^{-3}$, solar wind $U_0 = 0.75$ km/s, solar magnetic field $B_0 = 24$ G, initial temperature $T'_0 = 8.1$ kK, a single AW with frequency of 72 Hz and energy flux density $\approx 217$ kW/m$^2$.

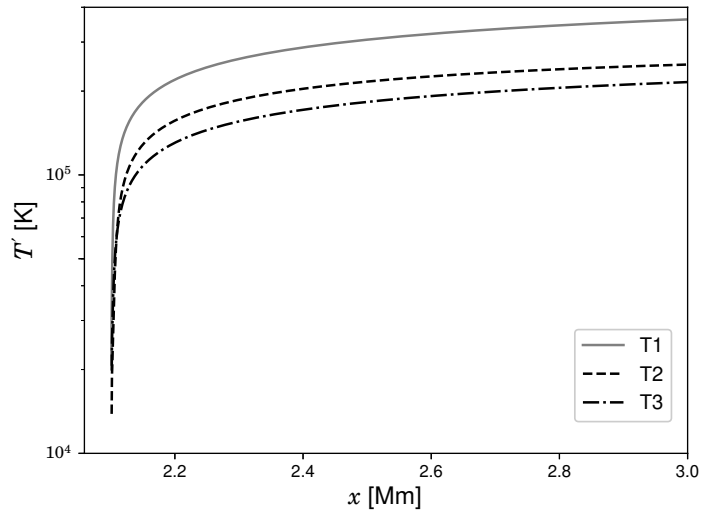

Figure 4.10: Comparison of 3 calculated solar temperature profiles. Profile T1 is shown in Fig. 4.1, while profiles T2 and T3 are calculated with different initial values given in the text of this section.

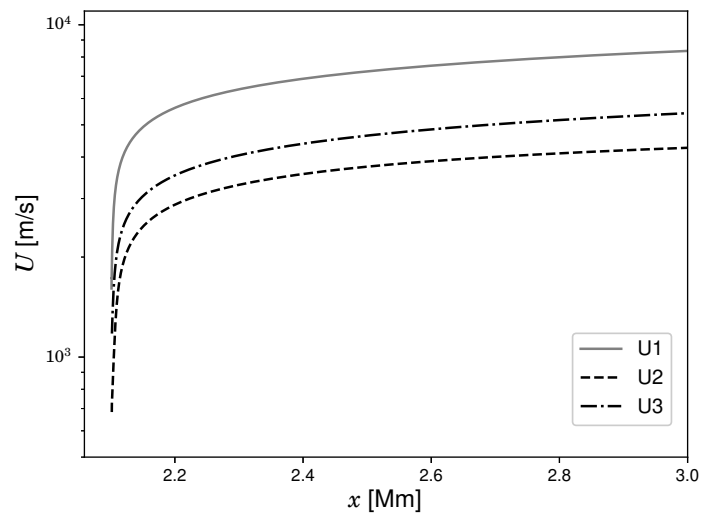

Figure 4.11: Comparison of 3 calculated solar wind profiles. Profile U1 is shown in Fig. 4.2, while profiles U2 and U3 are calculated with different initial values given in the text of this section.

The comparison of the solar temperature profiles is shown in Fig. 4.10 and the solar wind profiles in Fig. 4.11. The widths of the TR in the second and third calculations are almost equal $\approx 4$ km, twice larger than the width of the TR in the first calculation. Profile T2 reaches higher temperature than profile T3, which is to be expected since the density and the input wave energy flux density for T2 are both larger than for T3. Denser plasma can be heated more through viscous friction and this also explains the almost equal TR widths despite the larger input wave frequency for T3. Contrary to the temperature profiles, the solar wind profile U2 reaches lower velocity than profile U3. However, U2 starts significantly lower than U3 because the initial solar wind velocity for U3 is 50% larger than that for U2 and therefore in terms of final to initial solar wind velocity ratio, the solar wind increase in profile U2 is larger than the increase in profile U3. And just like the heating, the denser plasma can be accelerated more through viscous friction.

# List of Publications

# ACKNOWLEDGEMENTS

I am grateful to Iglika Dimitrova and Beka Nathan for the patience, help and support during the research, to Yana Maneva, Martin Stoev, Boian Lazov for the research in the early stages, to Emil Petkov, Aleksander Stefanov, Aleksander Petkov, Zlatan Dimitrov for the support and interest in this research, to Victor Danchev for being my devoted student, which significantly improved this thesis and to Milena Georgieva for the help and support during the final stages of the research. I am thankful also to the members of the jury for their fruitful comments and notes, to Yavor Shopov and Prof. Georgi Rainovski, for the discussion during the defense of the thesis, to Peter Todorov, Nikola Serafimov, Simona Ilieva, Angel Demerdjiev, Miroslav Georgiev, Georgi Kotev, Assoc. Prof. Theodora Bolyarova, Lozan Temelkov, Maksim Varonov, Josef, Maria and Avram Nathan for their interest and support.

---

[1]Monatsbericht der Königlichen Preussische Akademie der Wissenschaften zu Berlin

# SOURCE CODE IN FORTRAN

## A.1  Wynn-Epsilon Algorithm

```fortran
Subroutine LimesCross(N, S, rLimes, i_Pade, k_Pade, i_err, error
    , Nopt)
        implicit integer (i-n)
        implicit real(8) (a-h), real(8) (o-z)
        dimension s(0:N), a(0:N+2), b(0:N+1), c(0:N+1)
        Zero=0.d0
        One =1.d0
        rLimes=s(0)
        i_Pade=0
        k_Pade=0
        rNorm=Zero
        if (N<2) then
                i_err=1 ! N>=2
                Nopt=2
                rLimes=s(N)
                goto 137
        endif

        do i=0,N
                b(i) = s(i)
                c(i) = Zero
                rNorm=rNorm+dabs(s(i))
```

```
                    end  do
            if  (rNorm==0)  then
                        i_err=2  !  S_i=0
                        goto  137
            endif
            error= dabs(s(1)-s(0))
            do  i=2, N
                        eloc= dabs(s(i)-s(i-1))
                        if(eloc<error)  then
                                    error = eloc
                                    k_Pade=0
                                    i_Pade=i
                                    rLimes=s(i)
                                    i_err=3
                                    Nopt=i_Pade + k_Pade+1
                        endif
            end  do

            Nb=N-2  ! N boundary
            k=0
123         continue

            !       beginning  of  the  calculation  of  Pade  table  P_{
                k_Pade , i_Pade }
            !       k_Pade= power  of  the  denominator
            !       i_Pade= power  of  the  numerator
            !       Table  ordered  as  a  matrix  numbering

            do  i=0, Nb
                        Center= b(i+1)
                        if(k>=1)  then
                                    rNorth= a(i+2)
                                    rNC= rNorth-Center
                                    if(rNC==Zero)  then
                                                rLimes= Center
                                                k_Pade=k
                                                i_Pade=k_Pade+i
                                                Nopt=i_Pade + k_Pade +1
                                                error=zero
                                                i_err=4
                                                goto  137
                                    endif
```

```
          endif
          West= b(i)
          WC= West−Center
          if(WC==Zero) then
                  rLimes= Center
                  k_Pade=k
                  i_Pade=k_Pade+i +1
                  Nopt=i_Pade + k_Pade +1
                  error=zero
                  i_err=5
                  goto 137
          endif
          East= b(i+2)
          EC= East−Center
          if(EC==Zero)  then
                  rLimes= Center
                  k_Pade=k
                  i_Pade=k_Pade+i+2
                  Nopt=i_Pade + k_Pade +1
                  error=zero
                  i_err=6
                  goto 137
          endif
!         The new Wynn criterion Eq. (3.46)
          Wynn=One/WC + One/EC
          if(Wynn.NE.Zero) eloc=One/dabs(Wynn) ! in search
                  of local minimum
          denom=Wynn
          if(k>=1) denom=denom−One/rNC
          if(denom==Zero) then
                  i_err=7 ! 1/rLimes=0, i.e. divergent
                          result
                  Nopt=i_Pade + k_Pade +1
                  goto 137
          endif
          South= Center+One/denom
          C(i)=South
          SC=South−Center
!         Optimal Pade approximant
          if(eloc<error) then
                  error= eloc
                  i_err=0
```

```
                        k_Pade=k+1
                        i_Pade=k_Pade+i
                        Nopt=i_Pade + k_Pade +1
                        rLimes= South  ! = P(k_Pade,i_Pade)
                endif
        end do ! i cycle
        k=k+1
!       Change of the succesions
        do i=0, Nb
                a(i)=b(i)
                b(i)=c(i)
        enddo
!       Reduction of the boundary
        Nb=Nb−2
        if(Nb>=0) goto 123
!       Exit
137     continue
end Subroutine LimesCross
```

## A.2   MHD Calculation

### A.2.1   Kinetic coefficients

```
        module coefficients

        real(8), parameter :: Zero=0.d0
        real(8), parameter :: One=1.d0
        real(8), parameter :: Two=2.d0
        real(8), parameter :: pi=4.d0*datan(One)          ! pi
        real(8), parameter :: rmu0=4.d−07*pi              ! mu_0
        real(8), parameter :: vc=299792458.d0             ! c (
            velocity of light)
        real(8), parameter :: eps0=One/(rmu0*vc**2)       ! eps_0
        real(8), parameter :: hbar=1.0545718d−34          ! hbar (
            reduced planck constant)
        real(8), parameter :: qe=1.60217662d−19           !
            electron charge
        real(8), parameter :: e2=qe**2/(4.d0*pi*eps0)     ! Eq.
            (1.93)
        real(8), parameter :: alphaS=e2/(hbar*vc)
        real(8), parameter :: rkB=1.38064852d−23          ! k_B (
            Boltzmann constant)
```

```fortran
real(8), parameter :: rMe=9.10938356d-31        ! m (
    electron mass)
real(8), parameter :: rMp=1.6726219d-27         ! M (
    proton mass)
real(8), parameter :: rMav=rMp/(2.d0)           !
    average mass
real(8), parameter :: rI=.5d0*rMe*vc**2*alphaS**2
real(8), parameter :: rcp=2.5d0                 ! c_p
real(8), parameter :: gama=rcp/(rcp-One)        ! gamma
real(8), parameter :: az=7.297352d-03           ! fine-
    structure constant
real(8), parameter :: rC=.9d0                   ! C
real(8), parameter :: rMSun=1.988d+30           ! Mass
    of the Sun
real(8), parameter :: rSun=695.7d+06            ! Radius
    of the Sun
real(8), parameter :: gc=6.674d-11              ! G (
    Gravitational constant)
real(8), parameter :: gSun=gc*rMSun/rSun**2     ! Solar
    gravity acceleration

complex(8), parameter :: cZero=dcmplx(Zero,Zero)
    ! complex Zero
complex(8), parameter :: cOne=dcmplx(One,Zero)
    ! complex One

contains

subroutine statistics(rho, T, B0, vA2, P, alpha, rD, re,
    rL, sigma, rnum, eta, rnuk, Pr, rkap, ootp, eta2,
    rNuk2, taupp)
implicit real(8) (a-h,o-z), integer (i-n)

real(8), parameter :: sigea=1.d-20              ! Electron-atom
    collision cross section
real(8), parameter :: sigpa=1.d-20              ! Proton-atom
    collision cross section

if (T<Zero) then
        print *, 'Stopping at coefficients calculation
            because T=',T/rkB
        stop
```

```
        endif

        rnrho=rho/rMp              ! n (number density)
        va2=B0**2/(rmu0*rho)       ! Alfven velocity squared
        vTe=dsqrt(T/rme)           ! Electron thermal velocity
        vTp=dsqrt(T/rMp)           ! Proton thermal velocity

        rnq=(dsqrt(rme*T/(2.d0*pi))/hbar)**3
        Rn=rnrho/(rnq*dexp(-rI/T))

!       Machine epsilon
        epsmach = One
111     epsmach = epsmach/Two
        if (epsmach+One>One) goto 111
        epsmach=epsmach*1024
!       Ionization ( almost 1)
        alpha=One/dsqrt(One+1.5d0*Rn)
        do 10, i=1,5
                y=One/(One+(alpha+One)*Rn)
                f=alpha**2-y
                fp=Two*alpha+Rn*y**2
                alpha=alpha-f/fp
                error=dabs(alpha-dsqrt(y))
                if (error<epsmach) goto 11
10      continue
11      continue

        rn=alpha*rnrho             ! e/p density
        rna=(one-alpha)*rnrho      ! Neutrals density
        P=rnrho*(One+alpha)*T      ! Pressure
        re=e2/T                    ! Eq. (1.107)
        rD=One/dsqrt(8.d0*pi*re*rn)   ! Debye radius Eq.
           (1.102)
        rL=dlog(rD/re)             ! Coulomb logarithm Eq. (1.106)
        sigee=(rL/.6d0)*re**2      ! Electron cross section Eq.
           (1.108)
        sigpp=(rL/.4d0)*re**2      ! Proton cross section Eq.
           (1.108)
        rle=one/(rn*sigee+rna*sigea)   ! Electron mean free
           path Eq. (1.110)
        rlp=One/(rn*sigpp+rna*sigpa)   ! Proton mean free path
           Eq. (1.125)
```

```
tauee=rle/vTe      ! Electron tau Eq. (1.111)
taupp=rlp/vTp      ! Proton tau Eq. (1.126)
sigma=rn*qe**2*tauee/rme            ! Electric conductivity
    Eq. (1.114)
rnum=rme*vc**2/(4.d0*pi*e2*rn*tauee)      ! Magnetic
    diffusivity Eq. (1.116)
rkap=1.5d0*rle*vTe*rn    ! Heat conductivity Eq. (1.122)
rnuk=rlp*vTp      ! Kinematic viscosity Eq. (1.128)
eta=rho*rnuk      ! Dynamic viscosity Eq. (1.130)
Pr=rnuk/rnum      ! Magnetic Prandtl number Eq. (1.133)
```
`! Magnetic field dependence`
```
oocp=qe*B0/rmp  ! Proton cyclotron frequency Eq. (1.141)
ooce=qe*B0/rme  ! Electron cyclotron frequency
ootp=taupp*oocp  ! Proton omega tau << 1
oote=tauee*ooce  ! Electron omega tau << 1
eta2=eta/(One+rC*(oocp*taupp)**2) ! Dynamic viscosity
    Eq. (1.143)
rnuk2=eta2/rho  ! Kinematic viscosity
```

**end subroutine** statistics

**end module**

## A.2.2 Eigenvalues and eigenvectors

**module** eigen4

**implicit integer** (i−n), **complex**(8) (a−c,u,w), **real**(8) (d
    −h,o−t,v,x−z)

**complex**(8), **parameter** :: cIone=dcmplx(0.d+00,1.d+00)
    ! imaginary unit

**contains**

! Calculation of matrix K Eq. (2.20) via M Eq. (2.21)
**subroutine** minit(cMa,ndim,comg,va2,vu,rnuk,rnum,dux,rho,
    deta,dnum)
        **implicit integer** (i−n), **complex**(8) (a−c,u,w),
            **real**(8) (d−h,o−t,v,x−z)

        **dimension** cMa(ndim+1,ndim+1)

```
          cMa(1,3)  = -cIone                           ! M(1,3)
          cMa(2,4)  = -cIone                           ! M(2,4)
          cMa(3,1)  = -cIone*(comg+dux)/rnum           ! M(3,1)
          cMa(3,3)  = -cIone*(vu-dnum)/rnum            ! M(3,3)
          cMa(3,4)  =  cIone/rnum                      ! M(3,4)
          cMa(4,2)  = -cIone*comg/rnuk                 ! M(4,2)
          cMa(4,3)  =  cIone*va2/rnuk                  ! M(4,3)
          cMa(4,4)  = -cIone*(vu-deta/rho)/rnuk        ! M(4,4)
end subroutine minit

subroutine seigen(cMa, ndim, cev, cb, cu, cg, cw, nr)
          implicit integer (i-n), complex(8) (c), real(8)
              (a-b,d-h,o-z)

          dimension cMa(ndim+1,ndim+1), cev(0:ndim), cb(0:
              ndim), cu(0:ndim), cg(0:ndim), cw(0:ndim)
          dimension ev(2,ndim+1), tmp(ndim+1,ndim+1), evr(
              ndim+1,ndim+1), evi(ndim+1,ndim+1)

          ! ev(1,:) - eigenvalues real part
          ! ev(2,:) - eigenvalues imaginary part
          ! evr - eigenvectors real part
          ! evi - eigenvectors imaginary part

          ! From EISPACK
          call cg(ndim+1,ndim+1,real(cMa),aimag(cMa),ev
              (1,:),ev(2,:),1,evr,evi,tmp(1,:),tmp(2,:),tmp
              (3,:),eps)
          ! Right propagating wave
          nr=0
          if ( ev(1,ndim) > 0.) nr=1
          ! Assign the eigenvalues and eigenvectors
          do k=1,ndim+1
                cev(ndim-k+1) = dcmplx(  ev(1,k), ev(2,k)
                    ) ! Eigenvalue
                cb(ndim-k+1) = dcmplx( evr(1,k), evi(1,k)
                    ) ! Psi(1) magnetic field
                cu(ndim-k+1) = dcmplx( evr(2,k), evi(2,k)
                    ) ! Psi(2) velocity
                cg(ndim-k+1) = dcmplx( evr(3,k), evi(3,k)
                    ) ! Psi(3) magnetic field derivative
                cw(ndim-k+1) = dcmplx( evr(4,k), evi(4,k)
```

```
                        ) ! Psi(4) velocity derivative
            end do
    end subroutine seigen

    end module
```

### A.2.3 Energy-momentum density fluxes

```
    module energy

    use coefficients

    contains

    !!!    ####   ENERGY   ##### !!!

    ! Ideal wind Eq. (2.43)
    function fqidwind(ux, flux, T, alpha, qa, qp)
            implicit real(8) (a-b,d-h,o-z)

                    h = rcp*T/rMav
                    qp = flux*h           ! q_P
                    qa = flux*0.5d0*ux**2  ! q_U
            fqidwind = qa + qp ! q^{ideal}_{wind}
    end function fqidwind

    ! Ideal wave Eq. (2.46)
    function fqidwave(cb, cu, b02, ux, flux, qu, qb, qub)
            implicit complex(8) (c), real(8) (a-b,d-h,o-z)

                    qu = 0.25d0*flux*real(dconjg(cu)*cu)
                        ! q_u
                    qb = b02*ux*real(dconjg(cb)*cb)/(2.d0*rmu0
                        ) ! q_b
                    qub = -b02*real(dconjg(cb)*cu)/(2.d0*rmu0)
                        ! q_{ub}
            fqidwave = qu + qb + qub ! q^{ideal}_{wave}
    end function fqidwave

    ! Dissipatibe wave Eq. (2.47)
    function fqdisswave(cb, cu, cg, cw, b02, eta, rNum, qdu,
        qdb)
            implicit complex(8) (c), real(8) (a-b,d-h,o-z)
```

```
                         qdu = −0.5d0∗eta∗real( dconjg(cu)∗cw )
                                       ! q_eta
                         qdb = −b02∗rNum∗real(dconjg(cb)∗cg)/(2.d0
                               ∗rmu0)  ! q_rho
                 fqdisswave = qdu + qdb ! q^{diss}_{wave}
end function fqdisswave

! Dissipative wind Eq. (2.44)
function fqdisswind(ux, dux, rksi, dT, rkappa, qdUx,
   qdTx)
                 implicit real(8) (a−b,d−h,o−z)

                    qdUx = −rksi∗ux∗dux          ! q_ksi
                    qdTx = −rkappa∗dT             ! q_varkappa
                 fqdisswind = qdUx + qdTx
end function fqdisswind

! Eq. (2.94)
function cfgdot(ccb, ccu, ccg, ccw, cb, cu, cg, cw, fj,
   b02, ux, rho, eta, rnum)
                 implicit complex(8) (c), real(8) (a−b,d−h,o−z)

                    cbqid  = b02∗( ux∗dconjg(ccb)∗cb − (dconjg(
                             ccu)∗cb + dconjg(ccb)∗cu)/2.d0 )/(2.d0∗
                             rmu0)
                    cbqdis = −b02∗( rnum∗(dconjg(ccb)∗cg +
                             dconjg(ccg)∗cb)/2.d0 )/(2.d0∗rmu0)
                    cuqid  = 0.25d0∗fj∗dconjg(ccu)∗cu
                    cuqdis = −0.25d0∗eta∗( dconjg(ccu)∗cw +
                             dconjg(ccw)∗cu )
                    cfgdot = cbqid + cuqid + cbqdis + cuqdis
end function cfgdot

!!!     ####    MOMENTUM   #####  !!!

! Ideal wind Eq. (2.54)
function fpidwind(rho, u, T, alpha, pa, pp)
                 implicit real(8) (a−b,d−h,o−z)

                 pa= rho∗u∗∗2       ! Pi_U
                 pp= rho∗T/rMav     ! Pi_P
```

```
              fpidwind = pa + pp  ! Pi^{ideal}_{wind}
end function fpidwind

! Wave Eq. (2.52)
function fpwave(cb, b02)
        implicit complex(8) (c), real(8) (a-b,d-h,o-z)

        fpwave = 0.25d0*b02*real(dconjg(cb)*cb)/rmu0
end function fpwave

! Dissipative wind Eq. (2.51)
function fpdisswind(dux, rksi)
        implicit real(8) (a-b,d-h,o-z)

        fpdisswind = -rksi*dux
end function fpdisswind

end module
```

## A.2.4  Main program

```
program wkb4qn

use eigen4
use energy
use coefficients
use eps_algorithm
use Aitken
use rloss

implicit integer (i-n)
implicit complex(8) (c), real(8) (a-b,d-h,o-z)

integer, parameter :: n_d=3        ! Dimension
integer, parameter :: n_e=5        ! Number of
    extrapolation points
integer, parameter :: nlim=4
real(8), parameter :: dlim = dlog(1.001d+06)    !
    Attenuation limit
real(8), parameter :: pcor = 1.d-10     ! Correction
    accuracy
```

```
!       Arrays
```

```
      dimension cpsi(0:n_d,0:n_d,0:nlim+1), cev(0:n_d) ! Psi
          Eq. (2.20) and EV Eq. (2.96)
      dimension cMa(0:n_d,0:n_d)          ! Matrix M Eq. (2.21)
      dimension va2(0:n_d), rho(0:1)   ! v_A and rho
      dimension sU(0:n_e), sT(0:n_e), sx(0:n_e)        !
          Extrapolation
      dimension UU(-1:1), TT(-1:1)      ! Quasi Newton
      dimension Ux4(1:4), Tx4(1:4)      ! Derivatives
          calculation

      allocatable :: Tx(:), Ux(:), xi(:)        ! Temperature,
          wind and length
      allocatable :: rksidU(:), rkapdT(:)       ! ksi*d_x U and
          varkappa*d_x T

!     Common variables
      common /start/   cevp,u0,T0,B0,flux,h,cs,rhop,frlnp,rrl,
          nr
      common /waves/   qwave0,q0,qdisswnd0,pwave0,p0,pdisswnd0
      common /kin/     eta,rNum,rkap,rksi,zeta,rNuk,sigma,va,
          eta2,rNuk2,eta2p
      common /ev/      cpsi,cev
      common /derivs/  deta,drnum,drkap,drksi,deta2
      common /energy/  qdUx,qdTx,qu,qb,qub,qdu,qdb,qu0,qb0,qub0
          ,qdu0,qdb0,qg,pg,pgp,qr,qrl,frln

      np=nlim+1

!!!   INITIAL VALUES
      rnrho0=1.d+19 ! n_0
      T0=.77d+04*rKb    ! T_0
      u0=1.1d+03        ! U_0
      B0=1.d-03*3.      ! B_0
      heat=1.08d+02     ! Heating T_l/T_0
      freq=1.2d+02*.8 ! AW frequency
      omega=2.d0*pi*freq        ! AW ang. freq.
      cs=-dcmplx(Zero,omega)    ! AW complex ang. freq.
      B02=B0**2         ! B_0^2
      rho(0)=rnrho0*rMp         ! rho_0
      rho0=rho(0)               ! array init.

!!!   INITIAL KINETIC COEFFICIENTS
```

```
        call statistics(rho(0), T0, B0, vA2(0), P, alpha, rD, re
         , rL, sigma, rnum, eta, rnuk, Pr, rkap, oote, eta2,
         rNuk2, taupp)
!       Output to terminal of the initial values (subroutine
  omitted in the module)
        call printstat(rho(0), vA2(0), P, alpha, rD, re, rL,
         sigma, rnum, eta, rnuk, Pr, rkap)
        omega_c=(B02/rmu0)/eta
        va=dsqrt(va2(0))              ! Alfven velocity
        rk_c=omega_c/va
        print *,'omega_c=_',omega_c,'___k_c=_',rk_c,'_________
         lambda_c=_',2.d0*pi/rk_c

        xa=Zero                ! lower x boundary
        xb=1.d+07              ! upper x boundary
        h=va/freq*4.d0         ! step selection
        hmax=h*1.d+02          ! max step value
        hmin=h*1.d-01          ! min step value
        heps=1.d-02            ! h_epsilon Eq. (4.1)
        Na=int((xb-xa)/hmin)   ! Number of cycles for the loop

        allocate(Tx(0:Na+1))                 ! Temperature
        allocate(Ux(0:Na+1))                 ! Wind
        allocate(xi(0:Na+1))                 ! Height
        allocate(rksidU(0:Na+1))
        allocate(rkapdT(0:Na+1))

!!!     COMPUTER ERROR
        epsmach = One
111     epsmach = epsmach/Two
        if (epsmach+One/=One) goto 111
        reps=One/epsmach
        sqeps=dsqrt(epsmach)*1.37d+02

!!!     INITIAL PARAMETERS
        flux=rho(0)*u0  ! Eq. (2.2)
        dUx=Zero              ! Zero d_U
        dTx=Zero              ! Zero d_T
        q0=rho(0)*u0**3 ! rho_0*U_0^3 factor for d-less energy
        p0=rho(0)*u0**2 ! rho_0*U_0^2 factor for d-less momentum
        theta0=T0/(rMav*u0**2) ! Eq. (2.60)
        s2=gama*theta0          ! Eq. (2.65)
```

```
        beta=rho(0)*T0/rMav/(B02/(2.d0*rmu0))    ! Plasma beta
        Tx(0)=T0          ! initial T
        Ux(0)=u0          ! initial U
        Tx4(1)=Tx(0)      ! previous T
        Ux4(1)=Ux(0)      ! previous U

!       Matrix zeroed initially
        do i=0,n_d
         do j=0,n_d
                cMa(i,j) = Czero
         end do
        end do
!       Zero derivatives
        deta=Zero         ! d_x eta = 0
        deta2=Zero        ! d_x eta_2 = 0
        drnum=Zero        ! d_x num = 0
        dzeta=Zero        ! d_x zeta = 0
        do i=2,4
                Ux4(i)=U0
                Tx4(i)=T0
        enddo

!!!     EIGENVECTORS AND EIGENVALUES
        call minit(cMa,n_d,cs, va2(0), u0, rNuk2, rNum, dux, rho
           (0), deta2, drnum)          ! Iniitalize M
        call seigen(cMa,n_d,cev, cpsi(0,:,np), cpsi(1,:,np), cpsi
           (2,:,np), cpsi(3,:,np), nr )  ! Calculate EV
        call snorm(cpsi(:,:,np),n_d,flux,b02,u0,rho(0),eta2,rNum
           )                            ! Norm EV
!       EV output to terminal
        do i=0,n_d
                print*,"Eigenvector:", cev(i), freq, va/freq*1.d
                   -03
                print*,"Eigenvalues:"
                print*,"B_=",cpsi(0,i,np)
                print*,"U_=",cpsi(1,i,np)/va
                print*,"G_=",cpsi(2,i,np)*va/omega
                print*,"W_=",cpsi(3,i,np)/omega
                print*,cfgdot( cpsi(0,i,np), cpsi(1,i,np), cpsi
                   (2,i,np), cpsi(3,i,np), &
                        cpsi(0,i,np), cpsi(1,i,np), cpsi
                           (2,i,np), cpsi(3,i,np), &
```

```
                                  flux , b02 , u0 , rho ( 0 ) , eta2 , rNum
                                  )
            end do
            nl=abs ( nr −1)        ! nr is right wave , nl is left wave

! ! !        INITIAL FLUXES
            qidwv0 = fqidwave ( cpsi ( 0 , nr , np ) , cpsi ( 1 , nr , np ) , b02 , u0
                , flux , qu0 , qb0 , qub0 )
            qdisswv0 = fqdisswave ( cpsi ( 0 , nr , np ) , cpsi ( 1 , nr , np ) ,
                cpsi ( 2 , nr , np ) , cpsi ( 3 , nr , np ) , b02 , eta2 , rNum , qdu0 ,
                qdb0 )
            qw0 = qidwv0 + qdisswv0
            pw0 = fpwave ( cpsi ( 0 , nr , np ) , b02 )
!           No dissipation initially
            qdisswnd0=Zero     ! q^{ diss } _{wind }(0)=0
            pdisswnd0=Zero     ! Pi ^{ diss } _{wind }(0)=0

! ! !        AW power necessary for reaching the selected heating
            wa=One              ! lower boundary
            wb=1.d+06           ! upper boundary
!           Heating for wb AW power
136         hi =(qw0+qdisswnd0 ) / q0∗wb         ! Eq. (2.59)
            tau =(pw0+pdisswnd0 ) / p0∗wb        ! Eq. (2.59)
            Discr = ( s2 −One )∗∗2 − 2.d0∗hi ∗(gama∗∗2−One)+gama∗tau ∗(
                gama∗tau +2.d0∗(gama+s2 )  )  ! Eq. (2.64)
            if ( Discr >=Zero ) then
                    V=(  (gama−One )∗(2.d0∗hi+One ) + 2.d0∗s2  ) / (gama+
                        s2+gama∗tau+dsqrt ( Discr ))  ! Eq. (2.66)
                    theta=V∗(One+theta0+tau −V)        ! Eq. (2.67)
            else
                    print ∗ , ' Quadratic ␣equation ␣not ␣consistent , ␣DISCR
                        ␣=␣' , discr        ! Discr <0
                    stop     ! Cannot perform the calculation with
                        the given initial values
            endif
            Tf=theta ∗rMav∗u0∗∗2      ! Eq. (2.68)
!           If heating not enough , double upper boundary and go back
            if ( Tf/T0 − heat < Zero ) then
                    wb=wb∗2.d0
                    goto 136
            endif
!           Heating for wm AW power
```

```fortran
137       wm=wa+.5d0*(wb-wa)
          hi=(qw0+qdisswnd0)/q0*wm           ! Eq. (2.59)
          tau=(pw0+pdisswnd0)/p0*wm          ! Eq. (2.59)
          Discr = (s2-One)**2 - 2*hi*(gama**2-One)+gama*tau*( gama
     *       *tau+2.d0*(gama+s2) )            ! Eq. (2.64)
          if (Discr>=Zero) then
               V=( (gama-One)*(2.d0*hi+One) + 2.d0*s2 )/(gama+
     *            s2+gama*tau+dsqrt(Discr)) ! Eq. (2.66)
               theta=V*(One+theta0+tau-V)         ! Eq. (2.67)
          endif
          Tf=theta*rMav*u0**2        ! Eq. (2.68)
!         If clauses for checks
          if ( Tf/T0 - heat > Zero) wb=wm ! Heating is larger,
     *       decrease upper boundary
          if ( Tf/T0 - heat < Zero) wa=wm ! Heating is smaller,
     *       increase lower boundary
          if ( dabs(wb-wa) < 1.d-07) goto 138    ! Both
     *       boundaries almost coincide, continue
          if ( dabs(Tf/T0-heat) > 1.d-07*heat ) goto 137 !
     *       Selected heating has not been reached, go back
!         Initiate AW energy and momentum fluxes
138       w=wm*1.d+0
          qwave0=qw0*w
          pwave0=pw0*w
          qwave=qwave0
          qwavep=qwave
          pwave=pwave0
          rho(1)=rho(0)
!!!       Zeroing gravity variables
          qg=Zero
          pg=Zero
          qrp=Zero
          frlnp=Zero

          rrl=1.d-02        ! C_lambda
          omega_m=5.d0*(B02/(2.d0*rmu0))*e2**2*rL/dsqrt(rMp*Tf**5)
     *       ! omega_max at Tf Eq. (2.90)
          omega_0=5.d0*(B02/(2.d0*rmu0))*e2**2*rL/dsqrt(rMp*T0**5)
     *       ! omega_max at T_0 Eq. (2.90)
!         Initial output to terminal
100       format('The stiffness ratio is:',G10.5)
          print 100,va**3/(u0*(rNum+rNuk)*omega)
```

```
        print *, 'h␣=', h, qwave0, pwave0, 'beta␣=', beta, flux, omega_m,
            's^2␣=', s2
!       Initialising output files
        open(unit=91, file="wkb4-r.txt", action="write")
        open(unit=93, file="wkb4qn-q.txt", action="write")
        open(unit=94, file="wkb4qn-p.txt", action="write")
        open(unit=95, file="wkb4qn-v.txt", action="write")
        open(unit=96, file="wkb4qn-ue.txt", action="write")
        open(unit=97, file="wkb4qn-te.txt", action="write")
        open(unit=98, file="wkb4qn-ev.txt", action="write")
!!!     START
        call system_clock(i_start, i_rate) ! Start the clock
!       Zeroing eigenvector variables
        cevp=cZero
        cA=cZero
        cAp=cZero
        A=Zero
        etap=eta        ! previous eta = current eta
        eta2p=eta2      ! previous eta2 = current eta2
        rhop=rho(1)     ! previous rho = current rho
        zeta=Zero       ! Second viscosity = 0
        i=1             ! Next step
        y=Zero
        xi(0)=xa        ! First x
        do while (xi(i-1)<=xb)           ! Main loop for
            increasing x
333         xi(i)=xi(i-1)+h             ! Distance from Sun
            kc=0                        ! Recursions counter
!!!         EXTRAPOLATION START
!           If insufficient extr. points, constant T and U
            if (i<n_e+1) then
                Uv=Ux(i-1)
                Tv=Tx(i-1)
                goto 777
            endif
!           Extrapolation points shift
            do j=0,n_e
                sU(j)=Ux(i-j-1) ! array U
                sT(j)=Tx(i-j-1) ! array T
                sx(j)=xi(i-j-1) ! array x
            enddo !next j
```

```
          call sainterpol(sx,sU,n_e,xi(i)) ! Aitken for U
          call sainterpol(sx,sT,n_e,xi(i)) ! Aitken for T
          call LimesCross(n_e, sU, Uv, lu, mu, i_err_u,
             error_u, no_u)    ! Wynn-Epsilon for U
          call LimesCross(n_e, sT, Tv, lt, mt, i_err_t,
             error_t, no_t)    ! Wynn-Epsilon for T
          write(96,*) xi(i),Uv, error_u/epsmach, i_err_u, lu,
             mu, no_u          ! Epsilon U parameters save
          write(97,*) xi(i),Tv, error_t/epsmach, i_err_t, lt,
             mt, no_t          ! Epsilon T parameters save
!!!       EXTRAPOLATION END
777       continue

          call VT(xi(i),Un,Tn,Uv,Tv,Ux4,Tx4,rho(1),va2(1),
             alpha,rksidU(i),rkapdT(i),cA,A,i_vt,i,frl)
          if (i_vt==1) stop       ! Negative discriminant
          if (i_vt==2) goto 222   ! Attenuated wave

!!!       QUASI-NEWTON APPROXIMATION
          f=Un ! x
          g=Tn ! y
          deltaU=sqeps*Uv
          deltaT=sqeps*Tv
          do l=-1,1
                  if (l==0) cycle
                  Uc=Uv+l*deltaU
                  cAd=cAp
                  pgd=pgp
                  qrd=qrp
                  call VTd(xi(i),Un,Tn,Uc,Tv,Ux4,Tx4,rho
                     (1),cAd,Ad,i_vt,i,pgd,qrd)
                  if (i_vt==2) stop       ! Negative
                     discriminant
                  UU(1)=Un
                  TT(1)=Tn
          end do !next l
          dxf=(UU(1)-UU(-1))/(2.D0*deltaU) ! d_x f
          dxg=(TT(1)-TT(-1))/(2.D0*deltaU) ! d_x g
          do l=-1,1
                  if (l==0) cycle
                  Tc=Tv+l*deltaT
                  cAd=cAp
```

```
                              pgd=pgp
                              qrd=qrp
                              call VTd(xi(i),Un,Tn,Uv,Tc,Ux4,Tx4,rho
                                 (1),cAd,Ad,i_vt,i,pgd,qrd)
                              if (i_vt==2) stop        ! Negative
                                 discriminant
                              UU(1)=Un
                              TT(1)=Tn
                     end do !next l
                     dyf=(UU(1)-UU(-1))/(2.D0*deltaT) ! d_y f
                     dyg=(TT(1)-TT(-1))/(2.D0*deltaT) ! d_y g
                     det=(one-dxf)*(one-dyg)-dyf*dxg ! Eq. (3.41)
                     x=Uv       ! x_0
                     y=Tv       ! y_0
                     ! Eq. (3.42)
                     Un=Uv+( (one-dyg)*(f-x)+ dyf*(g-y) )/det
                     Tn=Tv+( dxg*(f-x)+ (one-dxf)*(g-y) )/det

!!!                  RECURSION CHECK
                     kc=kc+1
                     if (kc>1000) goto 456
                     if ( (dabs(Un-Uv)>pcor*dabs(Uv)) .or. (dabs(Tn-
                        Tv)>pcor*dabs(Tv)) ) then
                                    Tv=Tn    ! NEW TEMPERATURE IS OLD
                                    Uv=Un    ! NEW WIND IS OLD
                                    cA=cAp          ! THROW AWAY THE
                                       CURRENT WKB CALCULATION
                                    goto 777         ! BACK TO THE
                                       DERIVATIVES SECTION
                     endif
456                  continue
!!!                  THE VALUES
                     Ux(i)=Un
                     Tx(i)=Tn
                     dTxN=fderiv4(Tx(i),Tx4,h,i)
                     dUxN=fderiv4(Ux(i),Ux4,h,i)
                     rho(1)=flux/ux(i)
!                    Previous is current
                     rho(0)=flux/ux(i-1)
                     rhop=rho(1)

!                    Shift of arrays for derivatives
```

```
        do  j=4,2,-1
                  Ux4(j)=Ux4(j-1)
                  Tx4(j)=Tx4(j-1)
        enddo
        Ux4(1)=Ux(i)
        Tx4(1)=Tx(i)
!       Previous is current
        etap=eta
        eta2p=eta2

!!!     WKB WAVE
        cevp=cev(nr)
        cex=cdexp(-dcmplx(Zero,One)*cA/2.d0)
        do  j=0,n_d
                  cpsi(j,nr,np)=cpsi(j,nr,np)*cex
        enddo
!       Previous is current
        cAp=cA
        pgp=pg
        qrp=qr
        frlnp=frln

!!!     IDEAL FLUXES
        qidwnd = fqidwind( Ux(i), flux, Tx(i), One, qa,
           qp )
        pidwnd = fpidwind( rho(1), Ux(i), Tx(i), One, pa
           , pp )
        qidwnd0 = fqidwind( Ux(i-1), flux, Tx(i-1), One,
            qa0, qp0 )
        pidwnd0 = fpidwind( rho(0), Ux(i-1), Tx(i-1),
           One, pa0, pp0 )

!!!     WAVE ENERGY DERIVATIVE
        dqwave=fderiv2(qwave,qwavep,h)
        qwavep=qwave      ! Previous is current

!!!     TOTAL ENERGY AND MOMENTUM
        qtot = qwave + qdisswnd + qidwnd + qg + qrl
        ptot = pwave + pdisswnd + pa + pp + pg
        qold = qwave + qdisswnd + qidwnd0
        pold = pwave + pdisswnd + pidwnd0
```

```
! ! !                    TRANSITION  REGION
                         if  (  (dTxN/Tx(i)>dlnT_max)  .and. (i>10)  )
                            dlnT_max=dTxN/Tx(i)

! ! !                    BETA
                         beta=rho(1)*Tx(i)/rMav/(B02/(2.d0*rmu0))

! ! !                    HALL  MAGNETIC  FIELD
                         Bh=(qe*eta2)/(eps0*vc**2*rMp)

! ! !                    WRITE  TO  FILES
                         rMach=ux(i)/dsqrt(gama*Tx(i)/rMav)         ! Mach
                            number
                         write(91,*)  xi(i),  dreal(cpsi(0,nr,np)),  dreal(
                            cpsi(1,nr,np))/va,    dreal(cpsi(2,nr,np))*va/
                            omega, &
                                         dreal(cpsi(3,nr,np))/omega, &
                                         aimag(cpsi(0,nr,np)),  aimag(
                                            cpsi(1,nr,np))/va,  aimag(
                                            cpsi(2,nr,np))*va/omega, &
                                         aimag(cpsi(3,nr,np))/omega
                         write(95,*)  xi(i),  va,  Ux(i),  Tx(i)/rkB,  rho(1),
                            eta,  rNum,  dTxN,  dUxN,  cdabs(cpsi(1,nr,np))
                            **2, &
                                         h*dTxN/Tx(i),  alpha,  kc,  oote,  rMach
                                            ,  beta,  eta2,  Pr,  re,  rL,  taupp,
                                            frl ,  qr/(rho(1)*Tx(i))*rMp
                         write(94,*)  xi(i),  ptot,  pwave,  pdisswnd,  pidwnd
                            ,  pa,  pp,  pg,  dabs(ptot-pold)/ptot
                         write(93,*)  xi(i),  -A,  qtot,  qu,  qb,  qub,  qdu,
                            qdb,  qdUx,  qdTx,  qa,  qp,  qg,  qrl,  dabs(qtot-
                            qold)/qtot,-dqwave/rho(1)
                         write(98,*)  xi(i),  dxf,dxg,dyf,dyg

! ! !                    STEP  CORRECTION
                         h=heps*Tn/dabs(dTxN)       ! Eq. (4.1)
                         if  (  h<hmin  )  h=hmin
                         if  (  h>hmax  )  h=hmax
                         i=i+1
             enddo ! next i
222          Na=i-1  ! last iteration
!            Close the files
```

```fortran
        close(91)
        close(93)
        close(95)
        close(94)
        close(96)
        close(97)
        close(98)
        call system_clock(i_fin,i_rate) ! Stop the clock
!!!     END

        beta=rho(1)*Tx(Na)/rMav/(B02/(2.d0*rmu0))        ! beta
        omega_m=5.d0*(B02/(2.d0*rmu0))*e2**2*rL/dsqrt(rMp*Tx(Na)
     &    **5) ! omega_max Eq. (2.90)
!       Output to terminal
        print*,'h_=',h,qwave,pwave,'beta_=',beta,flux,omega_m
        write(*,*) 'T/T0_=',Tx(Na)/T0, 'U/U0_=',Ux(Na)/u0, rho0/
     &    rho(1)
        write(*,*) 'Calculation_finished_in',(i_fin-i_start)/
     &    real(i_rate),'seconds!'
        write(*,*) 'max(dlnT/dx)_=',dlnT_max,'________TR_
     &    length_=',One/dlnT_max*1.d-03,'_____Mach_=',Ux(Na)/
     &    dsqrt(gama*Tx(Na)/rMav)
        write(*,*) 'kappa/sigma*qe^2/Te_=',rkap/sigma*qe**2/Tx(
     &    Na),'____eta/rkap/sqrt(Me*Mp)*9/4_=',eta/rkap/dsqrt(
     &    rMe*rMp)*9./4.

!!!     QUADRATIC EQUATION CHECK WITH DIMENSIONAL AND
!       DIMENSIONLESS VARIABLES
!       wq=flux*( 0.5d0*( Ux(Na)**2-u0**2 ) + rcp*( Tx(Na)-T0 )/
!    rMav )
        wq=( ( V**2-One )/Two+rcp*(theta-theta0) )*flux*Ux(0)
     &    **2-flux*gSun*xi(Na)/q0
        wp=flux*( Ux(Na) - u0 + (Tx(Na)/Ux(Na) - T0/u0)/rMav )/
     &    pw0+pgp/pw0
!       wp=( ( V - One ) + (theta/V-theta0) )*flux*Ux(0)/pw0
!       Izglejda gcc version 4.6.3 predpochita da umnojava po
!    rMav vmesto da deli

!!!     ENERGY CONSERVATION !!!
        print*,'_W_=_',w
        print*,'Wq_=_',wq,'Diff_=_',dabs(w-wq)/w
!!!     MOMENTUM CONSERVATION !!!
```

```fortran
      print *, '_W_=_',w
      print *,'Wp_=_',wp,'Diff_=_',dabs(w-wp)/w

      contains

!     Eigenvector normalisation subroutine
      subroutine snorm(cpsi, n_d, flux, b02, ux, rho, eta,
     rNum)
            implicit integer (i-n)
            implicit complex(8) (c), real(8) (a-b,d-h,o-z)

            dimension cpsi(0:n_d,0:n_d), cev(0:n_d)

            do ic=0,n_d
                  scal = cfgdot( cpsi(0,ic), cpsi(1,ic),
                     cpsi(2,ic), cpsi(3,ic), &
                              cpsi(0,ic), cpsi(1,ic),
                                 cpsi(2,ic), cpsi(3,ic)
                                 , &
                              flux, b02, ux, rho, eta,
                                 rNum )
                  if (scal<Zero) scal=scal*(-One)
                  rNorm = dsqrt( One/scal )

                  do jc=0,n_d
                        cpsi(jc,ic) = cpsi(jc,ic)*rNorm
                  enddo
            enddo
      end subroutine snorm

!     Quadratic equation solution subroutine
      subroutine squadr(hi, tau, s2, theta0, V, i_err, Discr)
            implicit real(8) (a-b,d-h,o-z)

      Discr = (s2-One)**2 - 2.d0*hi*(gama**2-One)+gama
         *tau+( gama*tau+2.d0*(gama+s2) ) ! Eq. (2.64)
      if (Discr>=Zero) then
            V=( (gama-One)*(2.d0*hi+One) + 2.d0*s2 )
               /(gama+s2+gama*tau+dsqrt(Discr))
                        ! Eq. (2.66)
            i_err=0
      else
```

```fortran
                              print *,'Discr␣=',Discr, '␣␣␣␣␣␣␣hi␣=',hi
                                 ,'␣␣␣␣␣␣tau ',tau ,'␣␣␣␣␣␣s2␣=',s2
                              i_err=1              ! Discr<0
                    endif
              end subroutine

!        Subroutine for solution of the MHD equations
         subroutine VT(x,Un,Tn,Uv,Tv,Upa,Tpa,rho,va2,alpha,
              rksidxU,rkapdxT,cA,A,i_vt,iv,frl)
                    implicit integer (i-n)
                    implicit complex(8) (c), real(8) (a-b,d-h,o-z)

                    dimension cpsi(0:n_d,0:n_d,0:nlim+1), cev(0:n_d)
                    dimension cMa(0:n_d,0:n_d)
                    dimension Upa(1:4), Tpa(1:4)      ! Wind and
                        temperature derivatives

                    common /start/   cevp,u0,T0,B0,flux,h,cs,rhop,
                        frlnp,rrl,nr
                    common /waves/   qwave0,q0,qdisswnd0,pwave0,p0,
                        pdisswnd0
                    common /kin/     eta,rNum,rkap,rksi,zeta,rNuk,
                        sigma,va,eta2,rNuk2,eta2p
                    common /ev/      cpsi,cev
                    common /derivs/  deta,drnum,drkap,drksi,deta2
                    common /energy/ qdUx,qdTx,qu,qb,qub,qdu,qdb,qu0,
                        qb0,qub0,qdu0,qdb0,qg,pg,pgp,qr,qrl,frln

                    i_vt=0
                    do j=0,n_d
                         do k=0,n_d
                              cMa(j,k) = Czero
                         end do
                    end do

!!!              KINEMATICS
                    call statistics(rho, Tv, B0, vA2, P, alpha, rD,
                        re, rL, sigma, rNum, eta, rnuk, Pr, rkap,
                        oote, eta2, rNuk2, taupp)
                    va=dsqrt(va2)
                    rksi=4.d0*eta/3.d0+zeta
```

```
! ! !            DERIVATIVES
                 dTx=fderiv4(Tv,Tpa,h,iv)          ! d_x T
                 dUx=fderiv4(Uv,Upa,h,iv)          ! d_x U

                 deta=dsqrt(rMp*Tv)/(rL*e2**2)*Tv*dTx ! d_x eta
                 deta2=fderiv2(eta2,eta2p,h)         ! d_x eta2
                 drnum=−2.5d0*vc**2/(4.d0*pi)*e2*rL*dTx/(Tv**2*
                     dsqrt(Tv/rMe)) ! d_x num
                 drkap=4.5d0*Tv*dsqrt(Tv/rMe)/(2.d0*rL*e2**2)*dTx
                     ! d_x kappa
                 drksi=4.d0*deta/3.d0+dzeta ! d_x ksi
                 rksidxU=rksi*dUx          ! ksi*d_x U
                 rkapdxT=rkap*dTx          ! kappa*d_x T

! ! !            EIGENVECTORS AND EIGENVALUES
                 call minit(cMa, n_d, cs, va2, Uv, rNuk2, rNum,
                     dUx, rho, deta2, drnum )
                 call seigen(cMa, n_d, cev, cpsi(0,:,np), cpsi
                     (1,:,np), cpsi(2,:,np), cpsi(3,:,np), nr )
                 call snorm(cpsi(:,:,np),n_d,flux,b02,Uv,rho,eta2
                     ,rNum)

! ! !            WKB
                 cA=cA−h*(cev(nr)+cevp)
                 if ( dabs(aimag(cA)) > dlim ) then
                         i_vt=2
                         return
                 endif
                 A=aimag(cA)      ! k''(R)

! ! !            FLUXES
                 qidwv = fqidwave( cpsi(0,nr,np), cpsi(1,nr,np),
                     b02, Uv, flux, qu, qb, qub )
                 qdisswv = fqdisswave( cpsi(0,nr,np), cpsi(1,nr,
                     np), cpsi(2,nr,np), cpsi(3,nr,np), b02, eta2,
                     rNum, qdu, qdb )
                 qu=w*qu*dexp(A)          ! q_u
                 qb=w*qb*dexp(A)          ! q_b
                 qub=w*qub*dexp(A)        ! q_ub
                 qdu=w*qdu*dexp(A)        ! q_eta
                 qdb=w*qdb*dexp(A)        ! q_varrho
                 qwave = (qu+qb+qub+qdu+qdb)
```

```
                    ! q_wave Eq. (2.45)
    pwave = w*fpwave(cpsi(0,nr,np), b02)*dexp(A)
                    ! Pi_wave Eq. (2.52)
    qdisswnd = fqdisswind( Uv, dUx, rksi, dTx, rkap,
        qdUx, qdTx)
    pdisswnd = fpdisswind( dUx, rksi )
    qg=flux*gSun*x              ! q_g Eq. (2.75)
    pg=pgp+gSun*(rhop+rho)*h/Two     ! Pi_g Eq.
        (2.80)

!!!              RADIATIVE LOSS RATE
    Tk=dlog10(Tv/rKb)
    rn=rho/rMp
    frl=frloss(Tk)             ! mathcal{P} Fig. (2.1)
    frln=frl*rn**2/Uv
    qr=qrp+(frlnp+frln)*h/Two
    qrl=qr*Uv*rrl              ! q_r Eq. (2.85)

!!!              QUADRATIC EQUATION
    hi=(qwave0-qwave)/q0+(qdisswnd0-qdisswnd)/q0-qg/
        q0-qrl/q0              ! Eq. (2.59)
    tau=(pwave0-pwave)/p0+(pdisswnd0-pdisswnd)/p0-pg
        /p0                    ! Eq. (2.59)
    call squadr(hi, tau, s2, theta0, V, i_vt, discr
        )
    if (i_vt==1) then
            print*,'Stopping... ⎵⎵⎵⎵at⎵x⎵=',x,'⎵⎵⎵⎵⎵
                Mach⎵=',Uv/dsqrt(gama*Tv/rMav)
            print*,'T/T0=',Tv/T0,'⎵U/U0=',Uv/u0,'⎵
                dU⎵=',dUx,'⎵dT⎵=',dTx
            return
    endif
    Un= V*u0            ! Eq. (2.66)
    Tn= rMav*Un*( u0*(One+theta0+tau) - Un )
        ! Eq. (2.67)
    end subroutine VT

!       Subroutine VT for Quasi-Newton method
    subroutine VTd(x,Un,Tn,Uv,Tv,Upa,Tpa,rho,cAd,Ad,i_vt,iv,
        pgd,qrd)
                implicit integer (i-n)
                implicit complex(8) (c), real(8) (a-b,d-h,o-z)
```

```
          dimension cpsid(0:n_d,0:n_d), cevd(0:n_d), cev
             (0:n_d)
          dimension cMad(0:n_d,0:n_d)
          dimension Upa(1:4), Tpa(1:4)      ! Wind and
             temperature derivatives

          common /start/  cevp,u0,T0,B0,flux,h,cs,rhop,
             frlnp,rrl,nr
          common /waves/  qwave0,q0,qdisswnd0,pwave0,p0,
             pdisswnd0

          i_vt=0
          do j=0,n_d
                  do k=0,n_d
                          cMad(j,k) = Czero
                  end do
          end do

!!!       KINEMATICS
          call statistics(rho, Tv, B0, vA2d, Pd, alphad,
             rDd, red, rLd, sigmad, rNumd, etad, rNukd,
             Prd, rkapd, ooted, eta2d, rNuk2d, &
                            tauppd)
          zetad=Zero
          rksid=4.d0*etad/3.d0+zetad

!!!       DERIVATIVES
          dTxd=fderiv4(Tv,Tpa,h,iv)
          dUxd=fderiv4(Uv,Upa,h,iv)
          detad=dsqrt(rMp*Tv)/(rLd*e2**2)*Tv*dTxd
          deta2d=fderiv2(eta2d,eta2p,h)
          drnumd=-2.5d0*vc**2/(4.d0*pi)*e2*rLd*dTxd/(Tv
             **2*dsqrt(Tv/rMe))
          drkapd=4.5d0*Tv*dsqrt(Tv/rMe)/(2.d0*rLd*e2**2)*
             dTxd
          drksid=4.d0*detad/3.d0+dzetad
          rksidxUd=rksid*dUxd
          rkapdxTd=rkapd*dTxd

!!!       EIGENVECTORS AND EIGENVALUES
          call minit(cMad, n_d, cs, va2d, Uv, rNuk2d,
```

```
                    rNumd , dUxd , rho , deta2d , drnumd )
          call seigen(cMad, n_d, cevd, cpsid(0,:), cpsid
              (1,:), cpsid(2,:), cpsid(3,:), nrd )
          call snorm(cpsid , n_d , flux ,b02,Uv,rho , eta2d ,rNumd
              )

!!!          WKB
          cAd=cAd−h∗(cevd(nrd)+cevp)
          if ( dabs(aimag(cAd)) > dlim ) then
                  i_vt=2
                      return
          endif
          Ad=aimag(cAd)

!!!          FLUXES
          qidwvd = fqidwave( cpsid(0,nr), cpsid(1,nr), b02
              , Uv, flux , qud , qbd , qubd )
          qdisswvd = fqdisswave( cpsid(0,nr), cpsid(1,nr),
                cpsid(2,nr), cpsid(3,nr), b02 , eta2d , rNumd ,
                qdud , qdbd )
          qud=w∗qud∗dexp(Ad)
          qbd=w∗qbd∗dexp(Ad)
          qubd=w∗qubd∗dexp(Ad)
          qdud=w∗qdud∗dexp(Ad)
          qdbd=w∗qdbd∗dexp(Ad)
          qwaved = (qud+qbd+qubd+qdud+qdbd)
          pwaved = w∗fpwave( cpsid(0,nr), b02)∗dexp(Ad)
          qdisswndd = fqdisswind( Uv, dUxd, rksid , dTxd ,
                rkapd , qdUxd , qdTxd)
          pdisswndd = fpdisswind( dUxd, rksid )
          qgd=flux∗gSun∗x
          pgd=pgd+gSun∗(rhop+rho)∗h/Two

!!!          RADIATIVE LOSS RATE
          Tk=dlog10(Tv/rKb)
          rn=rho/rMp
          frld=frloss(Tk)
          frlnd=frld∗rn∗∗2/Uv
          qrd=qrd+(frlnp+frlnd)∗h/Two
          qrld=qrd∗Uv∗rrl

!!!          QUADRATIC EQUATION
```

```fortran
            hid=(qwave0−qwaved)/q0+(qdisswnd0−qdisswndd)/q0−
                qgd/q0−qrld/q0
            taud=(pwave0−pwaved)/p0+(pdisswnd0−pdisswndd)/p0
                −pgd/p0
            call squadr(hid, taud, s2, theta0, Vd, i_vt,
                discr)
            if (i_vt==1) then
                    print *,'QN Stopping ... Mach=',Uv/dsqrt
                        (gama*Tv/rMav)
                    print *,'T/T0=',Tv/T0,' U/U0=',Uv/u0,'
                        dU=',dUxd,' dT=',dTxd
                    return
            endif

            Un= Vd*u0
            Tn= rMav*Un*( u0*(One+theta0+taud) − Un )
      end subroutine VTd

!       Function for derivatives calculation
      function fderiv2(fn,fp,h)
              implicit real(8) (a−b,d−h,o−z)

              fderiv2=(fn−fp)/h
      end function fderiv2

      function fderiv4(fn,fpa,h,iv)
              implicit integer (i−n)
              implicit real(8) (a−b,d−h,o−z)

              dimension fpa(1:4)

              fderiv4=(fn−fpa(1))/h
      end function fderiv4

      end program
```